\documentclass{aa}  

\usepackage{natbib}
\bibpunct{(}{)}{;}{a}{}{,} % to follow the A&A style
\usepackage{hyperref}
\usepackage{graphicx}
\usepackage{scrextend}
\usepackage{appendix}
\usepackage{float}
\usepackage{siunitx}
\usepackage{bm}

\usepackage{txfonts}
\usepackage{url}
\usepackage[shortlabels]{enumitem}
\usepackage{color}

\makeatletter
\renewcommand\@biblabel[1]{}
\makeatother
\begin{document} 

  \title{A comparative analysis of denoising algorithms for extragalactic imaging surveys}
  \author{V. Roscani
          \inst{1}
          \and
          S. Tozza
          \inst{3}
          \and
          M. Castellano
          \inst{1}
          \and
		  E. Merlin
		  \inst{1}
          \and
          D. Ottaviani
          \inst{1}
          \and
          M. Falcone
          \inst{2}
          \and
          A. Fontana
          \inst{1}
          }

   \institute{INAF - Osservatorio Astronomico di Roma,
              Via Frascati 33, I-00040, Monteporzio, Italy.        \email{valerio.roscani@inaf.it}
         \and
             Department of Mathematics, Sapienza University of Rome,
             P. le Aldo Moro, 5, 00185 Rome, Italy            
         \and
             Department of Mathematics and Applications “Renato Caccioppoli”, University of Naples Federico II, Via Cintia, Monte S. Angelo, I-80126 Naples, Italy
             }

   \date{ }%Received September 15, 1996; accepted March 16, 1997}

 \abstract{}
 {We present a comprehensive analysis of the performance of noise-reduction (``denoising'') algorithms to determine whether they provide advantages in source detection, mitigating noise on extragalactic survey images.
 
}
 {The methods under analysis are representative of different algorithmic families: Perona-Malik filtering, Bilateral filter, Total Variation denoising, Structure-texture image decomposition, Non-local means, Wavelets, and Block-matching. 
 We tested the algorithms on simulated images of extragalactic fields with resolution and depth typical of the Hubble, Spitzer, and Euclid Space Telescopes, and of ground-based instruments. 
 After choosing their best internal parameters configuration, we assess their performance as a function of resolution, 
background level and image type, also testing their ability to preserve the objects fluxes and shapes. Finally, we analyze in terms of \textit{completeness} and \textit{purity} the catalogs extracted after applying denoising algorithms on a simulated Euclid Wide Survey VIS image, on real H160 (HST) and K-band (HAWK-I) observations of the CANDELS GOODS-South field.}
 {
 Denoising algorithms often outperform the standard approach of filtering with the Point Spread Function (PSF) of the image. Applying Structure-Texture image decomposition, Perona-Malik filtering, the Total Variation method by Chambolle, and Bilateral filtering on the Euclid-VIS image, we obtain catalogs that are both more pure and complete by 0.2 magnitudes than those based on the standard approach. The same result is achieved with the Structure-Texture image decomposition algorithm applied on the H160 image. The relative advantage of denoising techniques with respect to PSF filtering increases at increasing depth. 
 
 Moreover, these techniques better preserve the shape of the detected objects with respect to PSF smoothing.}
 {Denoising algorithms provide significant improvements in the detection of faint objects and 
 enhance the scientific return of current and future extragalactic surveys. 
 We identify the most promising denoising algorithms among the $20$ considered techniques.
  
 }

   \keywords{Techniques: image processing -- Methods: numerical, data analysis  -- Surveys}

   \maketitle
%
%________________________________________________________________

\section{Introduction}\label{sec:intro}

Measuring  the  amount  of  photons  that  we  receive  from  astronomical sources over a given range of wavelengths is the primary way to gather information about the Universe. From the advent of digital photography in the 1980's, charge coupled device (CCD) imaging is one of the primary ways by which we do so. Currently, CCD devices can reach 100 million pixels, with read noise as low as one electron, almost 100$\%$ quantum efficiency, and sensitivity from the X-rays to the near-infrared \citep[see e.g.][for a review]{Lesser2015}. 

Before being ready for the extraction of meaningful scientific content, astronomical images must be processed to, for instance, combine different observations into a single mosaic, correct for flat-field, transients, artifacts, and defects, subtract a global or local background, etc. Once these preparatory steps are completed, the quality of the image mainly depends on its resolution capability (which is proportional to $\lambda/D$, the ratio between the observed wavelength and the diameter of the telescope, in the case of diffraction-limited instruments, e.g. space observatories; or from the atmospheric seeing for ground-based facilities), and on its depth (i.e. magnitude at a given reference signal-to-noise ratio (SNR), which mainly depends on the duration of the observations (exposure time). Since increasing the latter is often unfeasible or too demanding, searching for alternative methods to increase the SNR is important. A possible solution can be the application of noise reduction (``denoising'') techniques. 

Wavelet transforms are a standard and popular tool used for denoising and detection of sources on astronomical images. The technique is extremely versatile, as it can be applied on a wide range of scientific cases (e.g. X-ray images: XMM-LSS survey \citep{xmm} and Fermi catalog \citep{fermi10gev,fermi100mev}, Cosmic Microwave Background maps \citep{starck2004}, N-body simulations \citep{romeo2003}, etc.). Other interesting applications are summarized in \cite{starck2006}. 
Among the different possible implementations, 
the widely used is the so-called "\textit{à trous}" algorithm \citep{atrous}. This algorithm is an undecimated wavelet transform (UWT), which is also isotropic, and because of this, it is very efficient for the detection of isotropic objects. It is therefore clear the reason of its popularity for astronomical image processing, where many objects are nearly isotropic (e.g. stars, galaxies, galaxy clusters) \citep{starlet}.

For what concerns extragalactic optical/near-infrared imaging, 
images are typically convolved with a PSF shaped kernel to enhance source detection  \citep[an application of the lemma by][]{neyman}; this is the most standard example of a denoising algorithm, since filtering reduces the noise variance, allowing real sources to raise above the background. In many familiar cases, the typical PSFs of telescopes are quite similar to 2D Gaussians, making the PSF filtering basically indistinguishable from a Gaussian filtering. However, in several non-astronomical applications of image analysis this approach is often outclassed by other, more refined methods, designed to be more efficient and to better preserve the borders and edges of the sources. 

A special mention is required for machine learning and deep learning techniques, lately becoming popular in the image processing field and in astronomy. An interesting application is proposed by \cite{cdl}, where a sparse dictionary learning algorithm has been tested on multiple astronomical images. On the other hand, autoencoder neural networks for image denoising seem to be very promising \citep{vincent2008,Vincent2010,Junyuan2012}. Some applications of autoencoders on different areas of astronomy can be found in the literature, even if they are primarily used for other purposes (e.g. spectral energy distribution recovery \citep{frontera2017}, gravitational waves denoising \citep{shen2017}, stellar cluster detection \citep{karmakar2018}).  
Although these techniques are undoubtedly highly appealing, 
the aim of this work is to perform a comprehensive 
comparison of "traditional" denoising algorithms based on non linear partial differential equations (PDEs) and variational methods. The comparison with other approaches based on machine learning will be developed in future works.

We compared several classes of denoising techniques, in order to find which ones yield the best improvements in source detection. To this aim, we have performed an extended set of tests. We considered many different noise reduction algorithms, roughly belonging to the following families: Perona-Malik (PM) filtering, Bilateral filter, Total Variation (TV) denoising, Structure-texture image decomposition, Block-matching, Non-local means, and Wavelets. 
Note that the numerical methods employed ranges from variational methods to PDEs-based techniques, also including some statistical methods.

We tested them using two different datasets. First, we focused on simulated images, created by state-of-the-art codes and prescriptions in order to mimic different realistic cases. This simplified environment has the advantages to allow a detailed analysis of the results, since the ``truth'' is perfectly known. 
For real images, we applied the algorithms giving the best results obtained on the simulated dataset to check if the improvement is confirmed. 
To our knowledge, this is the first attempt to extensively  compare a large number of denoising algorithms in an  astrophysical context.
In general, the performance of any of these methods depends on the kind of noise that affects the image. Here we are mainly interested in extragalactic imaging, and in particular we focus on the next generation of optical - near-infrared instruments and surveys such as Euclid \citep[][]{euclid}, LSST \citep{lsst}, DES \citep{des}, and WFIRST \citep{wfirst}. 
The paper is organized as follows: in Sect. \ref{sec:denoising-methods} we list and briefly describe all the denoising methods used in our tests, providing mathematical formulations and code information. In Sect. \ref{images} we present our datasets. In Sect.  \ref{subsec:quality_tests} we describe and discuss all the tests we carried out, showing the results we obtained in Sect. \ref{sec:results}. In Sect. \ref{sec:realimages} we apply the methods on real images from space and ground-based. Finally, Sect. \ref{conclusions} summarizes the main results, discussing also the possible future applications. Throughout this paper we adopt the AB magnitude system \citep{Oke83} and a $\Lambda CDM$ cosmology with $\Omega_m=0.3$, $\Omega_\Lambda=0.7$, $H_0=70~Km s^{-1} Mpc^{-1}$.

%__________________________________________________________________

%%%%%%%%%%%%%%%%%%%%%%%%%%%%%%%%%%%%%%%%%%%%
\section{Denoising techniques}\label{sec:denoising-methods}
The focus of this paper is on the comparison of different techniques proposed in the literature, when they are applied on astronomical images. As we said in the introduction,  these images have specific features. 
So an efficient denoising method is crucial to extract the information contained in the image and could be used as a preliminary step for other image processing problems, like the image segmentation and/or deblurring. 

In order to select the most efficient denoising approach for extragalactic survey images we have analyzed 
different classes of methods in order to cover the main families of noise reduction techniques, namely non-linear filtering (\ref{subsec:PM}), bilateral filter (\ref{subsec:bilateral}), TV denoising (\ref{subsec:TV}), image decomposition (\ref{subsec:str-tex-dec}), wavelets (\ref{subsec:wavelets}) and non-local means (\ref{subsec:NL}). \\
We briefly summarize the mathematical formulations of these techniques and we provide information on the codes for reproducibility. 

%%%%%%%%%%%%%%%%%%%%%%%%%%%%%%%%%%%%%%%%%%%%
\subsection{Gaussian smoothing}\label{subsec:Gaussian}
Let us consider the intensity function $I(x,y)$ of a noisy image, with $(x,y)\in \Omega$, where $\Omega \subset \mathbb{R}^2$ is the reconstruction domain. Let $I_{clean}$ be the desired clean image. 
An image with a Gaussian noise component is 
\begin{equation}
    I(x,y) = I_{clean}(x,y) + \eta
\end{equation}
where $\eta \sim N(\mu,\sigma)$ is the additive noise component.\\
Of course, we want to reconstruct $I_{clean}$ from $I$.\\

\noindent This filter uses a Gaussian function for calculating the transformation to apply to each pixel in the image. 
Mathematically, applying a Gaussian filter to an image corresponds to convolve the image with a Gaussian function. Since the Fourier transform of a Gaussian is another Gaussian, applying a Gaussian smoothing has the effect of reducing the image's high-frequency components; a Gaussian filter is then {\em a low-pass filter}. 
In two dimensions, it is the product of two Gaussian functions, one in each dimension, so that the low-pass Gaussian filter is
\begin{equation}\label{eq:gaussian-filter}
G_\sigma(x,y) := \frac{1}{2\pi\sigma^2}\exp^{-\frac{x^2 + y^2}{2\sigma^2}}
\end{equation}
where $x$ is the distance from the origin in the horizontal axis, $y$ is the distance from the origin in the vertical axis, and $\sigma$ is the standard deviation of the Gaussian distribution. \\
Filtering the image $I: \Omega \subset \mathbb{R}^2 \rightarrow \mathbb{R}$ with a "low-pass" Gaussian filter corresponds to process it with the heat equation \citep{Gabor65,LFB94}, that is solving the following linear PDE 
\begin{equation}\label{eq:Heat}
\begin{cases}
    \displaystyle\frac{\partial I}{\partial t}(x,y,t) = \nabla I(x,y,t) \quad&\forall(x,y,t)\in\Omega\times(0,T_C]\,,\\[1.5ex]
\displaystyle\frac{\partial I}{\partial\mathbf{\eta}}(x,y,t)=0\,, &\forall(x,y,t)\in\partial\Omega\times(0,T_C]\,,\\[1.5ex]
I(x,y,0)=I_0(x,y)\,,&\forall(x,y)\in\overline{\Omega}\,,
\end{cases}
\end{equation}
which has a diffusive effect on the initial datum $I_0$, for a small fixed time $T_C > 0$. 
The relation between the Gaussian filter \eqref{eq:gaussian-filter} and the problem \eqref{eq:Heat} is that the solution of the heat equation is a convolution with the Gaussian filter, i.e.
\begin{equation}
    I(x,y,t) = (G_\sigma(x,y) * I_0)(x,y)
\end{equation}
with $\sigma = \sqrt{2t}$.\\
It is well known that applying that filter does not preserve edges.
This edge blurring is due to the isotropic diffusion. 

We can get an improvement in three different ways:
\begin{enumerate}
    \item Modifying the heat equation (see Sect. \ref{subsec:PM})
    \item Making convolution "nonlinear" (see Sect. \ref{subsec:bilateral})
    \item Defining an optimization problem (see Sect. \ref{subsec:TV}).
\end{enumerate}

\noindent
{\em Code information}\\
In this paper we have used a simple Gaussian smoothing using a kernel that approximates a PSF of known Full Width at Half Maximum (FWHM) \citep{fwhm}, referring to it as "PSF", whereas with "Gaussian" we refer to the Gaussian filter with internal parameter $\sigma$. 
We made use of the "gaussian\_filter" routine implemented in the Python package Scipy\footnote{\url{https://docs.scipy.org/doc/scipy/reference/generated/scipy.ndimage.gaussian_filter.html}} \citep{scipy}, with $\sigma \approx \frac{FWHM_{pixel}}{2.355}$, easily obtained defining the Gaussian kernel radius $r = x^2 + y^2$,  where the kernel maximum is at $r = 0$ then $FWHM = 2\sqrt{2\ln{2\sigma}}$, see also \url{https://brainder.org/2011/08/20/gaussian-kernels-convert-fwhm-to-sigma/} for further details. 

%%%%%%%%%%%%%%%%%%%%%%%%%%%%%%%%%%%%%%%%%%%%
\subsection{Filters of Perona-Malik type}\label{subsec:PM}
An improvement of the simple Gaussian filter is obtained by modifying the heat equation. 
Following the PM model \citep{PM90}, we choose large values of $|\nabla I|$ as an indicator of the edge points of the image, in order to stop the diffusion at these  points. In this way we move from an isotropic to anisotropic diffusion as follows:
\begin{equation}\label{eq:PM}
\frac{\partial I}{\partial t} = div(\nabla I) \Rightarrow  \frac{\partial I}{\partial t} = div(g(|\nabla I|) \nabla I).
\end{equation}
The equation \eqref{eq:PM} must be complemented with  suitable boundary conditions (e.g. homogeneous Neumann boundary conditions) and an initial condition. 
Perona and Malik  pioneered the idea of anisotropic diffusion and proposed two functions for the diffusion coefficient (also called {\em edge-stopping functions}):
\begin{eqnarray}
g_1(|\nabla I|) &:= \displaystyle \frac{1}{1 + \Big(\frac{|\nabla I|}{K}\Big)^2}\\
g_2(|\nabla I|) &:= \exp\Big(-\Big(\frac{|\nabla I|}{K}\Big)^2\Big)
\end{eqnarray}
where $K$ is the gradient magnitude threshold parameter that decides the amount of diffusion to take place. \\
We also consider other three edge-stopping functions that have been proposed after the original work by Perona and Malik: \\
\citet{BSMH98} proposed an edge stopping function called Tukey's biweight function defined as:
\begin{equation}\label{def:c3}
g_3(|\nabla I|) := 
\left\{ \begin{array}{ll}
\frac{1}{2} \Big [ 1 - \Big(\frac{|\nabla I|}{K\sqrt{2}}\Big)^2  \Big ]^2  & \textrm{if $|\nabla I| \leq K\sqrt{2}$} \\
0 & \textrm{otherwise.}
\end{array} \right.
\end{equation}
\citet{GSZW12} proposed the following function:
\begin{equation} \label{def:c4}
g_4(|\nabla I|) := \frac{1}{1 + \Big(\frac{|\nabla I|}{K}\Big)^{\alpha(|\nabla I|)}}
\end{equation}
where
\begin{equation}
\alpha(|\nabla I|) := 2 - \frac{2}{1 + \Big(\frac{|\nabla I|}{K}\Big)^2}.
\end{equation}
\citet{Weickert98} proposed:
\begin{equation}\label{def:c5}
g_5(|\nabla I|) := 
\left\{ \begin{array}{ll}
1 - \exp(-3.31488*K^8 / (|\nabla I|)^8) & \textrm{if $|\nabla I| \neq 0 $} \\
1 & \textrm{otherwise.}
\end{array} \right.
\end{equation}

\noindent
{\em Code information}\\
This method has been implemented by us in C++ and it is available at:
\url{https://github.com/valerioroscani/perona-malik.git}. 

%%%%%%%%%%%%%%%%%%%%%%%%%%%%%%%%%%%%%%%%%%%%
\subsection{Bilateral filter} \label{subsec:bilateral}
The Bilateral filter is an edge-preserving denoising algorithm that was first introduced by \cite{TM98}. \\
It is defined as \citep[see also][]{BCCS12} 
\begin{equation}\label{eq:bilateral}
I(x) = \frac{1}{w} \sum_{x_i \in \Omega} I_0(x_i)f_r(\|I_0(x_i) - I_0(x)\|)g_s(\|x_i - x\|),
\end{equation}
where
\begin{equation}
w := \sum_{x_i \in \Omega}{f_r(\|I_0(x_i) - I_0(x)\|)g_s(\|x_i - x\|)}
\end{equation}
and
\begin{itemize}
    \item $I$ is the filtered image
    \item $I_0$ is the original input image to be filtered
    \item $x$ are the coordinates of the current pixel to be filtered
    \item $\Omega$ is the window centered in $x$, so $x_i \in \Omega$ is another pixel
    \item $f_r$ is the range kernel for smoothing differences in intensities (this function can be a Gaussian function)
    \item $g_s$ is the spatial (or domain) kernel for smoothing differences in coordinates (this function can be a Gaussian function).
\end{itemize}

It averages pixels based on their spatial closeness and on their radiometric similarity. Spatial closeness is measured by the Gaussian function of the Euclidean distance between two pixels and a certain standard deviation (sigma\_spatial). 
Radiometric similarity is measured by the Gaussian function of the Euclidean distance between two color values and a certain standard deviation (sigma\_color).\\

\noindent
{\em Code information}\\
We used the Python routine "denoise\_bilateral" available in the Python package \textsc{scikit-image}\footref{note-scikit}. 
We noticed that using our dataset, variations of the \textit{sigma\_spatial} were less effective than variations of \textit{sigma\_color}. We decided to set $\textit{sigma\_spatial}=3$ since it provides the best results.

%%%%%%%%%%%%%%%%%%%%%%%%%%%%%%%%%%%%%%%%%%%%
\subsection{Total Variation denoising}\label{subsec:TV}
Total-variation denoising (also known as total-variation regularization) is based on the principle that images with excessive and possibly spurious detail have high TV, defined as
\begin{equation}
    TV(u, \Omega) := \int_\Omega |\nabla u(x)| dx
\end{equation}
for a function $u \in C^1(\Omega)$  \citep[note that a similar definition can be given also for $L^1$ functions][]{KF57}. 
According to this principle, 
TV denoising tries to find an image with less TV under the constraint of being similar to the input image, which is controlled by the regularization parameter, i.e. tries to minimize $TV(I,\Omega)$.  
This minimization problem leads to the Euler-Lagrangian equation, which can be solved via the following evolutive problem:
\begin{equation}
    u_t = \frac{\partial}{\partial x} \Big( \frac{u_x}{\sqrt{u^2_x + u^2_y}}\Big) +     \frac{\partial}{\partial y} \Big(\frac{u_y}{\sqrt{u^2_x + u^2_y}}\Big) - \lambda (u-u_0), 
\end{equation}
for $t>0$ and $x,y \in \Omega$, with homogeneous Neumann boundary conditions and a given initial condition. 
TV denoising tends to produce “cartoon-like” images, that is, piecewise-constant images. The concept was pioneered by Rudin, Osher, and Fatemi in \citet{ROF92} and is today known as the ROF model. TV denoising is remarkably effective at simultaneously preserving edges whilst smoothing away noise in flat regions, even at low SNRs. \\

\noindent
{\em Code information}\\
We test the ROF method that was proposed by Chambolle in \citep{Chambolle2004} 
and the TV denoising using split-Bregman optimization \citep{GO09,Getreuer12,BushThesis11}. 
For the implementation of the two aforementioned methods we have used the Python routines "denoise\_tv\_chambolle" and "denoise\_tv\_bregman" belonging to the Python package \textsc{scikit-image} \footnote{\label{note-scikit}\url{https://scikit-image.org/docs/dev/api/skimage.restoration.html}} \citep{scikit}.

%%%%%%%%%%%%%%%%%%%%%%%%%%%%%%%%%%%%%%%%%%%%%
\subsection{Structure-texture image decomposition}\label{subsec:str-tex-dec}
A general approach to the denoising problem is based on the assumption that an image $I$ can be regarded as composed of a structural part $u$ (i.e. the objects in the image), and a textural part $v$ which corresponds to finest details plus the noise. 
Following the approach described in \citet{AGCO06}, such image decomposition technique is based on the minimization of a functional with two terms, one based on the total variation and a second one on a different norm adapted to the texture component. 
Given an image $I$ defined in a set $\Omega$, and let  $BV(\Omega)$ be the space of functions with limited total variation in $\Omega$ 
we can decompose $I$ into its two components by minimizing:
\begin{equation}
\inf \Big(\int_\Omega |\nabla u(x)| + \lambda \, ||v(x)||^p_X \, dx \Big)
\end{equation}
where $||\cdot||^p_X$ denotes the norm of a given space $X$ and the minimum is found among all functions $(u,v)\in BV(\Omega)\times X$ such that $u + v = I$. The parameter $p$ is a natural exponent, and $\lambda$ is the so-called splitting parameter which modifies the relative weights. The best decomposition is found at the $\lambda$ for which the correlation between $u$ and $v$ reaches a minimum. \\

%%%%%%%%%%%%%%%%%%%%%%%%%%%%%%%%%%%%%%%%%%%%
\noindent
{\em Code information}\\
In \citet{COFMPF15}, the authors proposed a C++ code named Astro-Total Variation Denoiser (ATVD), which implements three versions of the technique, based respectively on the $TV$-$L^2$ ($X = L^2(\Omega)$), $TV$-$L^1$ ($X = L^1(\Omega)$) and TVG \citep[$X$ being a Banach space as defined in][]{AGCO06} norms. 
Two thresholds are defined and used in the stopping criteria of the algorithms, called $\epsilon_{corr}$ and $\epsilon_{sol}$. \\
$\epsilon_{corr}$ defines the correlation algorithm  precision, whereas $\epsilon_{sol}$ defines the method precision (e.g. TVL2, TVG, TVL1). For all our tests, we will use $\epsilon_{corr} = 10^{-4}$ and $\epsilon_{sol} = 10^{-3}$, as suggested by the authors of %in
\citet{COFMPF15}. 

\begin{remark}

In the definition of the functional to be minimized one can also add a term taking into account some properties of the unknown image $f$ corresponding to the available image $g$. This is a typical situation when the physical image is modeled as  linear operator $A$  acting from a Hilbert space $X$ to a Hilbert space $Y$.  In this approach $X$ contains all the functions characterizing unknown objects and  $Y$ contains  the functions describing the corresponding measurable images, so that 
\begin{equation}
g=Af \, .
\end{equation}
A typical choice is to choose $X=Y=L^2(\Omega)$ and the inverse problem is then to minimize the functional
\begin{equation}
\int_\Omega \Vert Af-g\Vert_2^2 dx
\end{equation}
over $f\in X$. This problem is often ill-posed so a popular Tikhonov regularization is obtained adding another term $R(f)$ to the functional getting 
\begin{equation}
\int_\Omega \Vert Af-g\Vert_2^2 dx+ \mu R(f)
\end{equation}
where $\mu$ is a positive parameter to be tuned carefully. The term $R(f)$ can also be used to introduce a prior, e.g. the regularity of $f$ (based on Schwartz Theorem)  or the sparsity of $f$ (choosing $R(f)=\Vert f \Vert _1$). Imposing a morphological prior on the shapes, such has penalizing shapes different from ellipses, would require an enormous number of parameters in the case of astronomical images that usually include several sky objects and is a very challenging problem which goes beyond the scopes of this paper.
For the use of priors in other areas (e.g. in biomedical imaging), we refer the interested readers to \cite{BP06}, and for a general introduction to inverse problems in imaging to the book \cite{BB98}. 
\end{remark}

%%%%%%%%%%%%%%%%%%%%%%%%%%%%%%%%%%%%%%%%%%%%
\subsection{Wavelets} \label{subsec:wavelets}
The wavelets transform is the counterpart for images of the Fourier transform and the wavelets domain, which is a sparse representation of the image that can be thought of similarly to the frequency domain of the Fourier transform \citep{Valens99areally}. Being a  sparse representation means that most values are zero or near-zero and truly random noise is represented by many small values in the wavelet domain. Setting all values below some threshold to $0$ reduces the noise in the image, but larger thresholds also decrease the detail present in the image. \\
Let us recall the relation introduced in Sect. \ref{subsec:Gaussian}
\begin{equation}
    I = I_{clean} + \eta,
\end{equation}
where $\eta$ is the noise and $I_{clean}$ is the clean image (signal). The components of $\eta$ are independent and identically distributed (iid) as ${\cal N}(0,\sigma^2)$ and independent of $I_{clean}$. 
The goal is again to remove the noise obtaining an approximation $\widehat{I}$ of $I_{clean}$ minimizing the mean square error (MSE)
\begin{equation}
    \textit{MSE}(\widehat{I}) := \frac{1}{N} \sum_{j=1}^N (\widehat{I_j} - I_j)^2,
\end{equation}
where $N$ is the number of pixels. 
Let us denote by $Y={\cal{W}} I$ the matrix of wavelet coefficients of the image $I$ where $\cal{W}$ is the orthogonal wavelet transform operator, similarly $F = {\cal{W}} I_{clean}$ and $E = {\cal{W}} \eta$ (see 
\cite{VK95}, \cite{Mallat89} for more details on ${\cal{W}}$). 
The wavelet transform is based on the subbands (called details) at different scales usually indexed by $k\in {\cal K}, {\cal K} \subset \mathbb{N}$. 
The wavelet-thresholding method filters each coefficient $Y_j$ from the detail subbands $k\in {\cal K}$ with a threshold function to obtain $\widehat{X}$. 
The denoised approximation is $\widehat{I} = {\cal{W}}^{-1} \widehat{X}$, where ${\cal{W}}^{-1}$ is the inverse wavelet transform. 
Two thresholding techniques are frequently used. 
The \emph{soft-threshold} function 
\begin{equation}
    \varphi_T(x) := sgn(x) \max (|x|-T,0)
\end{equation}
which shrinks the argument $x$ to $0$ by the threshold $T$. 
The \emph{hard-threshold} function 
\begin{equation}
    \psi_T(x) := x \, {\bf 1}_{\{|x|>T\}}
\end{equation}
which sets the input to $0$ if is below (or equal) the threshold $T$. 
Note that the threshold procedure removes noise by thresholding \emph{only} the wavelet coefficients of the corresponding subbands, keeping the low resolution coefficients unaltered. \\

\noindent
{\em Code information}\\
We consider the two thresholding methods defined in the Python routine "denoise\_wavelet" \footref{note-scikit} \citep{CYV00,DJ94}, the first applies BayesShrink, which is an adaptive thresholding method that computes separate thresholds for each wavelet subband as described in \citet{CYV00}, the second is "VisuShrink", in which a single “universal threshold” is applied to all wavelet detail coefficients as described in \citet{DJ94}. This threshold is designed to remove all Gaussian noise at a given $\sigma$ with high probability, but tends to produce images that appear overly smooth.\\
In this work we decided to test several methods based on wavelet transforms. We selected the Meyer wavelet described in \citet{dmey} with VisuShrink thresholding method since, analyzing the application on our dataset, we found that it provides the best performance based on the analysis described in Sect.  \ref{subsec:quality_tests}. 
The list from which we took the Meyer wavelet can be found in \citet{pywavelets}. 
From now on, we will refer to this method as Orthogonal Wavelets. \\
We also consider other methods, based on multiscale wavelets decomposition (implemented in the C++ library called \textsc{sparse2D}, available at CosmoStat web page\footnote{\url{http://www.cosmostat.org/software/isap}}). The first is an isotropic UWT, based on the \textit{à trous} algorithm, better known as "Starlet transform", where 5 wavelets scales and an iterative hard thresholding technique are set. From now on, we will refer to this method as Starlet. The other two both use a biorthogonal UWT using a set of filters \citep[introduced for the JPEG 2000 compression standard][]{starck_wavelets} called "7/9 filters" and 5 wavelets scales. For the first method a 3$\sigma$ threshold is set, whereas for the second one a multi-resolution Wiener filter is performed. From now on, we will refer to these two methods as "b-UWT(7/9)" and "b-UWT(7/9)+Wiener", respectively. The optimal configurations for the methods implemented in \textsc{sparse2D} have been kindly suggested by the authors. We used the \texttt{mr\_filter} program with the following options:
\begin{itemize}
    \item Starlet: \texttt{t}=default, \texttt{f}=3, \texttt{n}=5 
    \item b-UWT(7/9): \texttt{t}=24, \texttt{f}=default, \texttt{n}=5 
    \item b-UWT(7/9)+Wiener: \texttt{t}=24, \texttt{f}=6, \texttt{n}=5
    \end{itemize}
\noindent where \texttt{t} is the type of multi-resolution transform, \texttt{f} is the type of filtering, and \texttt{n} is the number of scales. For further details about these 3 methods and the UWTs in general, see \cite{starck_wavelets}.

%%%%%%%%%%%%%%%%%%%%%%%%%%%%%%%%%%%%%%%%%%%%
\subsection{Non-local means} \label{subsec:NL}
The non-local means algorithm averages the value of a given pixel with values of other pixels in a limited proximity, under the condition that the patches centered on the other pixels are similar enough to the patch centered on the pixel of interest. 
This algorithm is defined by the formula \citep{BCM05}
\begin{equation}
    NL[I_{0}](x) = \frac{1}{C(x)} \int_\Omega \exp{(- g_\sigma^h(x))} \, I_{0}(y) \, dy
\end{equation}
where 
\begin{equation}
  g_{\sigma}^h(x) :=  \frac{G_\sigma * |I_{0}(x+.) - I_{0}(y+.)|^2) (0)}{h^2},
\end{equation}
$I_0$ is the original image, $x\in\Omega$, $C(x)$ is a normalizing constant, $G_\sigma$ is a Gaussian kernel with $\sigma$ denoting the standard deviation, and $h$ acts as a filtering parameter. 
The algorithm has been found to have excellent performance when used to denoise images with specific textures\footnote{\label{note-scikit_NL}\url{https://scikit-image.org/docs/dev/api/skimage.restoration.html##skimage.restoration.denoise_nl_means}}. 

We define by $size_{I}$ the image size in pixels, by $size_{p}$ the size of the patch in pixels, by $d_{p}$ the maximal distance in pixels where to search patches, by $n$ the image number of dimensions ($n=2,3$ depends if we consider 2D or 3D images). 
In its original version the computational complexity of the algorithm is proportional to:  $size_{I} * (size_{p}*d_{p})^{n}$ \citep{BCM05}. A new "fast" version is now preferentially used since its actual complexity is proportional to: $size_{I} * d_{p}^{n}$ \citep{DCCOJ08}.

Compared to the classic algorithm, in the fast mode the distances are computed in a coarser way, indeed all pixels of a patch contribute to the distance to another patch with the same weight, no matter their distance to the center of the patch. This approach can result in a slightly poorer denoising performance.

When the standard deviation $\sigma$ is given, the method gives a  more robust computation of patch weights. 
A moderate improvement to denoising performance can be obtained subtracting the known noise variance from the computed patch distances, that improves the estimates of patch similarity \citep{BCM11}.\\

\noindent{\em Code information}\\
In this work both the fast and slow version of the algorithm are tested. After a first selection of patch sizes and distances,  through the analysis described in Sect.  \ref{subsec:quality_tests}, we decided to set $size_{p}=5$ and $d_{p}=6$. 
For our numerical tests we used the routine "denoise\_nl\_means"\footref{note-scikit_NL}, implemented in the Python package \textsc{Scikit-image}. 

\subsection{Block-matching and 3D filtering}\label{bm3d}
"Block-matching and 3D filtering" (BM3D) is a 3 dimensional block-matching algorithm which "groups" similar 2D fragments with a \textit{Matching} method in the image. The \textit{Matching} method finds similar fragments to a \textit{reference} one, grouping fragments closer than a defined threshold. The matched fragments are then stored in 3D arrays called \textit{groups}. 
A "collaborative filtering" is performed to each group, 
which consists in a 3D linear transform, a shrink to reduce noise and an invert linear transform which produces 2D estimates of all the fragments. Once the estimates are obtained, they are aggregated to form an estimate of the whole image. For further details see \cite{bm3d_paper}. \\

\noindent
{\em Code information}\\
In this work we tested a C++ code of BM3D available at \url{https://github.com/gfacciol/bm3d}. For performance improvements and memory reasons, the input images have been cut in smaller overlapping tiles, re-aggregated in a single output image at the end of the process, as suggested by the authors.

\section{Test dataset}\label{images}
We first test the denoising algorithms on five different simulated images (Table \ref{table:table_images}), chosen as to reproduce the properties of a wide range of typical cases in terms of resolution, depth, pixel scale and wavelength:
\begin{itemize}
\item VIS: Euclid satellite visual band (wavelength: 550-900 nm)
\item NIR H: Euclid satellite near-infrared H band (wavelength: 1372-2000 nm)
\item EXT G: ground-based optical filter
\item H160: Hubble Space Telescope (HST) near-infrared  F160W band  \citep[e.g. CANDELS-wide][]{Guo2013}
\item IRAC: Irac-Spitzer 3.6$\mu$m channel. 
\end{itemize}

From now on, we refer to the simulated images, provided as input to the algorithms, as "Original", whereas we refer to the simulated images representing the true sky, without noise included, as "Noiseless". 
\begin{center}
\textbf{TEST IMAGES}
\begin{table}[ht]
\begin{tabular}{lllll}
\hline \hline
Filter     & PSF-FWHM & Pixel Scale & Mag Lim$^{(a)}$ \\ \hline
&arcsec&arcsec&\\ \hline 
VIS      & 0.2  & 0.1         & 25.25           \\ 
NIR H    & 0.3  & 0.3         & 24.01          \\ 
EXT G    & 0.8  & 0.2        & 25.93           \\ 
H160      & 0.15 & 0.06         & 27.23           \\ 
IRAC     & 1.6  & 0.6         & 25.40          \\ 
\hline 
HUDF (H160) & 0.15     &   0.06           &   $28.16/29.74^{(b)}$   \\
Ks (HAWK-I) &  0.4    &    0.1           &   $24.45/26.3^{(c)}$\\ \hline \hline
\end{tabular}
\caption{\label{table:table_images} $^{(a)}$: SNR=5;$^{(b)}$: limiting magnitude at the CANDELS and at the full HUDF depth, respectively; $^{(c)}$: images from \cite{ks_shallow}, and from the HUGS survey \citep{ks_deep}, respectively.}
\end{table}
\end{center}
VIS and NIR H reproduce the expected features of the visual and near-infrared bands in the forthcoming ESA satellite Euclid \citep{euclid}, and EXT G simulates a typical ground-based complementary optical observation for the Euclid Wide Survey. H160 is modeled after the detection band in recent deep surveys such as CANDELS \citep{Grogin2011,Koekemoer2011} and 3D-HST \citep{Skelton2014}, whereas IRAC simulates the features of the \textit{Spitzer} Channel 1 band in the CANDELS GOODS-South field \citep{Guo2013}.

The images have been simulated with \textit{SkyMaker} \citep{skymaker} on the basis of source catalogs generated by the Empirical Galaxy Generator (EGG) \citep{EGG} and they have been perturbed by  Gaussian noise in order to reach the limiting magnitudes reported in Table  \ref{table:table_images}. 
All the PSFs  are Gaussian except for the IRAC case where a real IRAC 3.6$\mu$m channel PSF has been used. The H160 and HAWK-I images are real observations whose tests are described in Sect. \ref{subsec:realimagesspace}-\ref{subsec:realgroundbased}.

\begin{figure*}[h!]
        \centering
        \includegraphics[width=17cm]{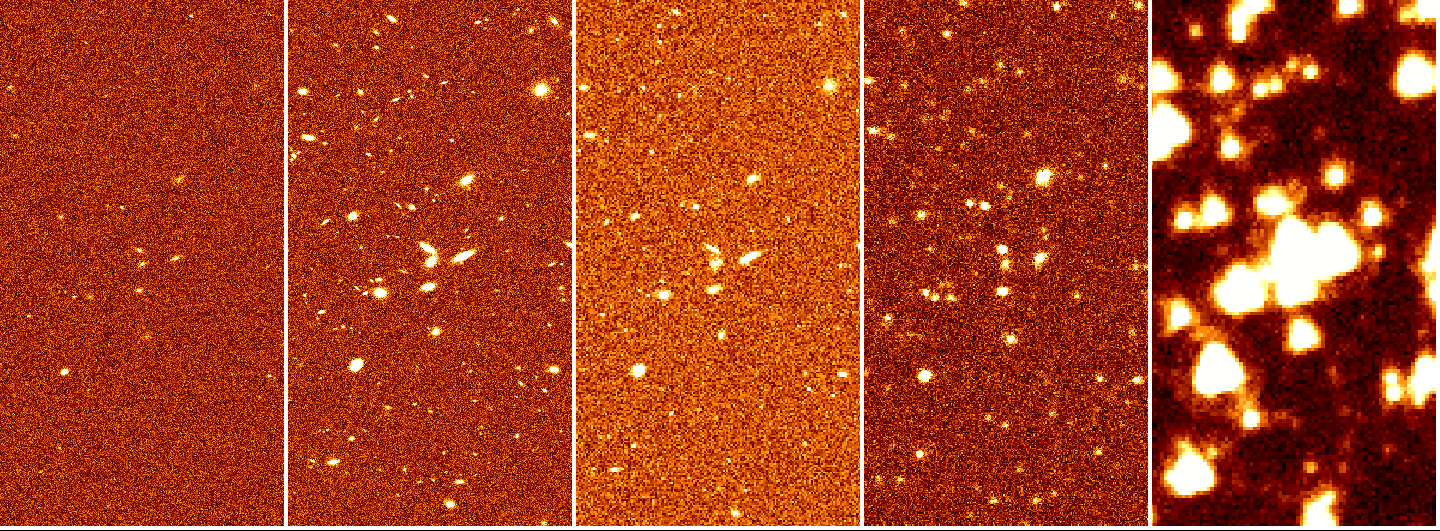}
        \caption{\label{fig:cl_crop} From left to right: Crops of the CL (Cluster) image central area for VIS, H160, NIR H, EXT G and IRAC}
\end{figure*}

\begin{figure*}[h!]
        \centering
        \includegraphics[width=17cm]{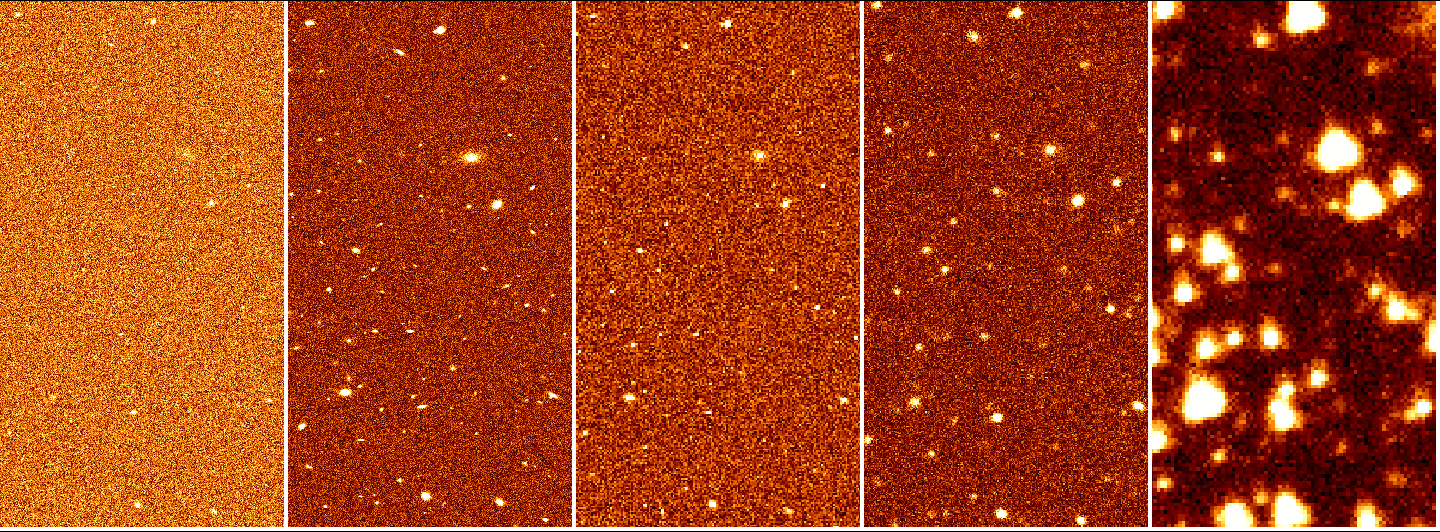}
        \caption{\label{fig:cm_crop} From left to right: Crops of the CM (Average field) image central area for VIS, H160, NIR H, EXT G and IRAC}
\end{figure*}

We can sort the simulated images in several different ways:
\begin{itemize}
\item Depth, from the deepest to the shallowest: $H160 > \textit{EXT G} > IRAC > VIS > \textit{NIR H}$
\item PSF, from the sharpest to the coarsest: $H160 > VIS > \textit{NIR H} > \textit{EXT G} >IRAC$
\item Pixscale, from the smallest to the largest: $H160 > VIS > \textit{EXT G} > \textit{NIR H} > IRAC$.
\end{itemize}
%%%%%%%%%%%%%%
For each simulated image, we cut three independent areas of the sky, which are the same for every band but differ in dimensions due to the different pixel scale. The regions are listed below:
\begin{itemize}
\item BG: centered on a big elliptical galaxy (see Fig. \ref{fig:bg_crop})
\item CL: centered on a cluster of galaxies (see Fig. \ref{fig:cl_crop})
\item CM: an average portion of the sky (see Fig. \ref{fig:cm_crop}).
\end{itemize}
The three regions have a dimension of:
\begin{itemize}
\item VIS: 1000x1000 pixels
\item NIR\_H: 333x333 pixels
\item EXT\_G: 500x500 pixels
\item H160: 1666x1666 pixels
\item IRAC1: 166x166 pixels.
\end{itemize}

After the analysis described in Sect. \ref{subsec:quality_tests}, commenting the results obtained in Sect. \ref{sec:results}, additional tests on real images (see Table \ref{table:table_images} for details) ground-based (HAWK-I) and from space (HST) are reported and analyzed in Sect. \ref{sec:realimages}. %________________________________________________________________
\section{Quality tests}\label{subsec:quality_tests}
The idea at the basis of the analysis is to first evaluate the algorithms through different tests, in order to apply only the most promising ones (with their best configurations) on real images. We organize our analysis on the five simulated images in different levels of testing.
A brief description of each step is given below:
\begin{figure*}
        \centering
        \includegraphics[width=17cm]{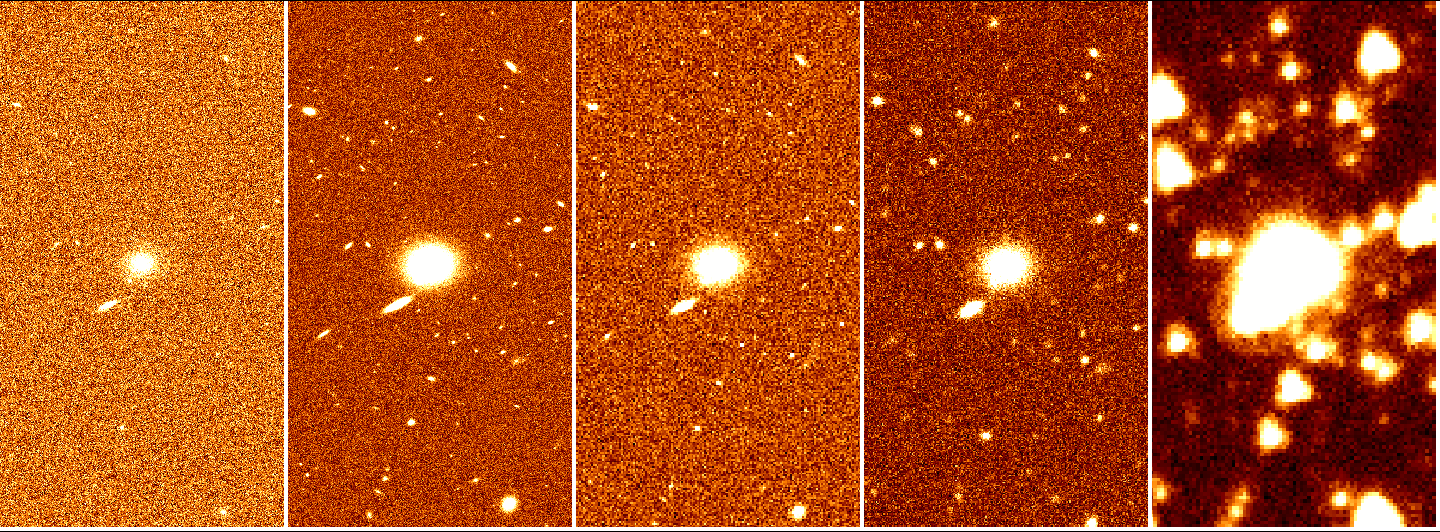}
        \caption{\label{fig:bg_crop} From left to right: Crops of the BG (Big Galaxy) image central area for VIS, H160, NIR H, EXT G and IRAC}
\end{figure*}
\begin{enumerate}

\item As a first step we compare the algorithms performance through three metrics: \textit{mean square error} (\textit{MSE}), \textit{structural similarity}   \citep[SSIM][]{1284395}
and \textit{CPU time}. The \textit{MSE} is defined as:
\begin{equation}
\textit{MSE} := \frac{\sum_{i=1}^{N}\left ( x_{i} - \widehat{x_i}\right )^{2}}{N}
\end{equation}
where $x_{i}$ is the i-th pixel in the denoised image and $\widehat{x_i}$ is the i-th pixel in the original image (without noise).
The SSIM is defined as:
\begin{equation}
\textit{SSIM} := \frac{(2\mu_x\mu_y+c_1)(2\sigma_{xy}+c_2)}{(\mu^2_x+\mu^2_y+c_1)(\sigma^2_x+\sigma^2_y+c_2)}
\end{equation}
where $\mu_x$ is the average of x,  $\mu_y$ is the average of y, $\sigma^2_x$ is the variance of x, $\sigma^2_y$ is the variance of y, $\sigma_{xy}$ is the covariance of x and y, $c_1$ and $c_2$ are constants proportional to the dynamic range of the pixel values.
The \textit{CPU time} is the computational time required by the algorithms to filter the image.
Through these tests we identify the main internal parameters of each algorithm and their ideal values.
\item We test the stability of the algorithms selected in the previous step when faced with non-stationary Gaussian noise. Furthermore, we test their performance as a function of the FWHM of the PSF and as a function of the background noise level.
\item We test the stability of the algorithms selected in the previous steps against variations of the main internal parameter value (identified in Step 1), measuring how the MSE varies as a function of the parameter values.
\item We test how the shapes of the objects is affected by the selected denoising algorithms checking if they preserve the FWHM of point-like objects, the \textit{ellipticity} and the FWHM of the galaxies profiles. 
\item We test the selected algorithms, studying two diagnostics,  
\textit{completeness} and \textit{purity}, which provide a quality estimate of the catalog produced after an ideal source detection, exploring a combination of \textsc{SExtractor} \citep{sextractor} detection parameters.
\item As last step, we test if the denoised images can be used also for photometry measurements, analyzing if the object fluxes are preserved after denoising.
\end{enumerate}
Finally, we apply the best performing algorithms of our selection on real images acquired from space and ground-based telescope, as described in Sect.~\ref{sec:realimages}.

%______________________________________________________________
\subsection{Implementation details} \label{subsec:implementation}
We compare the different images, following always the same procedure here described:
\begin{itemize}
\item The Original image is scaled to the range $[0,1]$ 
\footnote{The scaling is required only by PM methods which need values between 0 and 1 to work, and Bilateral, which needs only non-negative values for
its use. The scaling step has been introduced for the two methods mentioned above and applied to all the methods only for comparison reasons. We verified that for the other algorithms the results do not significantly change if the scaling is not applied. 
}
\item The Original image is filtered by the denoising algorithm providing the \textit{denoised} image
\item The \textit{denoised} image is scaled back to [$Original_{min}$,$Original_{max}$], where $Original_{min}$ and $Original_{max}$ are the minimum and maximum values in the Original image, using the following equation:
\begin{equation}
x^{i}_{Original}=(Original_{max}-Original_{min})*x^{i}_{[0,1]}+Original_{min}
\end{equation}
where $x^{i}_{Original}$ is the $i-th$ pixel in the Original image and $x^{i}_{[0,1]}$ is the $i-th$ pixel in the \textit{denoised} image scaled to [0,1]
\item \textit{MSE} and \textit{SSIM} are computed by comparing the \textit{denoised} image to the noiseless one.
\end{itemize}
In order to choose the best internal parameter for each denoising algorithm (a list of these parameters is in Sect. \ref{subsec:MSE_comparison}), we used different stopping criteria:
\begin{itemize}
\item \textbf{ATVD}: In this method the internal parameters are automatically optimized  by the algorithm, and a stopping rule is already implemented, through a minimization problem, as described in Sect. \ref{subsec:str-tex-dec} \\
\item \textbf{Perona-Malik}: In PM code we have a stopping rule composed by 3 conditions: 
%different stopping criteria, 
in the first one we compare at each time step $\textit{MSE}_{n}$ with $\textit{MSE}_{n-1}$ where $\textit{MSE}_{n}$ is the \textit{MSE} at the current time step, whereas $\textit{MSE}_{n-1}$ is the \textit{MSE} at the previous time step. The code continues running as $\textit{MSE}_{n-1}-\textit{MSE}_{n}>0$. The second condition concerns the number of iterations $n$: the code continues running until the number of iterations does not exceeds the maximum number of iterations NMAX, which is set to NMAX=500. The third condition is $\mid \frac{{MSE}_{n-1}- MSE_{n}}{MSE_{n-1}}\mid \le \epsilon$, with $\epsilon = 10^{-10}$ \\ %which is set to $10^{-4}$
\item \textbf{\textsc{sparse2D}}: Starlet and the two b-UWTs automatically estimate the noise background. 
For the optimal configurations of these three methods, we used the setting provided by the authors and reported at the end of Sect. \ref{subsec:wavelets}. 
Further information can be found in 
the documentation file available with the code on the CosmoStat webpage\footnote{\url{http://www.cosmostat.org/wp-content/uploads/2014/12/doc_iSAP.pdf}}.\\
\item \textbf{Other denoising algorithms}: For all the other denoising algorithms 
we get the optimal values of the main parameter(s) by minimizing the MSE with an iterative process. 
The stopping rule is reached when $\mid {MSE}_{n-1}- MSE_{n}\mid \le \epsilon$, setting $\epsilon = 10^{-10}$, and the number of iterations is lower than the maximum number of iterations NMAX, which is in this case set to NMAX=100. 
\end{itemize}
\section{Results}\label{sec:results}
In this section we analyze and comment in a separate and sequential way the results related to the quality tests, following the same order of the steps used in Sect. \ref{subsec:quality_tests}.
%%%%%%%%%%%%%%%%%%%%%%%%%%%%%%%%%%%%%%%%%%%%%%%%%%%%%%%%
\subsection{Ranking with \textit{MSE} and {SSIM}}\label{subsec:MSE_comparison}
In this test we use three metrics to constrain the performance of denoising methods: \textit{MSE}, \textit{SSIM} and \textit{CPU time} (Sect. \ref{subsec:quality_tests}).
We give priority to those algorithms that are able to minimize as much as possible the \textit{MSE}, preferring the fastest method (in terms of \textit{CPU time}) and the highest \textit{SSIM} in case of comparable \textit{MSE}. Following this criterion, in this step we identify the best configuration and the main parameters for every algorithm. 
These results are taken into account separately for all the simulated images presented in Sect. \ref{images}.

The main internal parameters identified for the different algorithms are listed below:
\begin{itemize}
    \item Orthogonal Wavelets: \texttt{sigma} - The noise standard deviation used for compute threshold(s)
    \item NL means: \texttt{h} - Cut-off distance in grey levels 
    \item TV Bregman: \texttt{weight} - Denoising weight, efficiency of denoising
    \item TV Chambolle: \texttt{weight} - Denoising weight, efficiency of denoising
    \item Gaussian: \texttt{sigma} - Standard deviation for Gaussian kernel
    \item Bilateral: \texttt{sigma\_color} - Standard deviation for gray value distance
    \item Perona-Malik: \texttt{T} - Number of iterations of the anisotropic diffusion
    \item ATVD (TVL1,TVL2,TVG): $\lambda$ - Structural-Texture splitting parameter
    \item BM3D: \texttt{sigma} - The noise standard deviation.
\end{itemize}
Further details for the algorithms implemented in Python and the measurement of \textit{MSE} and \textit{SSIM} can be found in the  \texttt{scikit-image} documentation\footref{note-scikit}.
The method used to identify the best internal parameter for each algorithm is described in Sect. \ref{subsec:implementation}. 
In \appendixname{ A} we show the best \textit{MSE} and CPU time values of every algorithm, for the different crops. The tables are organized to record the best \textit{MSE} and CPU time values obtained with the algorithms. The columns represent the different image simulated filters and the value indicated in bold is the lowest of the column. Tables \ref{tab:msebg}-\ref{tab:msecm}-\ref{tab:msecl} contain the \textit{MSE} values for the crops BG, CM and CL, respectively. 
Table \ref{tab:cpucm} contains the CPU time values for the crop CM, after fixing the optimal internal parameters. We remind that in the following "PSF filtering" amounts to filtering with a Gaussian whose FWHM is the same as the PSF-FWHM, whereas in the case of the "Gaussian filtering" the $\sigma$ (and thus the FWHM) is a free internal parameter.
We shortly summarize here the main results:
\begin{itemize}
\item TVL2, BM3D, Starlet, the two b-UWTs, PM, NLmeans slow, TV Chambolle always yield good performance, typically providing the lowest values of \textit{MSE}
\item TVL2, BM3D, Starlet, the two b-UWTs, PM, NLmeans slow, TV Chambolle always perform better than Gaussian filtering, with the only exception of the IRAC image (we discuss the IRAC situation below in Sect. \ref{subsec:irac}) 
\item the \textit{MSE} of all the methods is proportional to the pixel scale of the image, so that low sampling implies worse results 
\item in most cases (with the exception of IRAC, which we discuss below), the PSF filtering provides a larger (i.e. worse) value of the \textit{MSE} compared to the one provided by Gaussian filtering.
\item in some cases, the \textit{MSE} of the denoised image is larger (i.e. worse) than the one measured without denoising the image at all. Indeed some algorithms in the situations listed below tend to over-smooth the image, providing a worse MSE. This event occurs:
\begin{enumerate}[a)]
\item in VIS (CM) image, in the case of the PSF filtering
\item in all the H160 images for both the PSF filtering and TV Bregman
\item 2-4 times in NIR H  images, for methods NLmeans fast, Orthogonal Wavelets, TV Bregman and PSF filtering,  and only once for BM3D
\item only once in EXT G (CM), for the PSF filtering
\item 4 to 5 times in IRAC images, for NLmeans slow, NLmeans fast, TV Bregman, Orthogonal Wavelets, and PSF filtering.
\end{enumerate}
\item the \textit{SSIM} ranking typically reflects the \textit{MSE} ranking, pointing out the same group of best algorithms found in the \textit{MSE} ranking; even if some positions are swapped in few cases, the \textit{SSIM} values provided by the best algorithms are comparable ($\Delta \textit{SSIM}<\num{e-4}$ )
\item the CPU time table (Table \ref{tab:cpucm}) shows that Gaussian is the fastest algorithm among the ones we tested, followed by PSF and TV Bregman. The CPU time for the other algorithms differ from  1 to 4 order of magnitudes with respect to the Gaussian algorithm. 
However, the computational times are always manageable, at least for the cases of optimal performance. 
\end{itemize}

If we focus on the algorithms belonging to the same classes of methods, we can note that:
\begin{enumerate}
\item all the PM methods yield similar performance, (see Fig. \ref{fig:PM_mse_plot}), and  therefore we choose to only keep $g=g_1$ with the parameter $\textit{k}$ set to $\textit{k}=10^{-3}$ in the following steps
\item TVL2 performs clearly better than TVG and TVL1. In fact,  e.g. in BG, $1-\frac{mse}{mse_{Original}}$ value is always within 5\% from the value provided by the original image (no noise), with the exception of IRAC, where it drops to 0.2, which is still greater than the values provided by TVG and TVL1, as shown in Fig. \ref{fig:ATVD_mse_plot}
\item NLmeans slow performs slightly better than NLmeans fast for H160, VIS, and EXT G, ($1-\frac{mse}{mse_{Original}}$ differences are within 5\% in favor of NLmeans slow) and much better for NIR H and IRAC (where NLmeans fast performs worse than Original). See Fig. \ref{fig:other_mse_plot}
\item TV Chambolle performs better than TV Bregman in H160, NIR H, and IRAC. TV Bregman performs worse than Original, whereas for VIS and EXT G it performs 14\% and 3\% worse than TV Chambolle, respectively (see Fig. \ref{fig:other_mse_plot})
\item Bilateral is always within the best performing techniques  (see Fig. \ref{fig:other_mse_plot})
\item BM3D, the two b-UWTs and Starlet are always among the most efficient algorithms (see Fig. \ref{fig:new_mse_plot} and Tables \ref{tab:msebg}-\ref{tab:msecm}-\ref{tab:msecl} in \appendixname{ A}).
\end{enumerate}
Nevertheless, we keep Gaussian and PSF filtering for reference since they are widely used.
Hence, at the end of this first step we are left with $11$ methods: PM with edge-stopping function $g_1$ and $k=10^{-3}$, TVL2, BM3D, Starlet, b-UWT(7/9), b-UWT(7/9)+Wiener, Gaussian, PSF, NL means slow, Bilateral, and TV Chambolle. 
Following our experiments analysis we decide to discard $9$ algorithms: $4$ PM methods, TVG and TVL1, NL-means fast, TV Bregman, and Orthogonal Wavelets. 

\subsection{The IRAC results} \label{subsec:irac}
We note that the IRAC images do not follow the same trends as the other bands. 
In fact, whereas for all the other images there is always a small group of algorithms which perform better than all the others, for IRAC nearly all the denoising algorithms tend to have similar performance. We investigated the possibility that the number of pixels were not enough (166x166 pixels) to extract significant conclusions from these images and we tested the algorithms on an IRAC (CM) simulation with pixel scale 0.06 arcsec and size of 1000 $\times$ 1000 pixels. We noticed that, TV Chambolle, NL means slow and Gaussian provide the best performance (MSE $\propto\num{e-9}$), followed by BM3D, b-UWT(7/9), TV Bregman, TVL2, PM (g=g1 k=0.01) (MSE $\propto\num{1e-8}$ ), PSF (MSE $\propto\num{2e-8}$ ) and then Bilateral, PM (g=g1 k=0.001), b-UWT(7/9)+Wiener, Starlet and Orthogonal Wavelets (MSE $>\num{2e-8}$ ). 
After this small test we point out that again the IRAC band does not follow the trend defined in the other bands (even if the \textit{MSE} decreases for all the methods and the Original image), but with the increased number of pixels  TV Chambolle, NL means slow and Gaussian are the algorithms which provide the best performance. The low resolution of IRAC here plays a fundamental role, impacting on most of the algorithms performance. This aspect of the algorithms will be described later on, in Sect. \ref{subsec:fwhm_std}.
%%%%%%%%%%%%%%%%%%%%%%%%%%%%%%%%%%%%%%%%%%%%%%%%%%%%%%%%

\subsection{Stability against non-stationary Gaussian noise}\label{subsec:nonstationary}

In this section we discuss the results of tests on images with varying depths, i.e. obtained combining regions observed with different exposure times. To build the dataset, we used a noiseless simulated image $I_{2exp}$, which we mirrored along the $x$-axis to obtain a new image with two identical vertical halves; then, we added Gaussian noise with 
$\sigma = \sigma_{VIS}$ on the lower half $H_{l}$ and with $\sigma=2\sigma_{VIS}$ on the upper half $H_{u}$. 
In this way, the two haves of the image contain the very same objects with a different amount of observational noise, as if they had been observed with different exposure times. 
We applied the algorithms in their optimal configuration (see Sect. \ref{subsec:MSE_comparison}) on the mirrored version of the crops VIS (CM) and VIS (CL), and we calculated \textit{MSE} and \textit{SSIM} in $H_l$ and $H_u$ for both. We then compared these results with the ones obtained by the application of the algorithms on the mirrored image with stationary Gaussian noise, with $\sigma_{VIS}$ and $2\sigma_{VIS}$, we refer to these images as $I_{\sigma}$ and $I_{2\sigma}$, respectively.
From the results reported in Tables \ref{tab:cm_mirrored} and \ref{tab:clmirrored} in \appendixname{ B}, obtained on both VIS (CM) and VIS (CL), respectively, we notice that:
\begin{itemize}
    \item all the algorithms applied on $H_{l}$ produce \textit{MSE} values of the same order of magnitude of the ones obtained with $I_{\sigma}$
    \item nearly all of them applied on $H_{u}$ produce \textit{MSE} values of the same order of magnitude of the ones obtained with $I_{\sigma}$, with the exception of Starlet and the two b-UWTs, for which the application on $H_{u}$  produces a \textit{MSE} which is around 1 order of magnitude larger than the respective \textit{MSE} obtained with  $I_{2\sigma}$
    \item \textit{MSE} relative variation for BM3D, PSF, Gaussian, PM and TVL2 is $\propto10^{-3}$
    \item For Bilateral, NL means, and TV Chambolle the \textit{MSE} relative variation is $\propto10^{-2}$
    \item For Starlet, and the two b-UWTs methods, the \textit{MSE} relative variation is $\propto10^{-1}$
    \item For all the algorithms, \textit{SSIM} relative variation is <$10^{-2}$
\end{itemize}

We want to point out that \textit{MSE} achieved by methods like Starlet and the two b-UWTs on $H_{l}$ is $\propto 10^{-6}-10^{-7}$, meaning that \textit{MSE} relative variation ($\sim10^{-1}$) still implies small absolute variations ($\propto 10^{-6}-10^{-7}$). Finally, we can conclude that the algorithms tested are not influenced (or negligibly influenced in few cases) by images with non-stationary Gaussian noise.

\subsection{Stability against FWHM and depth variations}\label{subsec:fwhm_std} 

In the second part of this test, we compare the performance of the $11$ 
algorithms with respect to the variation of the FWHM and depth of the images. 
We consider two cases: 
\begin{itemize}
 \item A 1000x1000 pixels crop of the simulated VIS image convolved with different kernels, to degrade the resolution increasing the FWHM without changing the depth of the image (we considered the cases FWHM = 0.5, 1, 1.5 and 2.0'', with the original FWHM being 0.2'')
\item We decreased the depth of a 1000x1000 pixels crop of the simulated H160 image without changing the FWHM, by adding  Gaussian noise with increasing standard deviation $\sigma$ ($\times 1, 10, 20, 30$ and 40 times the original one) to the Noiseless image. 
\end{itemize}
The plots summarizing the results are shown in Figs. \ref{fig:fwhm_vis}- \ref{fig:noised_f160}, in \appendixname{ C}. We can note that:
\begin{enumerate}
\item The MSE calculated on the original image alone decreases at increasing FWHM due to the loss of information (i.e. small objects and details). All the  algorithms follow this trend while lowering the \textit{MSE} even more due to the effect of filtering (see Fig. \ref{fig:fwhm_vis})
\item The ratio between the MSE obtained by each algorithm  and the MSE computed on the original image ($\frac{mse}{mse_{Original}}$) increases at increasing FWHM, with the only exception of the Gaussian filtering which instead follows the opposite trend (see Fig. \ref{fig:fwhm_vis_2})
\item $\frac{mse}{mse_{PSF}}$ is weakly affected by variations of the FWHM for most of the denoising methods, with the exception of Gaussian 
(see Fig. \ref{fig:fwhm_vis_3})
\item As expected, the MSE increases at increasing background level (due to the increasing of $\sigma$ for the Gaussian noise) both in the original image and in the output denoised images for all the algorithms (see Fig. \ref{fig:noised_f160_2})
\item $\frac{mse}{mse_{PSF}}$ increases at increasing background level for all the methods  (see Fig. \ref{fig:noised_f160_3})
\item $\frac{mse}{mse_{Original}}$ decreases at increasing background level for all the methods 
(see Fig. \ref{fig:noised_f160}).
\end{enumerate}

Summarizing, we conclude that the best performance by any denoising algorithm are obtained on images with low SNR and high resolution (narrow FWHM). The best performance with respect to the PSF method are obtained by applying the denoising methods on image with high SNR, regardless of the PSF-FWHM. These results can be used to estimate the efficiency of the denoising algorithms in different situations, pointing out that when applied to high resolution images they provide the best improvements, whereas if applied on low SNR images (where there is the peak of performance), the improvements compared to the PSF are slightly less significant.
From these results, it would be very interesting to apply  these methods, as an alternative of the PSF filtering, on images with high resolution and high SNR.

%%%%%%%%%%%%%%%%%%%%%%%%%%%%%%%%%%%%%%%%%%%%%%%%%%%%%%%%
\subsection{Stability against variations of the parameters}
\begin{figure}
        \resizebox{\hsize}{!}{\includegraphics{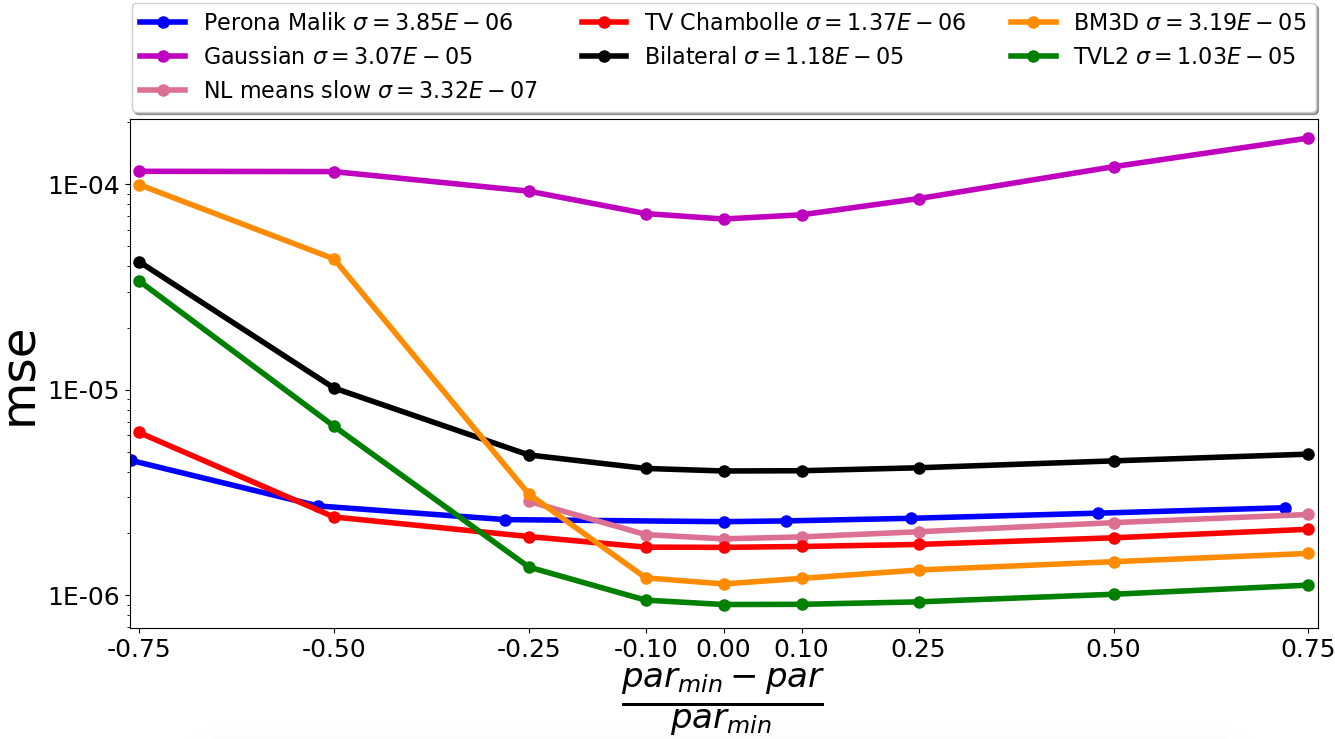}}
        \caption{\label{fig:stability_test} Step 3:  Stability against variations of the parameters. Each curve corresponds to a denoising algorithm. We plot the \textit{MSE} against the relative variation of the parameters, $\frac{par_{min}-par}{par_{min}}$. Obviously the absolute minimum of the curves is reached in 0 on the $x$-axis, corresponding to the ideal value of the parameter. In the upper panel we report the standard deviations $\sigma$ of the $mse_{mean}-mse$ distribution for each method}.
\end{figure}

In this test we analyze the selected methods by varying the values of those internal parameters that had been kept fixed to the optimal ones in the previous tests. 
The goal is to understand whether the performance are stable against sub-optimal parameter settings. 
We exclude from this analysis the three denoising methods belonging to the \textsc{sparse2D} package, for which we simply used the configuration provided by the authors, reported at the end of Sect. \ref{subsec:wavelets}, and 
the PSF filtering since it is just a particular case of the Gaussian filtering method. 
We perform the test on the VIS (CM) image and we change the main parameter value of each technique by $\pm 10\%,\pm 25\%, \pm 50\%$ and $\pm 75\%$  with respect to the value used for the \textit{MSE} analysis (see Subsect. \ref{subsec:MSE_comparison}). The results are shown in Fig. \ref{fig:stability_test}. 
We notice that most of the techniques tend to have similar performance when over-estimating the parameters, remaining relatively stable; on the contrary, under-estimating it significantly worsens the performance. However, all the algorithms (with the exception of BM3D) have a lower dispersion in \textit{MSE} compared to the Gaussian filtering (this is not evident in the plot because of the logarithmic $y$-axis scale, but we verified it numerically and we give the $\sigma$ values in the upper panel of the plot), meaning that they are generally more stable against the variation of the parameters. In addition, they yield a $\sim1$ order of magnitude lower mean when the parameters are below the optimal value, and by $\sim2$ order of magnitudes when they are above it. BM3D like the other algorithms under-performs when its main parameter is under-estimated, producing in this case a large dispersion of the order reached by the Gaussian.
%%%%%%%%%%%%%%%%%%%%%%%%%%%%%%%%%%%%%%%%%%%%%%%%%%%%%

\subsection{Conservation of the FWHM and ellipticity}\label{subsec:fwhm_conservation}
\begin{figure}[h!]
        \centering
        \resizebox{\hsize}{!}{\includegraphics[]{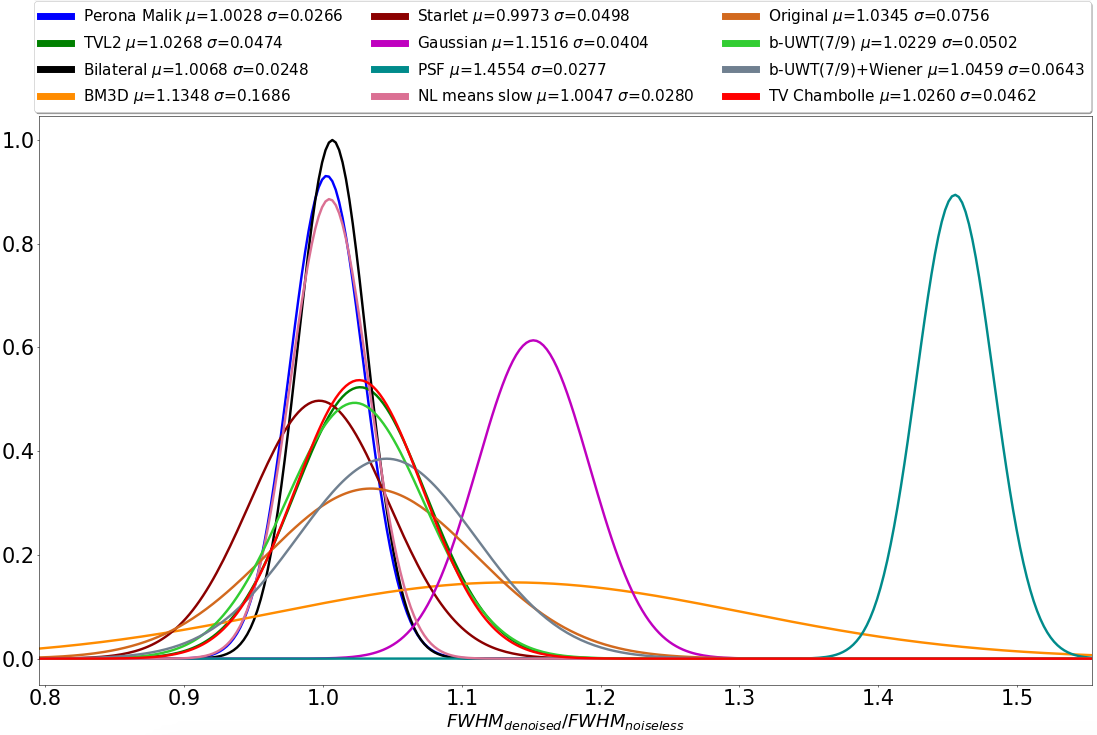}}
        \caption{\label{fig:FWHM_test_stars} Step 4: FWHM conservation test on stars. On the $x$-axis we plot the $FWHM_{denoised}/FWHM_{noiseless}$, where $FWHM_{noiseless}$ is the FHWM of the objects measured on the Noiseless image. $\mu$ and $\sigma$ are the mean and the standard deviation of the distribution of $FWHM_{denoised}/ FWHM_{noiseless}$}.
\end{figure}

\begin{figure*}[h!]
        \centering
        \includegraphics[width=17cm]{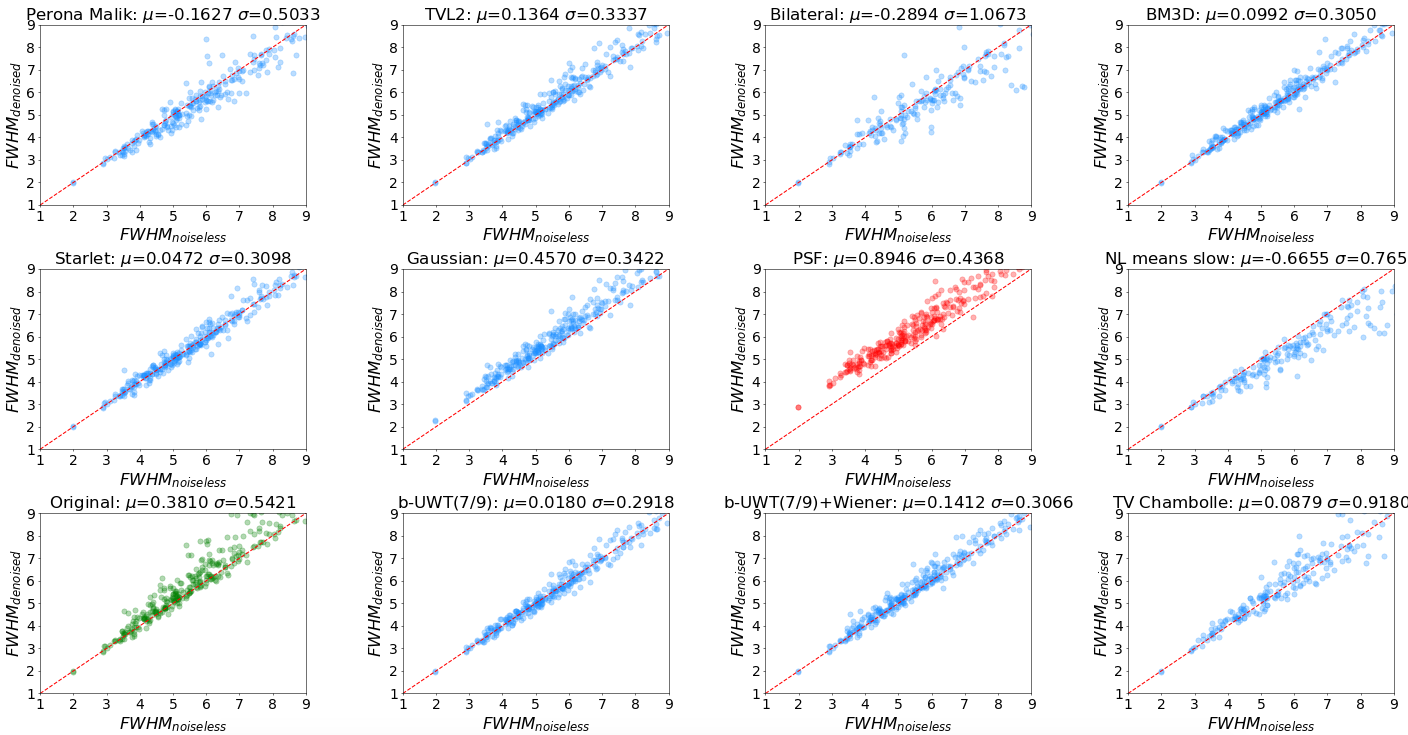}
        \caption{\label{fig:FWHM_test_galaxies} Step 4: FWHM conservation test on galaxies. On the $x$-axis we plot the FWHM of the objects measured on Noiseless image $FWHM_{noiseless}$, whereas on the $y$-axis we plot the FWHM measured on the Original image after the application of the denoising algorithms $FWHM_{denoised}$. $\mu$ and $\sigma$ are the mean and the standard deviation of the distribution of $FWHM_{denoised} - FWHM_{noiseless}$.}
\end{figure*}

The optimal denoising approach should not significantly alter size and shape of the detected sources so to enable a meaningful scientific analysis.
We thus tested the selected methods by measuring the FWHM of the detected sources with \textsc{SExtractor}, and comparing the measured values to the ones obtained on original, unfiltered images. We perform this test on the simulated VIS image described before, which is mainly populated by galaxies, and on a specific rendition of the simulated VIS image populated by stars distributed on a grid. 
The results are shown in Figs. \ref{fig:FWHM_test_stars}- \ref{fig:FWHM_test_galaxies}. 
Whereas for the stars in Fig. \ref{fig:FWHM_test_stars} the PSF filtering tends to smooth all the detected object as much as of $\sim50\%$ of the FWHM, {the other algorithms} have a much lower impact (the FWHM is degraded by less than 20\% of the original value), even though BM3D has a significant dispersion. Similarly, for the galaxies in Fig. \ref{fig:FWHM_test_galaxies}, the PSF filtering causes again a small offset, whereas most of the other methods tend to better preserve the FWHM. 

\begin{figure*}[h!]
        \centering
        \includegraphics[width=17cm]{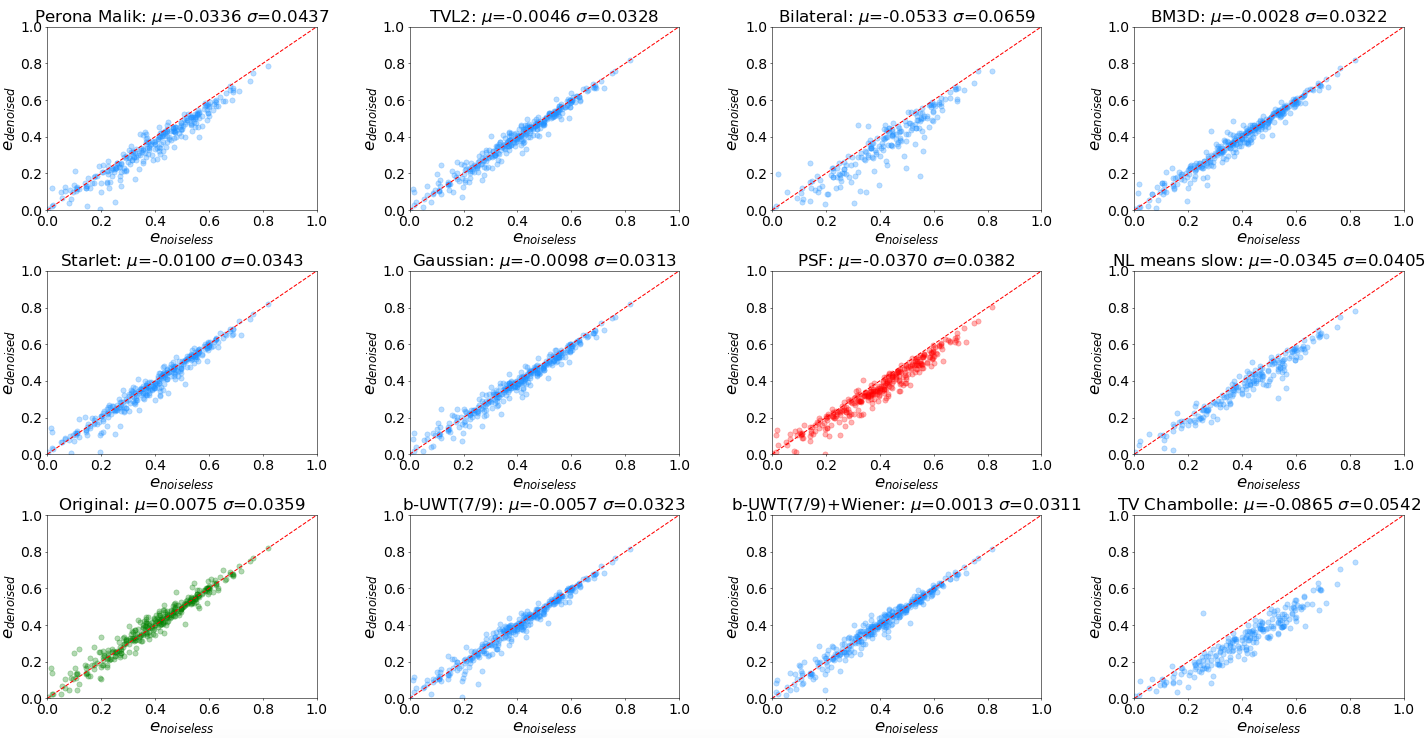}
        \caption{\label{fig:e_test_galaxies} Step 4: Ellipticity conservation test on galaxies. On the $x$-axis we plot the ellipticity $e$ of the objects measured on Noiseless image $e_{noiseless}$, whereas on the $y$-axis we plot the $e$ measured on the Original image after the application of the denoising algorithms $e_{denoised}$. $\mu$ and $\sigma$ are the mean and the standard deviation of the distribution of $e_{denoised} - e_{noiseless}$.}
\end{figure*}

For a comprehensive comparison, another quantity has been taken into consideration. The \textit{ellipticity} of the galaxies has been measured before and after the application of the denoising algorithms, using the parameter \texttt{ELLIPTICITY} from \textsc{SExtractor}. Looking at Fig. \ref{fig:e_test_galaxies}, it is possible to notice that most of the algorithms do not modify the ellipticity at an alarming level (even if in some cases the dispersion is far from being optimal, e.g. for Bilateral and TV Chambolle). Nevertheless, the PSF is one of the most performing method for this test.

From these tests, we can conclude that most of the tested algorithms preserve the shape of the sources similarly (in the case of the \textit{ellipticity}) or even better (in the case of the FWHM) than the PSF filtering. 

%%%%%%%%%%%%%%%%%%%%%%%%%%%%%%%%%%%%%%%%%%%%%%%%%%%%%

\subsection{Completeness and purity} \label{subsec:completeness-purity}

In this test - perhaps the crucial one - we checked the quality of the catalogs of sources extracted from the denoised images. We analyze two diagnostics, both relevant to assess the performance of the detection process: namely, the \textit{completeness} and the \textit{purity} as defined below.
We extract the catalogs running \textsc{SExtractor} in \textit{dual} image mode using a denoised image as detection band and the original image as measurement band so to perform a cross-correlation between the extracted and the true catalogs of sources both in terms of position and flux.

We used the simulated VIS 5000x5000 pixels image, searching for the best \textsc{SExtractor} parameters configuration for every denoised image. We have thus tested a large number of possible combinations of the two parameters which control the detection, i.e. \texttt{DETECT\_THRESH} (from a minimum value of 0.2 to a maximum of 6.0, with steps of 0.1) and \texttt{DETECT\_MINAREA} (with values: 3,6,9,12,15,30), considering only the combinations for which the quantity $\texttt{DETECT\_THRESH}*\sqrt{\texttt{DETECT\_MINAREA}}>1$, which provides a selection of objects with a global significance of at least 1-$\sigma$.
The number of detection parameters combinations which fulfill this requirement is $\sim 350$. We point out that the best algorithms configurations, used for this test and obtained by \textit{MSE} minimization, do not differ significantly from the best configurations found in the VIS (CM) image (Sect. \ref{subsec:quality_tests}). \\
We introduce some notations:
\begin{itemize}
\item $n_{detected}$ is the total number of detected objects, which includes both real and spurious detections indiscriminately
\item $n_{simulated}$ is the number of simulated objects in the image
\item $n_{spurious}$ is the number of spurious detections, as identified by the \emph{spurious sources identification approach} described in the following.
\end{itemize}

The \emph{spurious sources identification approach} that we define for this work is related to the \textsc{SExtractor} cross-correlation, when an association catalog is provided: we denote by $C_{R_{assoc}}$  the circle centered on the simulated object original position with radius $R_{assoc}$, which is the maximal distance allowed for the association made by \textsc{SExtractor}.  We set it to $6$ pixels (i.e. $3 \times FWHM$). 
Then, we tag an object as spurious if one of the following two conditions holds: 
\begin{itemize}
\item is outside $C_{R_{assoc}}$
\item (is inside $C_{R_{assoc}}$) AND  ($|mag_{measured}-mag_{simulated}|> 1.0$)  AND ($mag_{aperture}-mag_{simulated}|> 1.0$),
\end{itemize}
where $mag_{measured}$ is \textsc{SExtractor} \texttt{MAG\_AUTO} (an estimation of the total magnitude of the source), $mag_{simulated}$ is the true magnitude of the simulated object, and $mag_{aperture}$ is \textsc{SExtractor} \texttt{MAG\_APER} corresponding to the magnitude within a circular aperture with diameter of 6 pixels. 

\begin{figure*}[h]
        \centering
        \includegraphics[width=17cm]{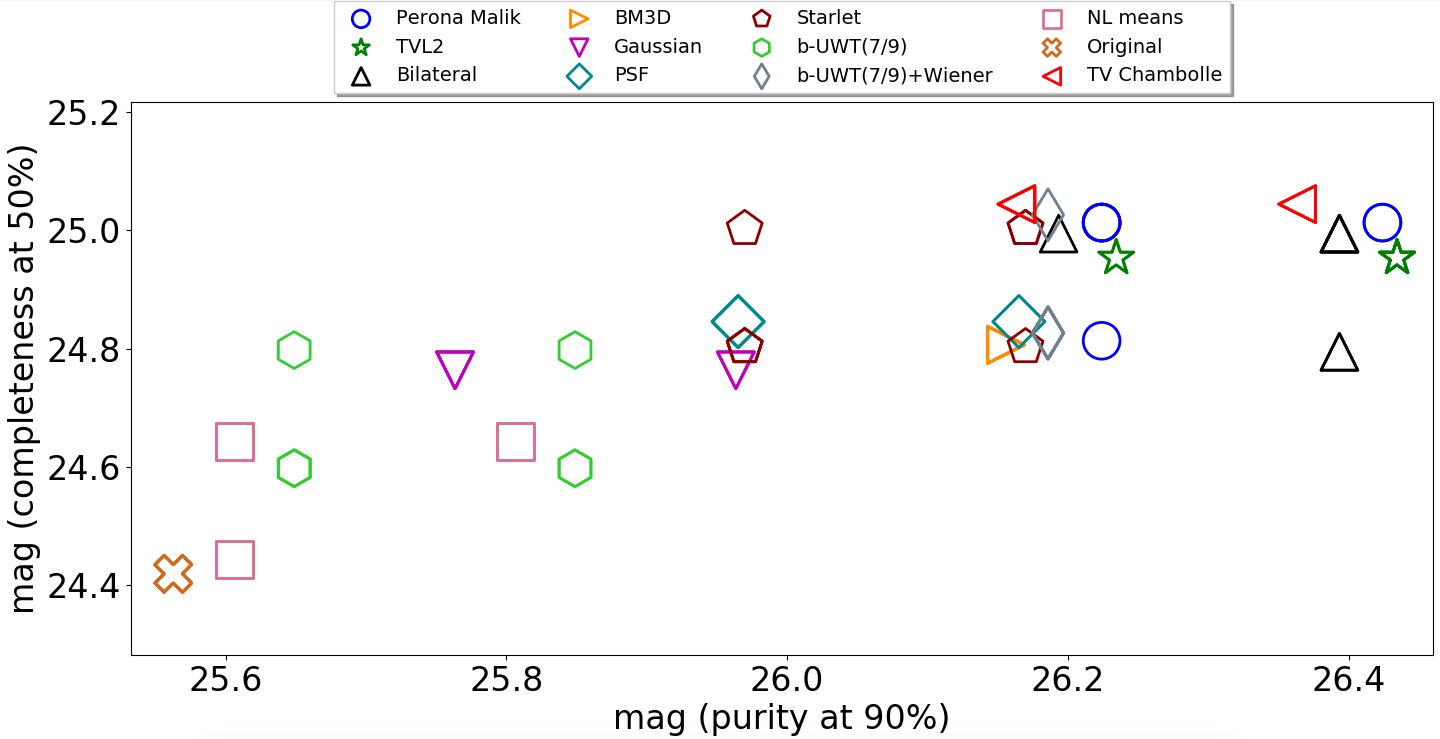}
        \caption{\label{fig:completeness-purity} Step 5: Completeness and purity test. We extracted catalogs on the VIS simulated image processed with the denoising algorithms, using different configurations of \textsc{SExtractor}. We plot the magnitude at which the completeness drops below 50\% against the magnitude at which the purity drops below 90\%. Each symbol corresponds to a different denoising method, which can be present multiple times in the plot due to different combinations of detection parameters. The positions of the symbols are slightly randomized to improve readability.}
\end{figure*}

Finally, we can now define the two diagnostics:  \begin{equation}
completeness :=\frac{n_{detected}-n_{spurious}}{n_{simulated}}
\end{equation}
\begin{equation}
purity := 1-\frac{n_{spurious}}{n_{simulated}}
\end{equation}

where $purity = purity_{assoc}$, determined by the association approach defined above.
We measure completeness and purity in 0.2 magnitudes bins. In Fig. \ref{fig:completeness-purity} we plot the magnitudes at which the completeness drops below 50\% against the one at which the purity drops below 90\%. Each symbol corresponds to a different denoising technique, and repetitions of the same symbols correspond to different combinations of the detection parameters for the same algorithm. For readability, a maximum of 5 combinations per algorithm (corresponding to the best ones) are shown in the figure. 

We note that all the methods improve the detection. BM3D performs like the PSF, whereas  methods like Starlet and b-UWT(7/9)+Wiener produce   remarkable results, with a 0.2 increment in completeness with respect to the PSF, but the best performance are reached by TVL2, Perona-Malik, TV Chambolle, and Bilateral.
Indeed, these 4 methods reach the completeness threshold 0.6 mag deeper, and the purity threshold 0.8 magnitude deeper than the non-denoised run. Moreover, they improve the detection compared to the PSF smoothing, reaching 0.2 magnitudes deeper in both completeness and purity.

%%%%%%%%%%%%%%%%%%%%%%%%%%%%%%%%%%%%%%%%%%%%%%%%%%%%%%%%

\begin{figure*}[h]
        \centering
        \includegraphics[width=17cm]{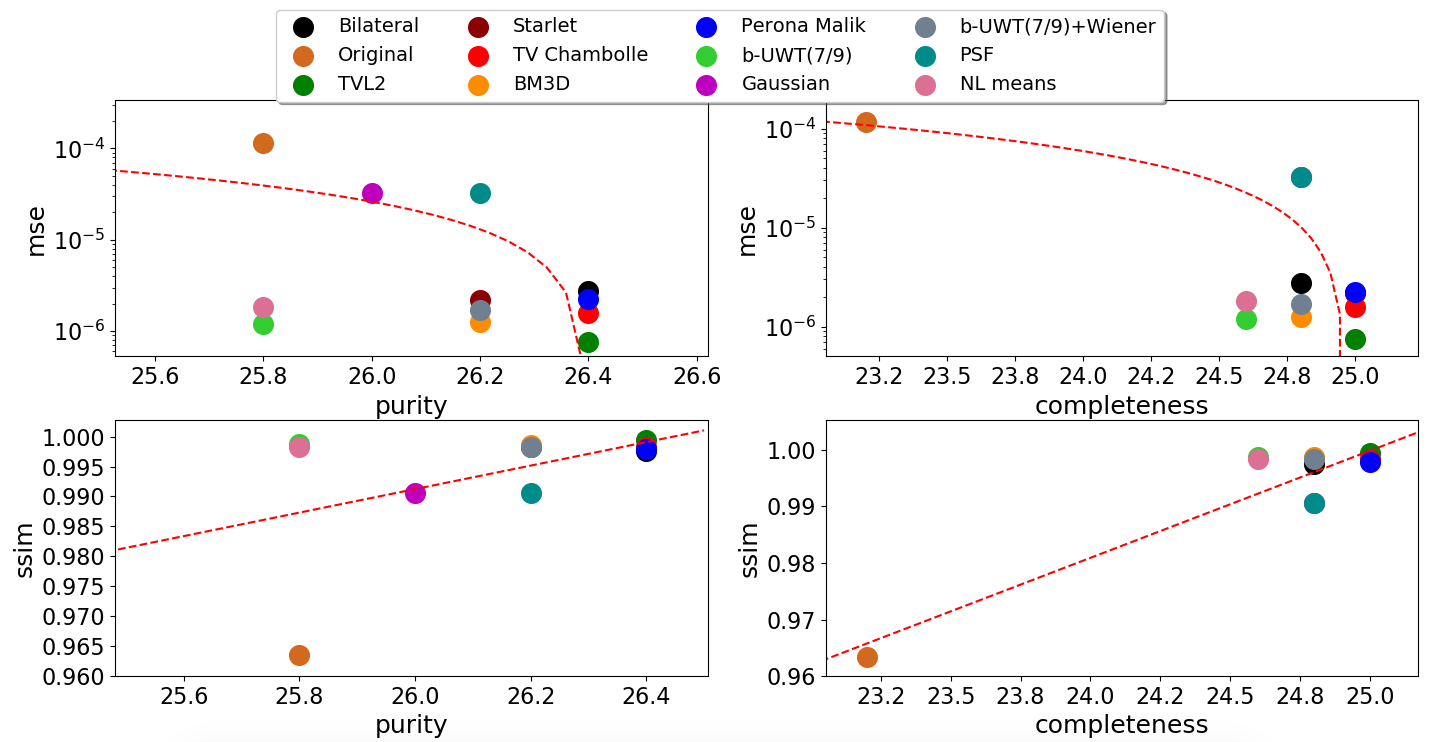}
        \caption{\label{fig:SSIM-MSE-completeness-purity} Step 5: Correlation between \textit{MSE} or \textit{SSIM} and purity or completeness. On the $x$-axis we plot the magnitudes at which completeness (purity) reaches 50\% (90\%), whereas on the $y$-axis we plot the metrics \textit{MSE} or \textit{SSIM}. Dashed lines are the linear best-fitting.}
\end{figure*}

It is tempting to consider the \textit{MSE} and \textit{SSIM} measured on the VIS images used for the completeness and purity analysis, searching for a possible correlation between the metrics and the diagnostics. In Fig. \ref{fig:SSIM-MSE-completeness-purity} the plots are produced using the results shown in Fig. \ref{fig:completeness-purity}. We find no or weak correlation between \textit{MSE} (\textit{SSIM}) and purity, whereas a stronger correlation exists between \textit{MSE} (\textit{SSIM}) and completeness.

We show the snapshots of a sample of objects detected by the different methods in the VIS image in \appendixname{ D}. These snapshots give a visual match of objects detected in the denoised images. We only show the best performing algorithms results compared to the Original, PSF-filtered and the Noiseless images. For VIS the algorithms are: PM, TVL2, Bilateral, and TV Chambolle, followed by BM3D, Starlet, b-UWT(7/9), and b-UWT(7/9)+Wiener.

\subsection{Conservation of the flux}

\begin{figure*}
        \includegraphics[width=17cm]{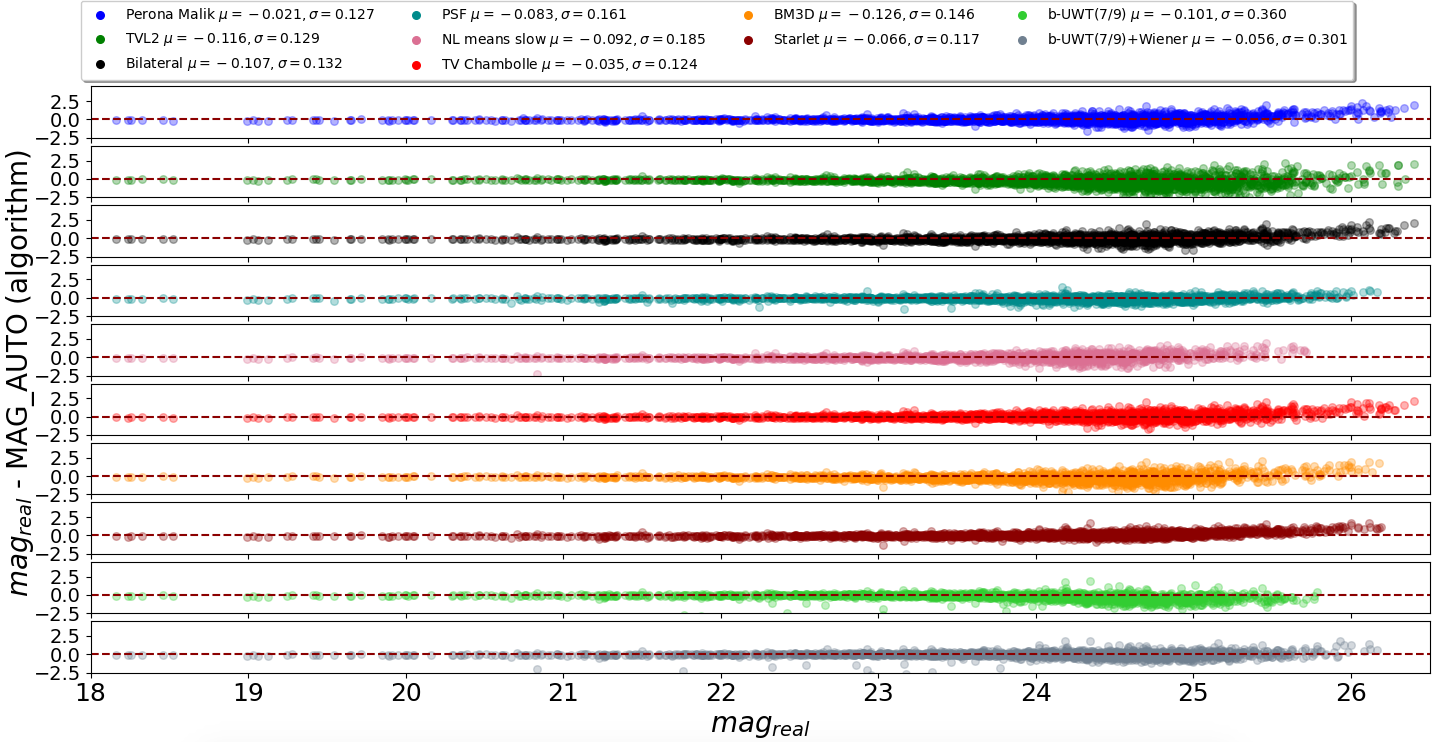}
        \caption{\label{fig:flux_conservation_all} Step 6: Flux conservation distribution for objects with magnitude within 19 and 23. 
        On the x-axis the real objects magnitude $mag_{real}$. On the y-axis, the difference between the magnitude measured \texttt{MAG\_AUTO} and $mag_{real}$.
        Only the detected objects within the purity and completeness thresholds (Sect. \ref{subsec:completeness-purity}) are considered. $\mu$ and $\sigma$ are the distribution mean and the standard deviation values, respectively.}
\end{figure*}

\begin{figure}
        \resizebox{\hsize}{!}{\includegraphics{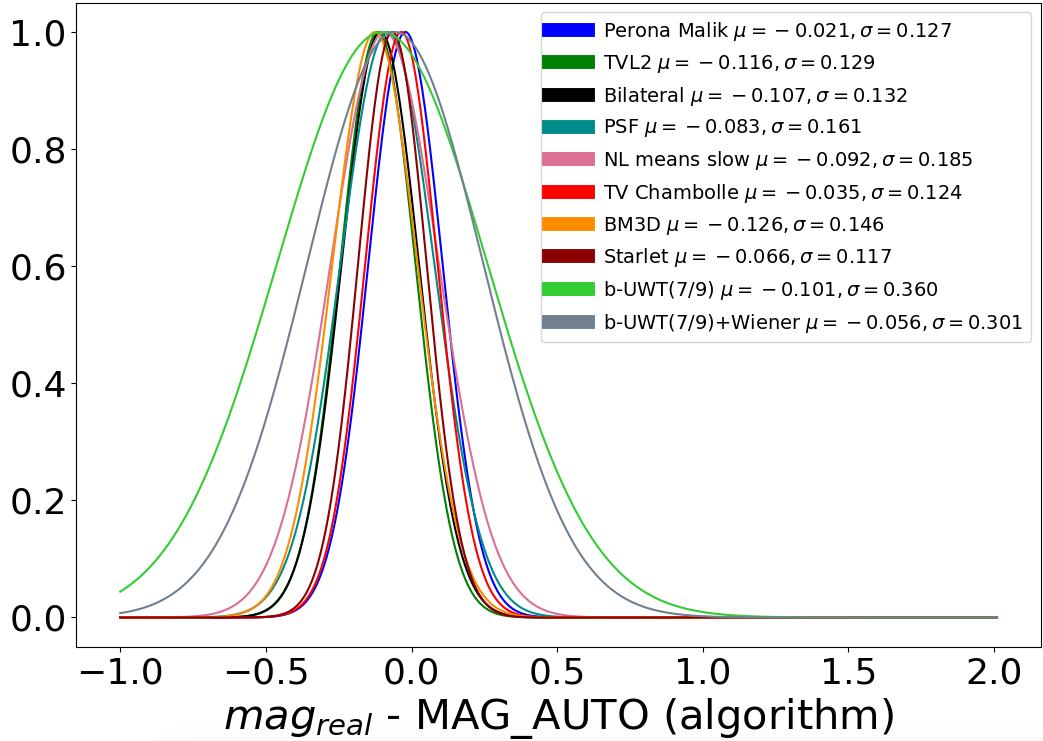}}
        \caption{\label{fig:flux_conservation_all_hist} Step 6: Flux conservation distribution for objects with magnitude within 19 and 23. On the x-axis, the difference between the magnitude measured \texttt{MAG\_AUTO} and the real objects magnitude from the catalog ($mag_{real}$). On the y-axis the \texttt{MAG\_AUTO} - $mag_{real}$ probability distribution function. Only the detected objects within the purity and completeness thresholds (Sect. \ref{subsec:completeness-purity}) are considered. $\mu$ and $\sigma$ are the distribution mean and the standard deviation values.}
\end{figure}

In this final test, we compare the total fluxes (by using \textsc{SExtractor} \texttt{MAG\_AUTO}) measured on the simulated denoised images with the true input fluxes for objects with magnitude within 19 and 23. 
We do not consider here the Gaussian filtering but only the PSF filtering, since the last one is the reference method for this analysis. The results are shown in Figs. \ref{fig:flux_conservation_all}-\ref{fig:flux_conservation_all_hist}. 
The standard deviation of the difference between measured and true magnitudes is $\sim$0.16 for PSF-filtered images. All denoising methods show similar performance with the exception of b-UWT(7/9) ($\sigma_{mag}$=0.36) and b-UWT(7/9)+Wiener ($\sigma_{mag}$=0.30).
We conclude that denoising algorithms preserve the overall calibration of the input images and they enable a photometric accuracy comparable to the one usually achieved on images filtered with the PSF.
 

\section{Test on real images}\label{sec:realimages}
After having analyzed the performance of denoising techniques on a series of simulated images, testing their performance with stationary and non-stationary Gaussian noise,  we test the algorithms on real images, using the HST $H160$ observations of the GOODS-South Field and a crop of the HAWK-I survey. 
Notably, the analysis of real observations provides us with a straightforward test of more realistic situations, in particular concerning the presence of non-Gaussian noise and/or correlated noise.

\subsection{Space telescope images}\label{subsec:realimagesspace}
\begin{figure}
        \resizebox{\hsize}{!}{\includegraphics{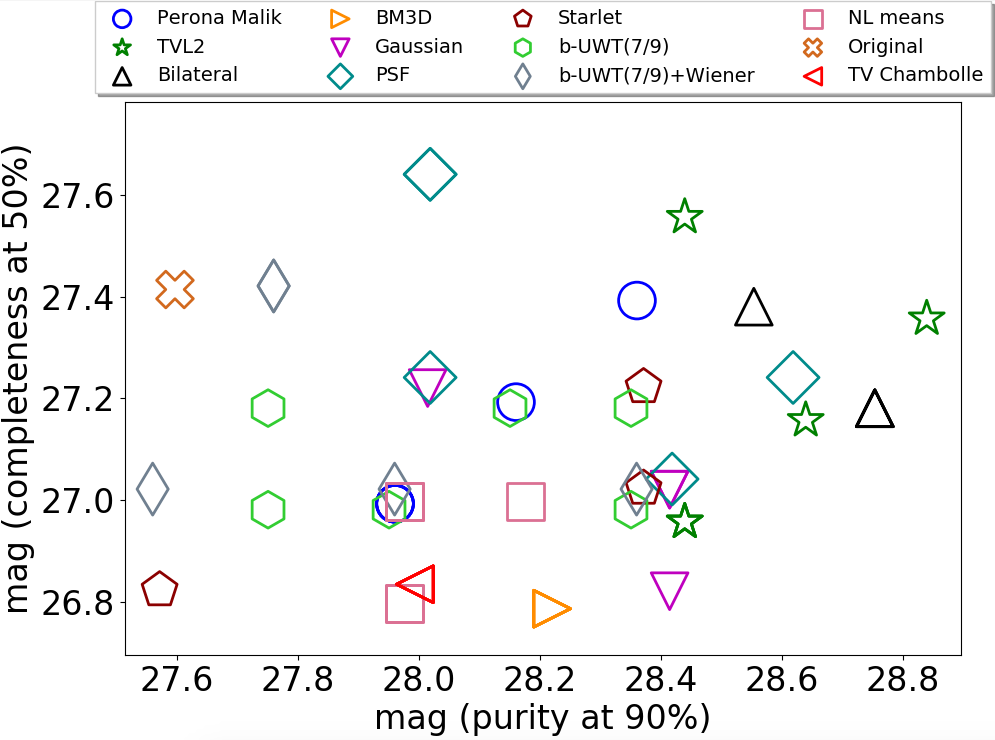}}
        \caption{\label{fig:real_images} Space telescope real Images Completeness \& Purity (GSDEEP and HUDF09). On the x-axis the magnitude at which the purity drops below 90\%, on the y-axis the magnitude at which the completeness drops below 50\%. Each symbol is referred to a different denoising method, which can be present multiple times in the plot due to different combinations of detection parameters, see text for details. The positions of the symbols are slightly randomized to improve readability.}
\end{figure}

We use two images of the area of the Hubble Ultra Deep Field: one at the full depth released with the official CANDELS mosaics that includes all WFC3 observations of that region (HUDF09, reaching H160=29.74 at SNR=5), the second, shallower one at the depth obtained with WFC3 observations of the CANDELS Treasury Program alone (GSDEEP, H160=28.16 at SNR=5) \citep{Koekemoer2011,Grogin2011}. We will use the former, deeper image as "true sky", against which we will compare the performance of denoising techniques on the shallower image. Using an analysis similar to that in Sect. \ref{subsec:completeness-purity}, we take as reference catalog the one obtained running \textsc{SExtractor} on HUDF09 with conservative detection parameters. The goal is again to check completeness and purity. We use again an association radius $R_{assoc}$ of $3 \times FWHM$, now corresponding to $7.5$ pixels. We identify an object as spurious using the same criteria used in Sect. \ref{subsec:completeness-purity} with a $mag_{aperture}$ within a circular aperture with diameter of $7.5$ pixels. The resulting plot, visible in Fig. \ref{fig:real_images}, is similar to the one obtained on the simulated image (Fig. \ref{fig:completeness-purity}). Clearly, TVL2 outperforms all the other algorithms, in particular the PSF by 0.2 mag in completeness and purity (or alternatively 0.4 magnitudes more complete and 0.2 magnitudes less pure); Bilateral performs better than the PSF filtering, by an amount of 0.2 magnitudes in completeness.  Perona-Malik provides a 0.2 mag more complete and 0.2 less pure catalog, being in this way an alternative of the PSF filtering. Starlet and b-UWT(7/9) perform slightly worse than the PSF.

As done in Sect. \ref{subsec:completeness-purity}, we show the snapshots of a sample of objects detected by the different methods in the GSDEEP image in \appendixname{ E}. 
The algorithms reported are: PM, TVL2, Bilateral, NL means, BM3D, Starlet, b-UWT(7/9), and b-UWT(7/9)+Wiener.

\subsection{Ground-based images} \label{subsec:realgroundbased}
\begin{figure}[h!]
        \resizebox{\hsize}{!}{\includegraphics{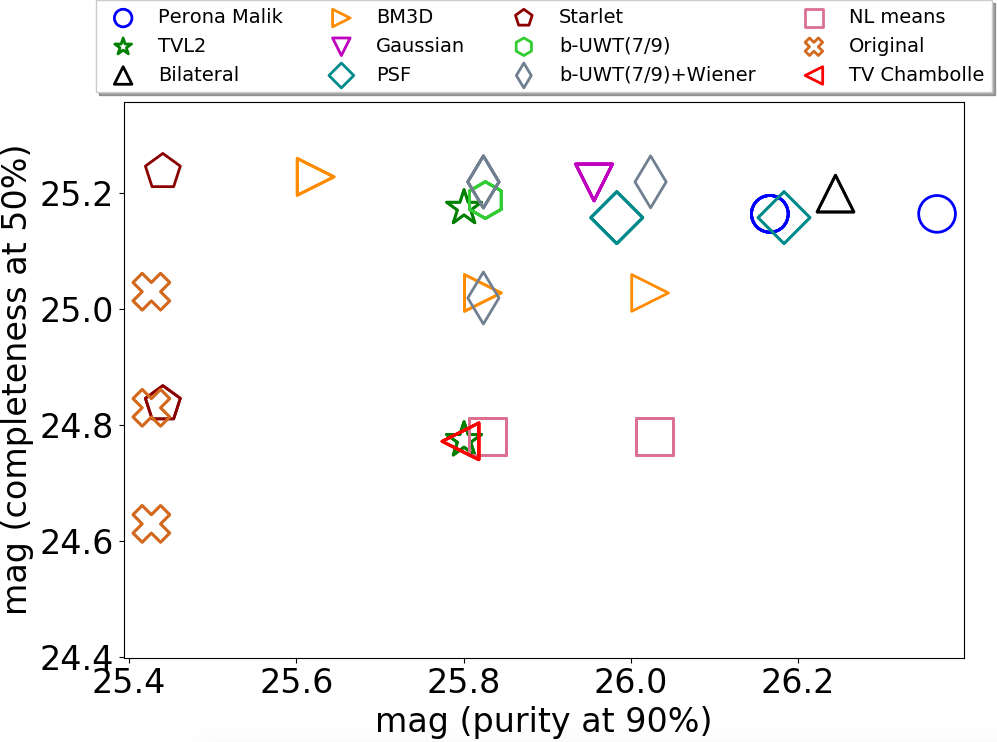}}
        \caption{\label{fig:real_images_ground} Ground-based real Images Completeness \& Purity (HAWK-I and HAWK-I UDS). On the x-axis the magnitude at which the purity drops below 90\%, on the y-axis the magnitude at which the completeness drops below 50\%. Each symbol is referred to a different denoising method, which can be present multiple times in the plot due to different combinations of detection parameters, see text for details.  The positions of the symbols are slightly randomized to improve readability.}
\end{figure}

We repeat the same tests described above on two Ks-band observations of the Goods-South field acquired with the HAWK-I imager at the VLT: a shallower observation of the field presented in  \citet{ks_shallow} and the $\sim$2 magnitude deeper observation released by the HUGS Survey \citep[][see Table~\ref{table:table_images}]{ks_deep}. As done before, we use the deepest image as "true sky" and we apply the algorithms to reduce the noise on the shallow image. We use again the association radius $R_{assoc}$ of $3\times FWHM$ corresponding to 11.25 pixels, with the relative $mag_{aperture}$ (11.25 pixels diameter), identifying an object as spurious using again the same criteria already used in Sect. \ref{subsec:completeness-purity} and Sect. \ref{subsec:realimagesspace}. %\ref{sec:realimages}. 
The resulting plot (see Fig. \ref{fig:real_images_ground}) shows that these algorithms improve the image detection compared to not making denoising at all (same result obtained in Sect. \ref{subsec:completeness-purity} and Sect. \ref{subsec:realimagesspace}), whereas they do not provide significant improvements compared to the PSF. Indeed, only PM creates a catalog of 0.2 magnitudes more pure at the same completeness. These results are in agreement with those in Sect. \ref{subsec:fwhm_std}, where we noticed that all these methods give the best with high resolution images (see Fig. \ref{fig:fwhm_vis_2}), such as VIS and H160. Indeed the lower resolution of the ground-based images impacts the algorithms performance. In the same way the methods, and mainly the PSF, perform better for lower SNR images (e.g. HAWK-I compared to VIS and H160), as shown in Figs. \ref{fig:noised_f160_3}-\ref{fig:noised_f160}, resulting in less significant improvements from the methods compared to the PSF.

\subsection{Artifacts and visual inspection}\label{subsec:artifacts}

The analysis made in 
\ref{subsec:completeness-purity} only considers the objects SNR and magnitude to classify the detections as \textit{correct} or \textit{spurious}. The algorithms perform denoising following different strategies,  and artifacts related to the family of these methods can be created. 
Several crops denoised by different tested methods are reported in \appendixname{ D-E}. 
By visually inspecting such snapshots, 
we can get an idea of the features and the artifacts produced. 
As a general comment, looking at the figures reported in \appendixname{ D}, TVL2 seems to be the best method, closer to the noiseless image reported in the last column of all the figures, followed by   
Starlet, b-UWT(7/9), and b-UWT(7/9)+Wiener, which also produce good results. 
PM, TV Chambolle, and BM3D are often very similar from a qualitative point of view. 
Most of the time the flux of the objects detected by TVL2 is highly concentrated (see e.g. Figs. \ref{fig:vis144}, \ref{fig:vis2446}), %,\ref{fig:gsdeep509}),
but in some cases the objects flux is distributed on a larger area, as visible e.g. in Fig. \ref{fig:vis2569}.  Starlet produces in some cases bright isotropic objects, which could lead to misleading morphological information (see e.g. Figs. \ref{fig:vis153}, \ref{fig:vis3616}).
Looking at Fig. \ref{fig:vis3924}, PSF seems to detect wrongly two objects, differently from the other denoising methods. 
This qualitative analysis is confirmed by the corresponding quantitative analysis made, as visible in Fig. \ref{fig:completeness-purity}.\\
\noindent When more than one object is present in the figures, the different schemes sometimes are not able to recognize all the objects (see e.g. Figs. \ref{fig:vis1845}, \ref{fig:vis2569}, \ref{fig:vis379}, \ref{fig:vis4495}, \ref{fig:vis4712}). 
Moreover, fluctuations of the background around the objects are not completely removed, 
and the smoothing can create artifacts (see e.g.  Fig. \ref{fig:vis4712}, where an elliptical galaxy seems to appear as a spiral galaxy after denoising). 
b-UWT(7/9), and b-UWT(7/9)+Wiener are susceptible to background fluctuations and create visual artifacts (see Figs. \ref{fig:vis402}, \ref{fig:vis799}). 

Inspecting the snapshots from real images, a qualitative analysis is inconclusive, so we have to stick to the quantitative analysis in Fig. \ref{fig:real_images}. In fact, looking at the crops reported in \appendixname{ E}, PM and Bilateral are often so similar that it is difficult to distinguish them, although from Fig. \ref{fig:real_images} the differences between them are clear (PM performs worse than Bilateral). 
Analogous remark can be made for BM3D, which  produces images similar to PM and Bilateral, as visible e.g. in Fig. \ref{fig:gsdeep1772}. 
TVL2 is again the most promising method, visually close to the HUDF09 also thanks to the automatic optimization of its internal  parameters, and this is confirmed by the results reported in Fig. \ref{fig:real_images}. 
Images from Starlet, b-UWT(7/9), and b-UWT(7/9)+Wiener are quite similar to each other, even though those from Starlet present less visual artifacts (see Figs. \ref{fig:gsdeep100}, \ref{fig:gsdeep1772}, \ref{fig:gsdeep2051}, \ref{fig:gsdeep607}).
Differently from the other methods, NL means produces nearly squared patches of uniform flux, which effectively reduce noise fluctuations (see Figs. \ref{fig:gsdeep73},  \ref{fig:gsdeep909}), but also could lead to artifacts (see e.g. Fig. \ref{fig:gsdeep2014}, where the galaxy in the bottom-right corner assumes a nearly boxy-shape).

Even if artifacts are generated, and the morphology of the objects detected could be altered, we tested the efficiency of the algorithms in preserving the FWHM and the ellipticity of the objects in Sect. \ref{subsec:fwhm_conservation}, proving that on the average the shapes are preserved. 
A more detailed assessment of the effect on galaxy morphology is beyond the scope of the present work. However, we note that a promising improvement in this direction can be obtained by classifying objects using a convolutional neural network \citep[e.g. one of the classifiers proposed in ][]{tuccillo17,Khalifa2017,barchi2020machine}, in order to assess whether their morphological class is preserved after denoising.

%______________________________________________________________
\section{Summary and conclusions} \label{conclusions}
The goal of this work is to make an extensive comparison between a number of denoising algorithms, aimed at identifying the best choice to improve the detection of faint objects in astronomical extragalactic images (e.g. considering the typical cases of HST and Euclid). To this purpose, we performed a large set of tests on simulated images. We also tested the methods on real images: from space and ground-based, collecting really interesting hints on many situations. 

We chose to test a significant number of denoising algorithms based on traditional techniques (mainly, PDEs and variational methods), 
leaving a more complete comparison, including machine learning techniques, for future works. In particular, we point out that ATVD-TVL2, Bilateral, Perona-Malik,  TV Chambolle, Starlet, and b-UWT(7/9)+Wiener are the most interesting to use among all, since they provide good performance in the different tests proposed, closely followed by BM3D.
Even if most of these methods are quite unusual for the astronomical community, they are very well-known in many other fields. They are known to outperform a straightforward PSF/Gaussian filtering, which is the standard choice in astronomy. We therefore considered these techniques as the reference ones, against which we tested all the other methods.

As a first test, we considered the two metrics \textit{MSE} and \textit{SSIM} (defined in Sect. \ref{subsec:quality_tests}) and checked which methods yield the best performance with respect to them. 
We compared their performance again through \textit{MSE} and \textit{SSIM} in relation to depth, resolution and type of image. We tested the algorithms ability to preserve the FWHM, in order to understand if they can preserve the shape of the objects, useful in case photometric measurements on the denoised image are needed. We tested their stability using the \textit{MSE}, varying the ideal parameter of a fixed percentage, with the goal of having a hint on their reliability, in case the best parameter is chosen wrongly. We have also tested possible detection improvements through two diagnostics, $completeness$ and $purity$, which are used to measure the fraction of real detections on the total number of objects in the image. 
Finally, we applied these methods on real images (CANDELS-GS-deep and a crop of HAWK-I). We summarize below the key points of the analysis performed in this paper:
\begin{itemize}
    \item From \textit{MSE} and \textit{SSIM} we noticed that 11 algorithms are always on top of rankings, especially for VIS and H160 images, which are of main interest for the detection in future surveys. These algorithms are: PM with edge-stopping function $g_1$ and $k=10^{-3}$, TVL2, BM3D, Starlet, b-UWT(7/9), b-UWT(7/9)+Wiener, Gaussian, PSF, NL means slow, Bilateral, and TV Chambolle
    \item From the stability test discussed in Sect. \ref{subsec:nonstationary}, the algorithms tested are not influenced (or negligibly influenced in few cases) by images with non-stationary Gaussian noise
    \item From the PSF and Depth variation test we noticed 
    that most of the methods are performing better with narrower FWHM. Whereas all the methods perform better with the noise level increasing (in terms of Gaussian noise standard deviation), their improvements are more significant  compared to the PSF, with high SNR images
    \item From the FWHM conservation test, we noticed that most of the algorithms tend to not smooth the image, in terms of FWHM increments. On the contrary, the PSF smoothing provides an offset in the FWHM measurement, for both, stars-only image and galaxies-image. On the other hand, the ellipticity is well preserved by the PSF and most of the algorithms
    \item From the completeness and purity test we found a small number of algorithms which provide 0.2 magnitudes more pure and complete catalogs than the PSF filtering, these are TVL2, Perona-Malik, TV Chambolle and Bilateral (whereas Starlet and b-UWT(7/9)+Wiener provide a 0.2 increment only for completeness)
    \item From Flux conservation test we found that most of the algorithms have similar performance to the PSF filtering, preserving the overall calibration of the input images
    \item From real image test (H160) we found that TVL2 outperforms all the other algorithms, and it is the only one that performs better than the PSF of 0.2 magnitudes in completeness and purity, whereas Bilateral produces only a 0.2 more pure catalog
    \item From real image test (HAWK-I) we found that only Perona-Malik outperforms the PSF filtering, by 0.2 magnitudes in purity, the other methods performs worse/similarly to the PSF.
    \item From the visual inspection performed in Sect. \ref{subsec:artifacts} on simulated and real images, 
    the best methods seems to be TVL2, PM, Bilateral, and Starlet, although they also generates artifacts like the other methods tested. The visual inspection confirms the results visible in Fig. \ref{fig:completeness-purity} and Fig. \ref{fig:real_images}. 
\end{itemize}

The results we obtained demonstrate that denoising algorithms should be considered valuable tools in presence of Gaussian noise, which is a good approximation of the noise in optical and near-infrared extragalactic surveys, as they clearly improve the detection of faint objects. 
Structure-texture image decomposition, Total Variation denoising, Perona-Malik, Bilateral filtering, and Undecimated Wavelets Transform methods are of particular interest.  
While further specific tests are needed to define for each survey the optimal approach, and its parameter set, among the above mentioned ones,  
our investigation on a small but reasoned reference set of simulated and real extragalactic images shows that the scientific return of ongoing and future surveys can be significantly enhanced by the adoption of these denoising methods in place of standard filtering approaches. 
Moreover, in our opinion the use of the increasingly popular machine learning techniques, possibly combined with the best methods coming out from our analysis, 
has the potential to further improve on the performance of "traditional" denoising techniques described here.
Also learning approaches for adding morphological priors \citep[see e.g.][]{PFS10} could be useful, subject of future works. 

\begin{acknowledgements}
We would like to thank the referee, Prof. Jean-Luc Starck, for the many useful comments and suggestions that helped us improving our work. 
This research has been carried on within the INdAM-INAF project FOE 2015 “OTTICA ADATTIVA”.  
The authors S. Tozza and M. Falcone are members of the INdAM Research Group GNCS, and gratefully acknowledge its financial support to this research.
\end{acknowledgements}

%\newpage
\bibliographystyle{aa} % style aa.bst
\bibliography{aanda} % your references Yourfile.bib
%-------------------------------------------------------------------
\newpage
\appendix
\onecolumn
\section{MSE comparison tables and plots}
\begin{figure*}[ht]
        \sidecaption
        \includegraphics[width=12cm]{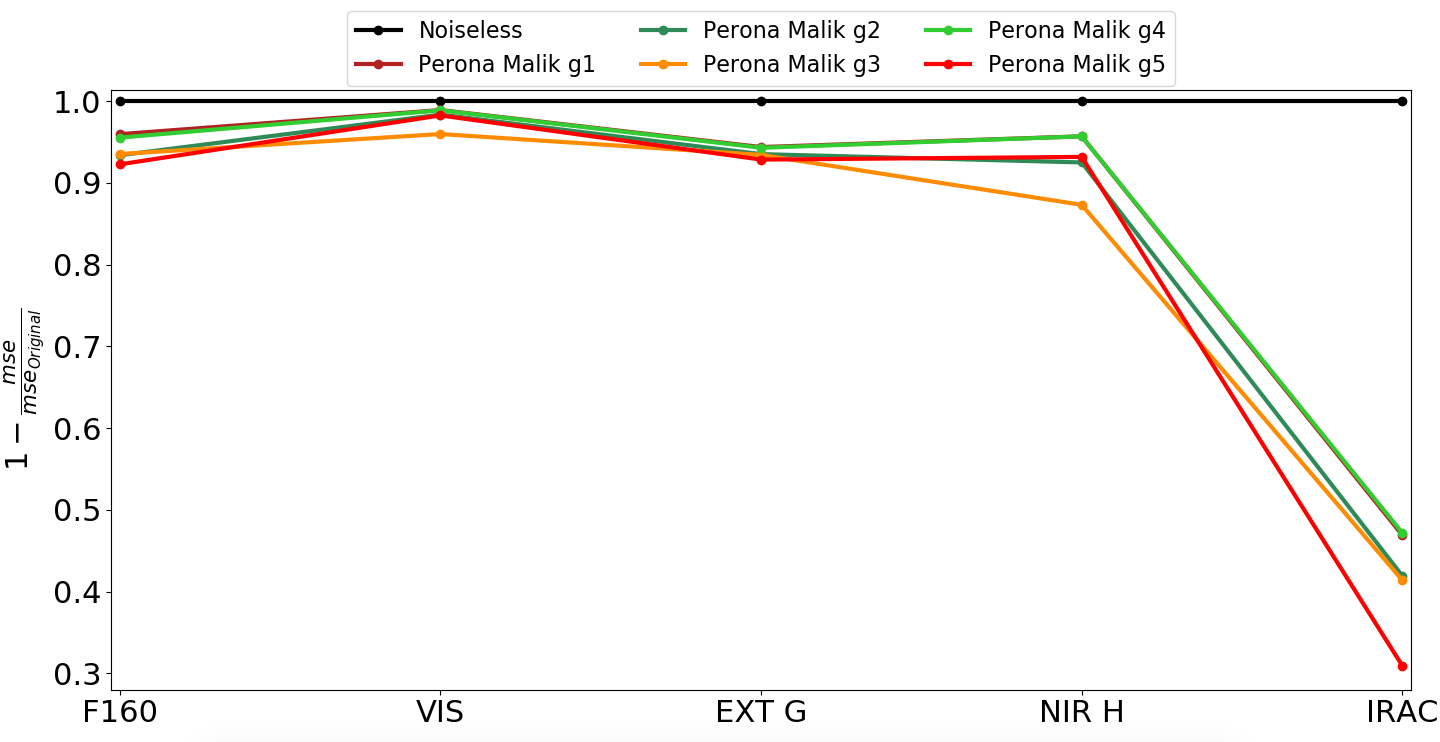}
        \caption{\label{fig:PM_mse_plot} {Step 1: MSE comparison between Perona-Malik functions on CM. On the x-axis, all the simulated CM crops in the different bands, whereas on the y-axis $1-\frac{mse}{mse_{Original}}$.}}
\end{figure*} 
\begin{figure*}[ht]
        \sidecaption
        \includegraphics[width=12cm]{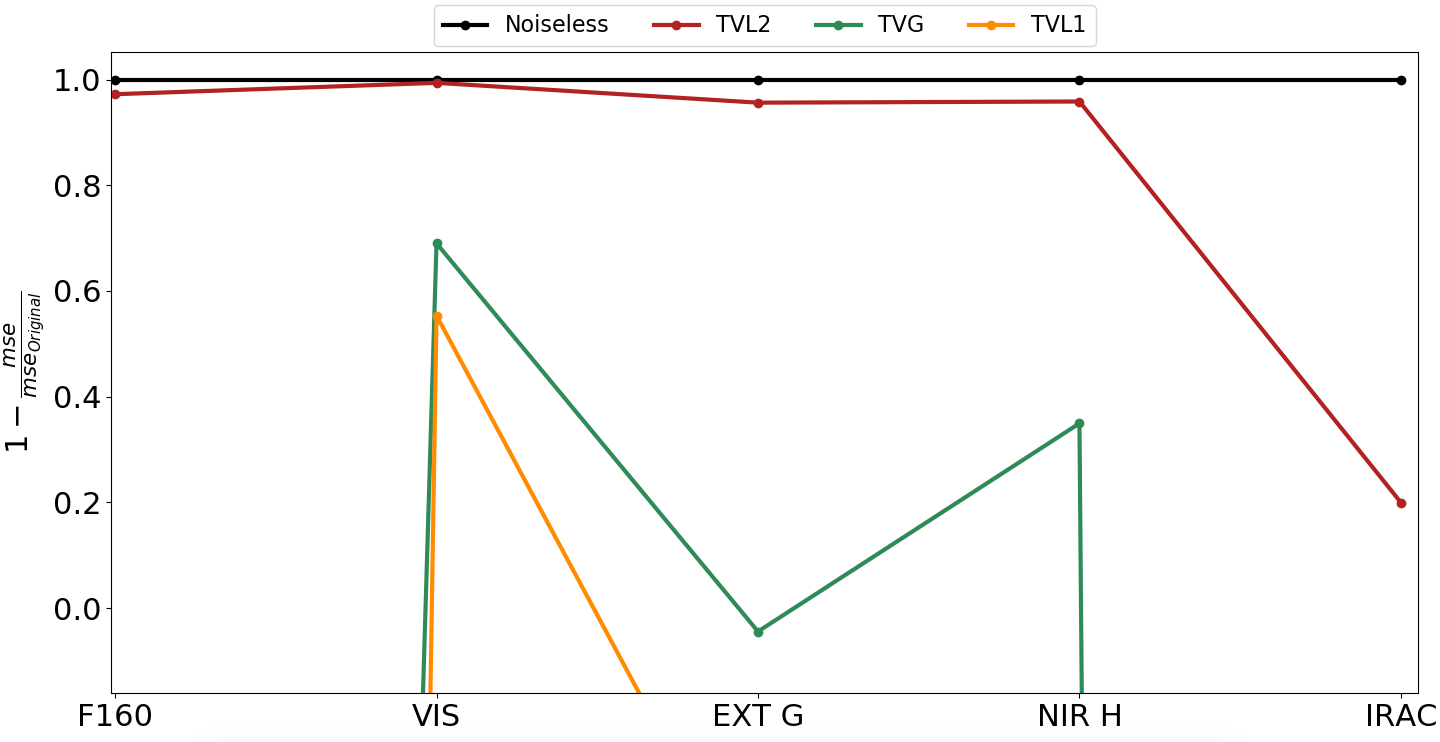}
        \caption{\label{fig:ATVD_mse_plot} Step 1: MSE comparison between ATVD algorithms on BG. On the x-axis, all the simulated BG crops in the different bands, whereas on the y-axis $1-\frac{mse}{mse_{Original}}$.}
\end{figure*}
\begin{figure*}[ht]
        \sidecaption
        \includegraphics[width=12cm]{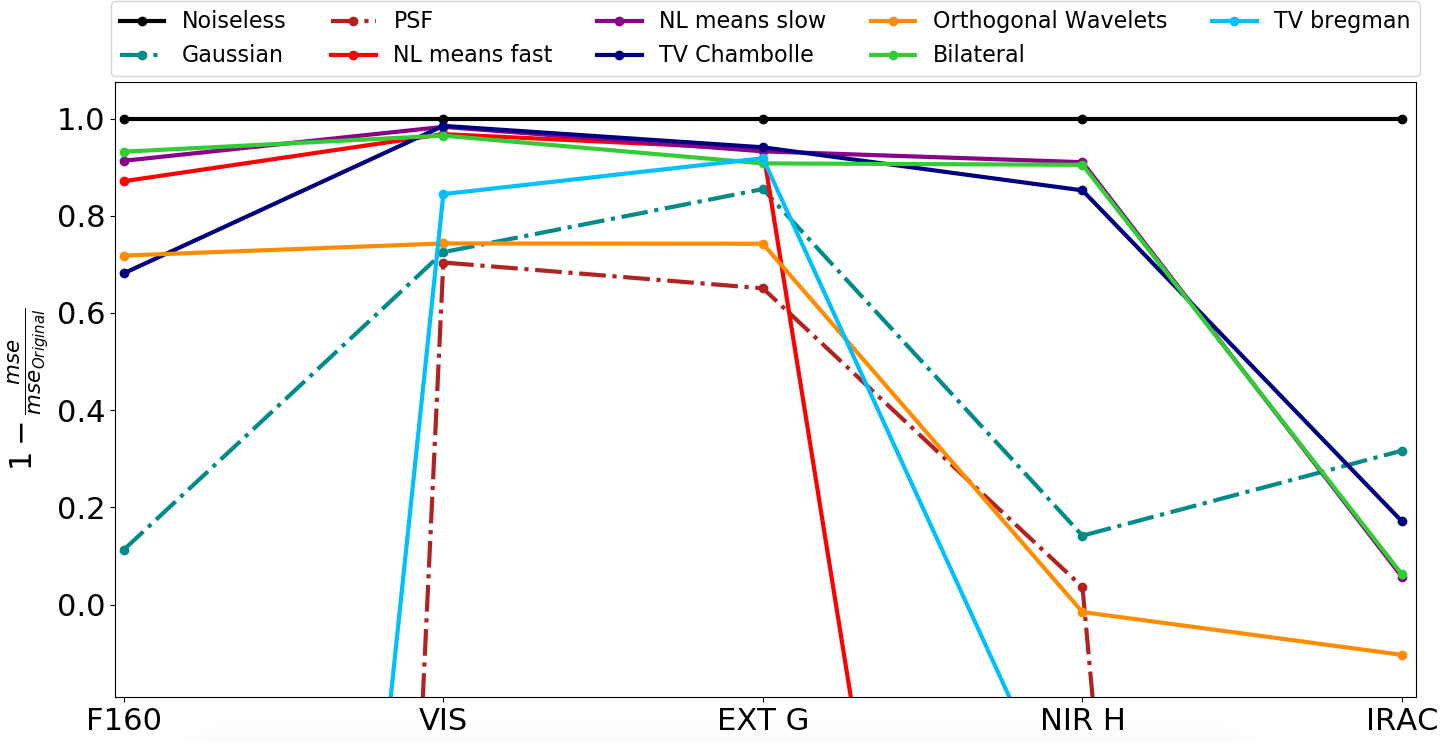}
        \caption{\label{fig:other_mse_plot} Step 1: MSE comparison between the other algorithms excluding ATVD and PM on CL. On the x-axis, all the simulated CL crops in the different bands, whereas on the y-axis $1-\frac{mse}{mse_{Original}}$.}
\end{figure*}

\begin{figure*}[ht]
        \sidecaption
        \includegraphics[width=12cm]{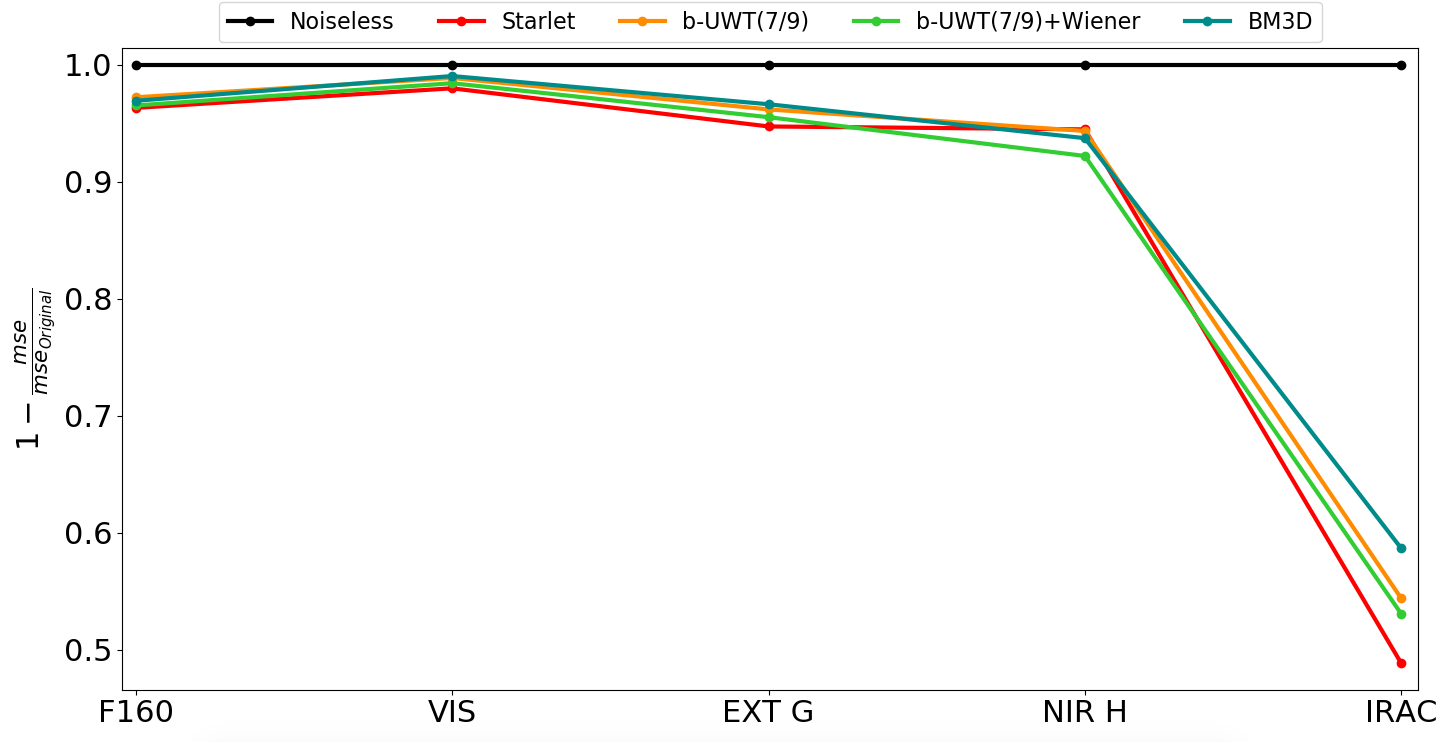}
        \caption{\label{fig:new_mse_plot} Step 1: MSE comparison between BM3D, Starlet and the two b-UWTs on CM. On the x-axis, all the simulated CM crops in the different bands, whereas on the y-axis $1-\frac{mse}{mse_{Original}}$.}
\end{figure*}
%
%
% the \\ insures the section title is centered below the phrase: Appendix A
\begin{table*}[ht]
\sisetup{detect-weight,mode=text,group-minimum-digits = 4}
\centering
\begin{tabular}{llllll
S[table-format=3.0]@{\,}c@{\,}l@{\,}c@{\,}l
  S[table-format=5.0]
  S[table-format=8.0]
  S[table-format=4.0]@{\,}c@{\,}S[table-format=2.0]@{}>{${}}c<{{}$}@{}S[table-format=2.0]}
\hline
\hline
name          & $MSE_{H160}$ & $MSE_{VIS}$ & $MSE_{EXT G}$ & $MSE_{NIR H}$ & $MSE_{IRAC}$ \\ \hline
TVL2          & \num{5.320e-8} & \bf{\num{7.286e-7}} & \num{3.672e-7} & \bf{\num{1.964e-4}}  & \num{7.051e-5} \\
TVL1          &  \num{7.297e-5}&\num{5.155e-5}       & \num{1.325e-5} & \num{4.409e-2} & \num{4.429e-2} \\
TVG           &  \num{3.922e-5}&\num{3.576e-5}       & \num{8.760e-6} & \num{3.071e-3} & \num{6.370e-3} \\
PM g=1 k=\num{1e-3}&  \num{9.490e-8}&\num{1.108e-6}       & \num{6.776e-7} & \num{2.528e-4} & \num{5.473e-5} \\
PM g=2 k=\num{1e-3}&  \num{9.790e-8}&\num{8.941e-5}       & \num{8.147e-6} & \num{1.926e-3} & \num{7.073e-5} \\
PM g=3 k=\num{1e-3}&  \num{9.930e-8}&\num{1.122e-4}       & \num{8.390e-6} & \num{3.640e-3} & \num{7.855e-5} \\
PM g=4 k=\num{1e-4}&  \num{7.960e-8}&\num{8.291e-5}       & \num{7.641e-6} & \num{2.760e-4} & \num{6.288e-5} \\
PM g=5 k=\num{1e-3}&  \num{1.077e-7}&\num{8.843e-5}       & \num{8.131e-6} & \num{1.775e-3} & \num{7.157e-5} \\
PSF  &  \num{8.557e-6}&\num{1.833e-5}       & \num{1.615e-6} & \num{3.492e-3} & \num{1.174e-3} \\
Original         &  \num{1.912e-6}&\num{1.154e-4}       & \num{8.390e-6} & \num{4.722e-3} & \num{8.811e-5} \\
TV Bregman    &  \num{3.778e-6}&\num{5.908e-6}       & \num{6.147e-7} & \num{1.960e-3} & \num{2.620e-4} \\
Gaussian      &  \num{1.301e-6}&\num{1.724e-5}       & \num{9.434e-7} & \num{3.469e-3} & \num{6.955e-5} \\
NL means slow &  \num{8.940e-8}&\num{1.513e-6}       & \num{5.992e-7} & \num{3.066e-4} & \num{6.454e-5} \\
NL means fast &  \num{1.019e-7}&\num{1.628e-6}       & \num{5.750e-7} & \num{5.551e-3} & \num{1.389e-4} \\
Bilateral     &  \num{1.109e-7}&\num{3.754e-6}       & \num{7.972e-7} & \num{4.612e-4} & \num{6.297e-5} \\
TV Chambolle  &  \num{1.876e-7}&\num{2.112e-6}       & \num{4.914e-7} & \num{2.894e-4} & \num{5.444e-5} \\
Orthogonal Wavelets      &  \num{5.119e-7}&\num{2.962e-5}       & \num{2.117e-6} & \num{4.776e-3} & \num{9.583e-5} \\ 
BM3D & \bf{\num{4.830e-8}}& \num{1.358e-6} &\bf{\num{2.999e-7}} & \num{2.974e-4} & \bf{\num{4.474e-5}}\\
Starlet & \num{6.760e-8} & \num{2.224e-6}& \num{5.061e-7}& \num{2.978e-4}& \num{5.211e-5}\\
b-UWT(7/9) & \num{4.900e-8}& \num{1.116e-6}& \num{3.707e-7}& \num{2.822e-4}& \num{4.771e-5}\\
b-UWT(7/9)+Wiener & \num{6.390e-8} & \num{1.645e-6}& \num{4.610e-7}& \num{3.780e-4}& \num{4.907e-5}\\
\hline \hline
\end{tabular}
\caption{MSE table of BG crops. The lowest MSE value per band is indicated in bold}
\label{tab:msebg}
\end{table*}
\begin{table*}[ht]
\centering
\sisetup{detect-weight,mode=text,group-minimum-digits = 4}
\begin{tabular}{llllll
S[table-format=3.0]@{\,}c@{\,}l@{\,}c@{\,}l
  S[table-format=5.0]
  S[table-format=8.0]
  S[table-format=4.0]@{\,}c@{\,}S[table-format=2.0]@{}>{${}}c<{{}$}@{}S[table-format=2.0]}
\hline
\hline
Name          & $MSE_{H160}$ & $MSE_{VIS}$ & $MSE_{EXT G}$ & $MSE_{NIR H}$ & $MSE_{IRAC}$ \\ \hline
TVL2          & \num{5.590e-8} & \bf{\num{8.746e-7}} & \num{3.268e-7} & \bf{\num{1.808e-4}} & \num{5.699e-5} \\
TVL1          &  \num{5.878e-4}&\num{1.075e-3}       & \num{5.485e-5} & \num{2.390e-1} & \num{8.910e-2} \\
TVG           &  \num{2.977e-5}&\num{3.147e-5}       & \num{4.477e-5} & \num{3.672e-3} & \num{5.752e-3} \\
PM g=1 k=\num{1e-3}&  \num{1.248e-7}&\num{2.277e-6}               & \num{4.646e-7} & \num{2.625e-4} & \num{3.886e-5} \\
PM g=2 k=\num{1e-3} &  \num{1.257e-7}&\num{1.889e-6}       & \num{4.310e-6} & \num{3.503e-4} & \num{4.969e-5} \\
PM g=3 k=\num{1e-3} &  \num{1.228e-7}&\num{8.549e-6}       & \num{7.173e-6} & \num{1.386e-3} & \num{5.755e-5} \\
PM g=4 k=\num{1e-4} &  \num{8.470e-8}&\num{1.281e-6}       & \num{3.914e-6} & \num{2.011e-4} & \num{4.697e-5} \\
PM g=5 k=\num{1e-3} &  \num{1.468e-7}&\num{2.006e-6}       & \num{8.287e-6} & \num{3.186e-4} & \num{5.055e-5} \\
PSF  &  \num{4.853e-5}&\num{1.532e-4}               & \num{9.444e-6} &\num{6.024e-3}  & \num{5.052e-3} \\
Original         &  \num{1.902e-6}&\num{1.116e-4}               & \num{8.288e-6} & \num{4.679e-3} & \num{7.319e-5} \\
TV Bregman    &  \num{2.170e-5}&\num{1.084e-4}               & \num{2.264e-6} & \num{1.023e-2} & \num{6.225e-4} \\
Gaussian      &  \num{1.835e-6}&\num{6.773e-5}               & \num{1.913e-6} & \num{4.373e-3} & \num{7.094e-5} \\
NL means slow &  \num{1.201e-7}&\num{1.883e-6}               & \num{4.885e-7} & \num{7.916e-4} & \num{1.559e-4} \\
NL means fast &  \num{1.990e-7}&\num{5.729e-6}               & \num{4.655e-7} & \num{6.982e-3} & \num{1.814e-4} \\
Bilateral     &  \num{1.104e-7}&\num{4.027e-6}               & \num{7.348e-7} & \num{3.630e-4} & \num{4.779e-5} \\
TV Chambolle  &  \num{4.964e-7}&\num{1.698e-6}               & \num{4.723e-7} & \num{6.008e-4} & \num{5.398e-5} \\
Orthogonal Wavelets     &  \num{5.303e-7}&\num{2.999e-5}               & \num{2.128e-6} & \num{4.732e-3}& \num{7.795e-5} \\
BM3D & \num{5.880e-8}& \num{1.290e-6} & \bf{\num{2.807e-7}}& \num{2.941e-4}& \bf{\num{3.025e-5}}\\
Starlet & \num{7.040e-8}& \num{2.343e-6}& \num{4.380e-7}& \num{2.589e-4}& \num{3.744e-5}\\
b-UWT(7/9) & \bf{\num{5.310e-8}}& \num{1.319e-6}& \num{3.174e-7}& \num{2.654e-4}& \num{3.334e-5}\\
b-UWT(7/9)+Wiener & \num{6.650e-8}& \num{1.837e-6}& \num{3.723e-7}& \num{3.653e-4}&\num{3.435e-5}\\
\hline \hline
\end{tabular}
\caption{MSE table of CM crops. The lowest MSE value per band is indicated in bold}
\label{tab:msecm}
\end{table*}

\begin{table*}[ht]
\centering
\sisetup{detect-weight,mode=text,group-minimum-digits = 4}
\begin{tabular}{llllll
S[table-format=3.0]@{\,}c@{\,}l@{\,}c@{\,}l
  S[table-format=5.0]
  S[table-format=8.0]
  S[table-format=4.0]@{\,}c@{\,}S[table-format=2.0]@{}>{${}}c<{{}$}@{}S[table-format=2.0]}
\hline
\hline
Name          & $MSE_{H160}$ & $MSE_{VIS}$ & $MSE_{EXT G}$ & $MSE_{NIR H}$ & $MSE_{IRAC}$ \\ \hline
TVL2          & \num{7.070e-8}  & \bf{\num{8.958e-7}} & \num{3.443e-7} & \bf{\num{2.451e-4}} & \num{3.633e-4} \\
TVL1          &  \num{2.754e-4}&\num{1.707e-4}       & \num{1.471e-5} & \num{1.121e-1} & \num{5.820e-2} \\
TVG           &  \num{4.424e-5}&\num{3.336e-5}       & \num{1.387e-5} & \num{4.519e-3} & \num{1.013e-2} \\
PM g=1 k=\num{1e-3}&  \num{1.463e-7}&\num{1.752e-6}               & \num{5.923e-7} & \num{3.433e-4} &  \num{3.312e-4}\\
PM g=2 k=\num{1e-3} &  \num{1.437e-7}&\num{1.865e-5}       & \num{7.553e-6} & \num{4.436e-4} & \num{3.433e-4} \\
PM g=3 k=\num{1e-3} &  \num{1.431e-7}&\num{6.034e-5}       & \num{8.342e-6} & \num{1.413e-3} & \num{3.501e-4} \\
PM g=4 k=\num{1e-4} &  \num{1.029e-7}&\num{1.076e-5}       & \num{6.809e-6} & \num{2.893e-4} & \num{3.397e-4} \\
PM g=5 k=\num{1e-3} &  \num{1.668e-7}&\num{1.432e-5}       & \num{7.471e-6} & \num{4.163e-4} & \num{3.439e-4} \\
PSF  &  \num{2.728e-5}&\num{3.421e-5}               & \num{2.913e-6} & \num{4.576e-3} & \num{2.952e-3} \\
Original         &  \num{1.915e-6}&\num{1.155e-4}               & \num{8.342e-6} & \num{4.748e-3}& \num{3.640e-4} \\
TV Bregman    &  \num{1.233e-5}&\num{1.795e-5}               & \num{6.815e-7} & \num{7.211e-3} & \num{6.886e-4} \\
Gaussian      &  \num{1.699e-6}&\num{3.175e-5}               & \num{1.211e-6} & \num{4.074e-3} & \bf{\num{2.487e-4}} \\
NL means slow &  \num{1.664e-7}&\num{1.976e-6}               & \num{5.627e-7} & \num{4.264e-4} & \num{3.432e-4} \\
NL means fast &  \num{2.470e-7}&\num{3.702e-6}               & \num{5.260e-7} & \num{1.975e-2} & \num{4.581e-4} \\
Bilateral     &  \num{1.307e-7}&\num{4.055e-6}               & \num{7.697e-7} & \num{4.521e-4} & \num{3.413e-4} \\
TV Chambolle  &  \num{6.098e-7}&\num{1.740e-6}               & \num{4.938e-7} & \num{7.015e-4} & \num{3.017e-4} \\
Orthogonal Wavelets     &  \num{5.399e-7}&\num{2.968e-5}               & \num{2.149e-6} & \num{4.821e-3} & \num{4.016e-4} \\
BM3D & \bf{\num{6.130e-8}}& \num{1.103e-6} & \bf{\num{2.777e-7}}& \num{1.031e-1}&\num{3.199e-4}\\
Starlet & \num{8.400e-8}& \num{2.387e-6}& \num{4.358e-7}& \num{3.596e-4}&\num{3.311e-4}\\
b-UWT(7/9) & \num{6.610e-8}& \num{1.333e-6}& \num{3.284e-7}& \num{3.611e-4}&\num{3.248e-4}\\
b-UWT(7/9)+Wiener & \num{8.080e-8}& \num{1.856e-6}& \num{3.831e-7}& \num{4.791e-4}&\num{3.265e-4}\\
\hline \hline
\end{tabular}
\caption{MSE table of CL crops. The lowest MSE value per band is indicated in bold}
\label{tab:msecl}
\end{table*}

\begin{table*}[ht]
\centering
\sisetup{detect-weight,mode=text,group-minimum-digits = 4}
\begin{tabular}{lccccc
S[table-format=3.0]@{\,}c@{\,}l@{\,}c@{\,}l
  S[table-format=5.0]
  S[table-format=8.0]
  S[table-format=4.0]@{\,}c@{\,}S[table-format=2.0]@{}>{${}}c<{{}$}@{}S[table-format=2.0]}
\hline
\hline
Name          & $T_{H160}$ & $T_{VIS}$ & $T_{EXT G}$ & $T_{NIR H}$ & $T_{IRAC}$ \\ \hline
              & s & s & s & s & s \\
              \hline
TVL2          & 9.047& 3.588& 0.688& 0.132& 0.0938\\
TVL1          & 5.408& 0.814& 0.339& 0.020& 0.0031\\
TVG           & 8.910& 5.177& 0.334& 0.134& 0.0914\\
PM g=1 k=\num{1e-3}& 6.353& 5.283& 2.127& 0.362& 0.063\\
PM g=2 k=\num{1e-3}& 10.519& 22.826& 0.248& 8.347& 0.159&\\
PM g=3 k=\num{1e-3}& 6.847& 30.578& 12.136& 5.544& 0.142&\\
PM g=4 k=\num{1e-4}& 88.506& 323.720& 109-356& 50.417& 6.512& \\
PM g=5 k=\num{1e-3}& 6.778& 16.001& 51.941& 11.955& 0.125&\\
PSF  &0.071& 0.019& 0.008& 0.008 & 0.001\\
Original         & n.a.& n.a.& n.a.& n.a.& n.a.\\
TV Bregman    & 0.222& 0.093& 0.021& 0.025& 0.011\\
Gaussian      & \bf{0.055}& \bf{0.017}& \bf{0.005}&\bf{0.002} & \bf{0.001}\\
NL means slow & 75.94& 27.75& 6.863& 3.344& 1.058\\
NL means fast & 7.514& 3.104& 1.118& 0.448& 0.208\\
Bilateral     & 37.09& 13.46& 3.602& 1.567& 0.489\\
TV Chambolle  & 10.61& 0.7668& 0.109& 0.587& 0.034\\
Orthogonal Wavelets      & 0.827& 0.329& 0.091& 0.038& 0.012\\ 
BM3D & 25.08& 10.499 & 2.836& 1.536& 0.655\\
Starlet & 7.671& 2.752& 0.702& 0.314& 0.083\\
b-UWT(7/9) & 9.013& 3.265& 0.812& 0.350& 0.095\\
b-UWT(7/9)+Wiener & 33.15& 12.01& 2.983& 1.318& 0.337\\
\hline \hline
\end{tabular}
\caption{CPU Time table of CM crops after fixing the optimal internal parameters for each method.The lowest time value per band is indicated in bold}
\label{tab:cpucm}
\end{table*}
% the \\ insures the section title is centered below the phrase: Appendix B
\newpage
\onecolumn
\section{Non-stationary Gaussian noise MSE comparison table}
\begin{table*}[ht]
\centering
\sisetup{detect-weight,mode=text,group-minimum-digits = 4}
\begin{tabular}{lllllr
S[table-format=3.0]@{\,}c@{\,}l@{\,}c@{\,}l
  S[table-format=5.0]
  S[table-format=8.0]
  S[table-format=4.0]@{\,}c@{\,}S[table-format=2.0]@{}>{${}}c<{{}$}@{}S[table-format=2.0]}
\hline
\hline
Name                & $MSE_{H_{l}}$ & $MSE_{I_{\sigma}}$ & $MSE_{H_{u}}$ & $MSE_{I_{2\sigma}}$ \\ \hline

TVL2                &7.622e-07 &\bf{7.654e-07} &2.477e-05 &2.500e-05 \\
PM g=1 k=\num{1e-3} &9.277e-05 &9.412e-05 &4.352e-04 &4.343e-04 \\
PSF                 &1.398e-05 &1.398e-05 &5.239e-05 &5.242e-05 \\
Original            &1.153e-04 &1.152e-04 &4.604e-04 &4.604e-04 \\
Gaussian            &4.575e-05 &4.574e-05 &1.822e-04 &1.823e-04 \\
NL means slow       &1.110e-04 &1.114e-04 &4.603e-04 &4.426e-04 \\
Bilateral           &1.129e-04 &1.130e-04 &4.581e-04 &4.505e-04 \\
TV Chambolle        &6.246e-05 &6.387e-05 &3.428e-04 &3.411e-04 \\
BM3D                &1.141e-04 &1.141e-04 &4.592e-04 &4.556e-04 \\
Starlet             &\bf{5.647e-07} &8.924e-07 &1.907e-05 &2.571e-06 \\
b-UWT(7/9)          &8.316e-07 &7.797e-07 &\bf{8.632e-06} &\bf{2.255e-06} \\
b-UWT(7/9)+Wiener   &7.954e-07 &1.009e-06 &1.957e-05 &3.191e-06 \\
\hline \hline
\end{tabular}
\caption{MSE values for $H_{l}$,$I_{\sigma}$,$H_{u}$,$I_{2\sigma}$ related to the VIS (CM) mirrored crop. See Sect. \ref{subsec:nonstationary} for further details. The best results are in bold}
\label{tab:cm_mirrored}
\end{table*}
\nopagebreak
\begin{table*}[ht]
\centering
\sisetup{detect-weight,mode=text,group-minimum-digits = 4}
\begin{tabular}{lllllr
S[table-format=3.0]@{\,}c@{\,}l@{\,}c@{\,}l
  S[table-format=5.0]
  S[table-format=8.0]
  S[table-format=4.0]@{\,}c@{\,}S[table-format=2.0]@{}>{${}}c<{{}$}@{}S[table-format=2.0]}
\hline
\hline
Name                & $MSE_{H_{l}}$ & $MSE_{I_{\sigma}}$ & $MSE_{H_{u}}$ & $MSE_{I_{2\sigma}}$ \\ \hline

TVL2                &8.105e-07 &\bf{7.933e-07} &2.463e-05 &2.500e-05 \\
PM g=1 k=\num{1e-3} &9.174e-05 &9.331e-05 &4.339e-04 &4.329e-04 \\
PSF                 &1.429e-05 &1.432e-05 &5.246e-05 &5.242e-05 \\
Original            &1.152e-04 &1.151e-04 &4.604e-04 &4.597e-04 \\
Gaussian            &4.578e-05 &4.577e-05 &1.821e-04 &1.818e-04 \\
NL means slow       &1.107e-04 &1.111e-04 &4.603e-04 &4.411e-04 \\
Bilateral           &1.126e-04 &1.127e-04 &4.579e-04 &4.495e-04 \\
TV Chambolle        &6.136e-05 &6.303e-05 &3.401e-04 &3.388e-04 \\
BM3D                &1.140e-04 &1.140e-04 &4.591e-04 &4.548e-04 \\
Starlet             &\bf{7.482e-07} &2.170e-06 &4.874e-05 &7.682e-06 \\
b-UWT(7/9)          &1.048e-06 &1.117e-06 &\bf{1.638e-05} &\bf{3.472e-06} \\
b-UWT(7/9)+Wiener   &1.038e-06 &1.588e-06 &3.805e-05 &5.305e-06 \\
\hline \hline
\end{tabular}
\caption{MSE values for $H_{l}$,$I_{\sigma}$,$H_{u}$,$I_{2\sigma}$ related to the VIS (CL) mirrored crop. See Sect. \ref{subsec:nonstationary} for further details. The best results are in bold}
\label{tab:clmirrored}
\end{table*}

\newpage

\twocolumn
\section{PSF and depth comparison plots}
\begin{figure}[h!]
        \resizebox{\hsize}{!}{\includegraphics{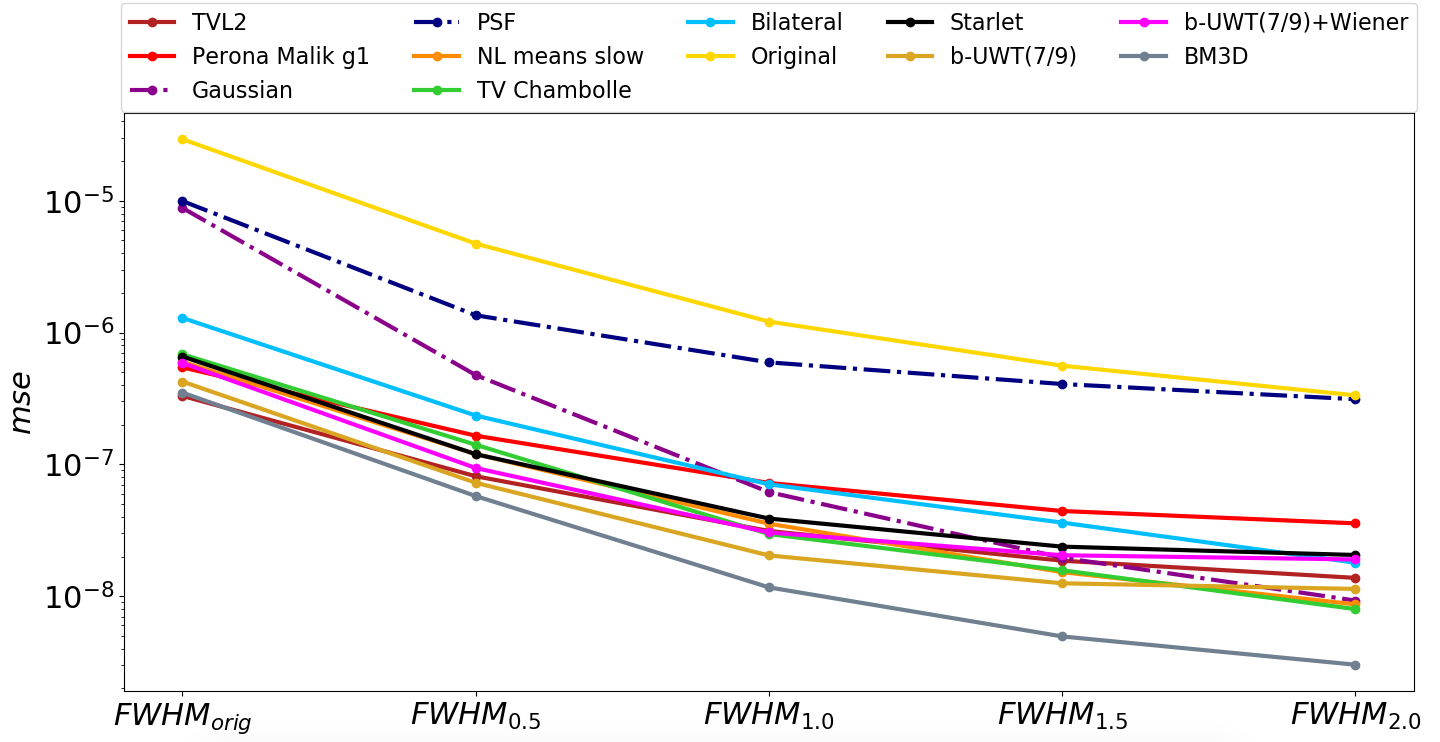}}
        \caption{\label{fig:fwhm_vis} VIS FWHM variation comparison plot. On the x-axis the VIS images with FWHM equal to the original value, 0.5, 1.0, 1.5 and 2.0 arcsecs, whereas on the y-axis $mse$.}
\end{figure}
\begin{figure}[h!]
        \resizebox{\hsize}{!}{\includegraphics{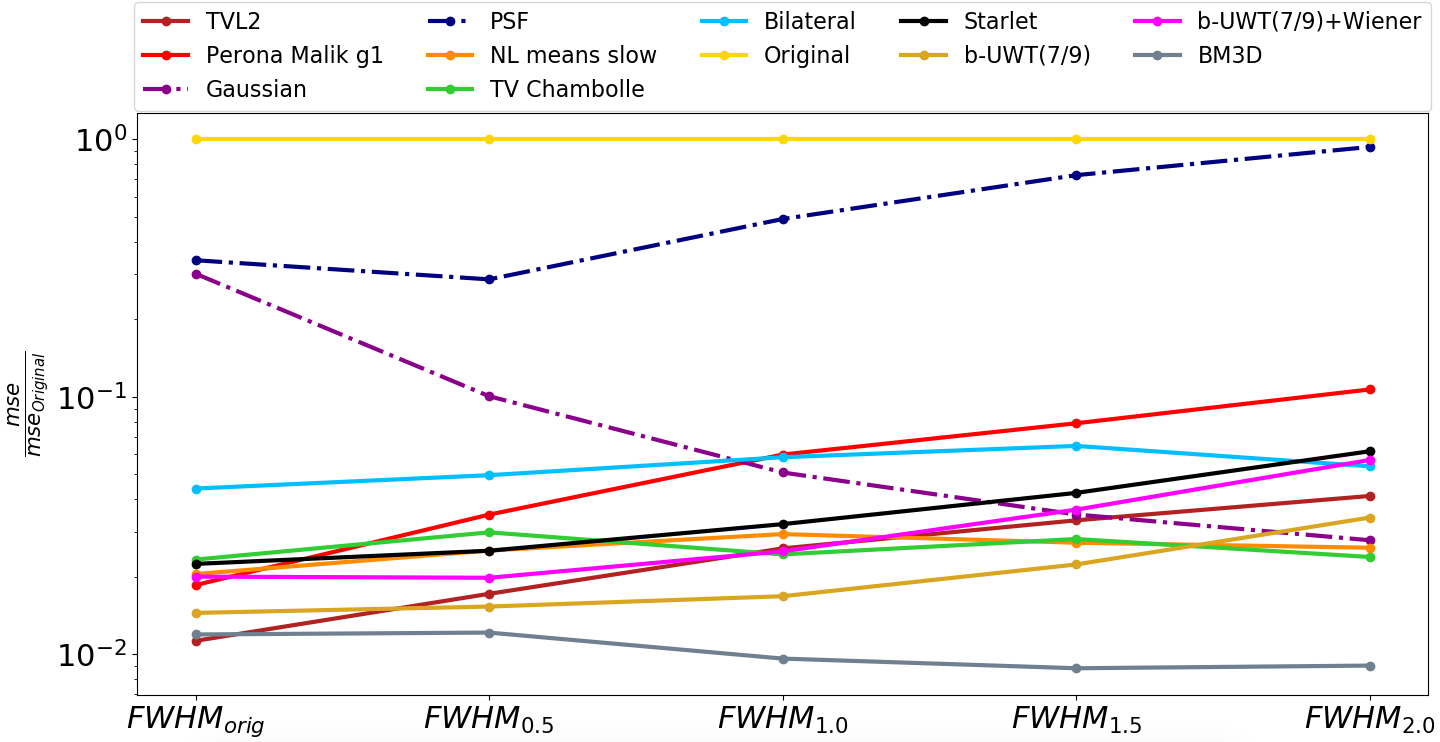}}
        \caption{\label{fig:fwhm_vis_2} VIS FWHM variation comparison plot. On the x-axis the VIS images with FWHM equal to the original value, 0.5, 1.0, 1.5 and 2.0 arcsecs, whereas on the y-axis $\frac{mse}{mse_{Original}}$.}
\end{figure}
\begin{figure}[h!]
        \resizebox{\hsize}{!}{\includegraphics[width=17cm]{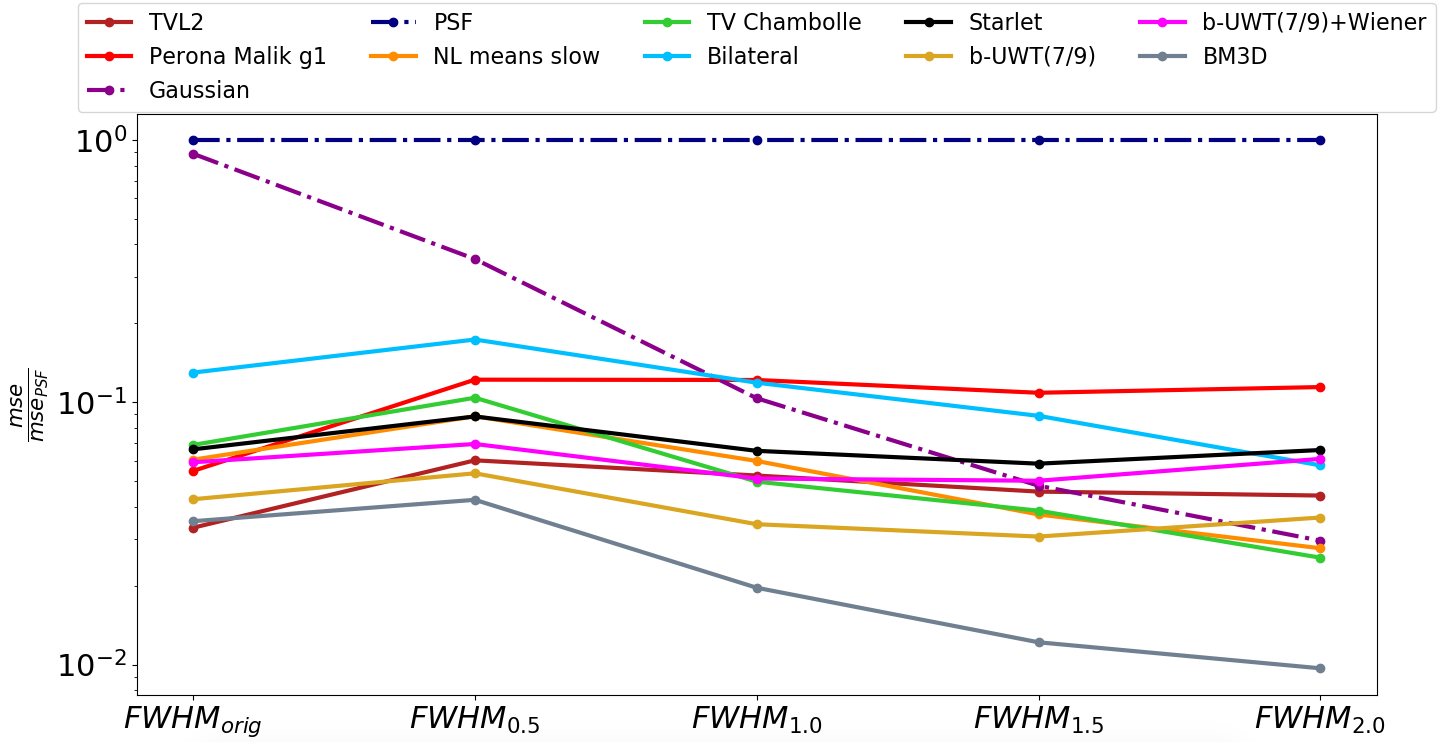}}
        \caption{\label{fig:fwhm_vis_3} VIS FWHM variation comparison plot. On the x-axis the VIS images with FWHM equal to the original value, 0.5, 1.0, 1.5 and 2.0 arcsecs, whereas on the y-axis $\frac{mse}{mse_{PSF}}$.}
\end{figure}
\begin{figure}
        \resizebox{\hsize}{!}{\includegraphics{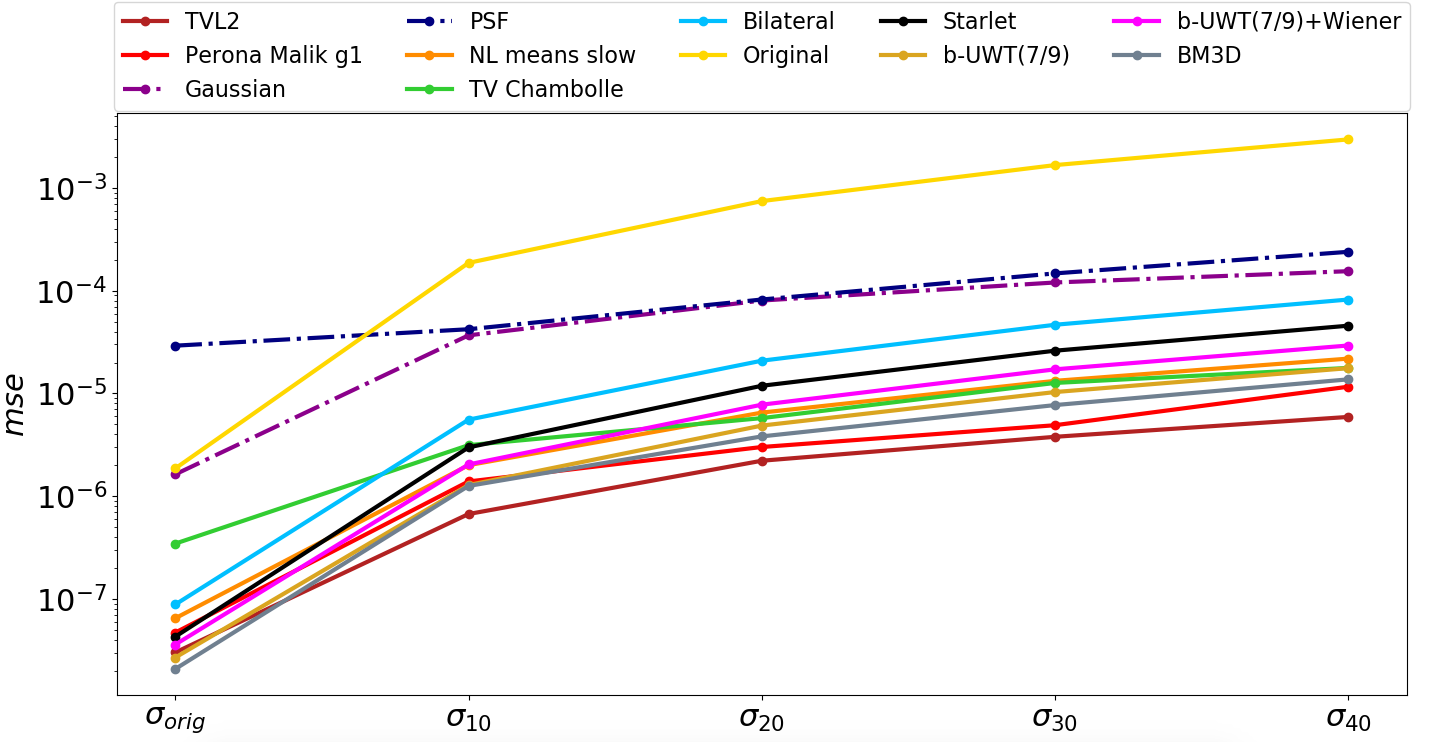}}
        \caption{\label{fig:noised_f160_2} H160 depth variation comparison plot. On the x-axis the H160 images with Gaussian noise standard deviation equal to 1, 10, 20, 30 and 40 times the original value, whereas on the y-axis $mse$.}
\end{figure}
\begin{figure}
        \resizebox{\hsize}{!}{\includegraphics{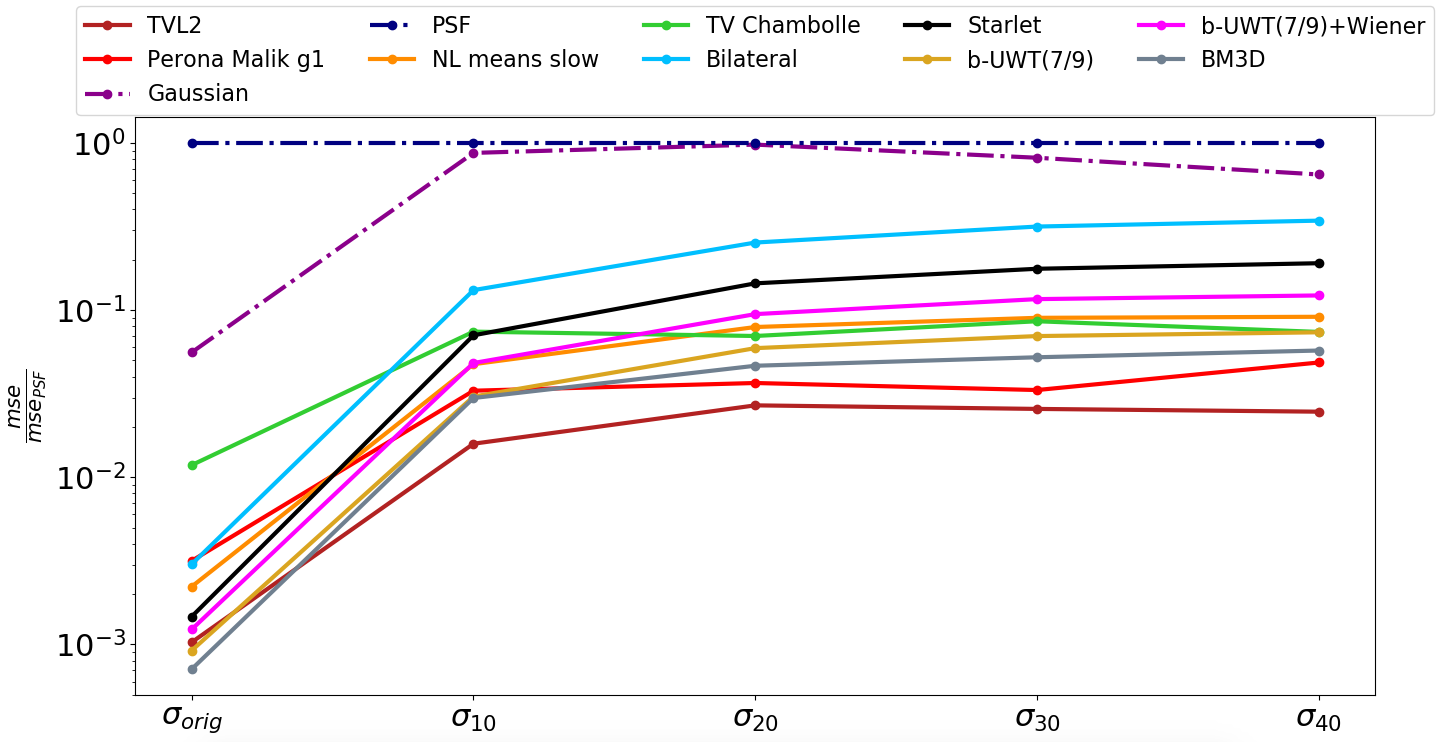}}
        \caption{\label{fig:noised_f160_3} H160 depth variation comparison plot. On the x-axis the H160 images with Gaussian noise standard deviation equal to 1, 10, 20, 30 and 40 times the original value, whereas on the y-axis $\frac{mse}{mse_{PSF}}$.}
\end{figure}

\begin{figure}
        \resizebox{\hsize}{!}{\includegraphics{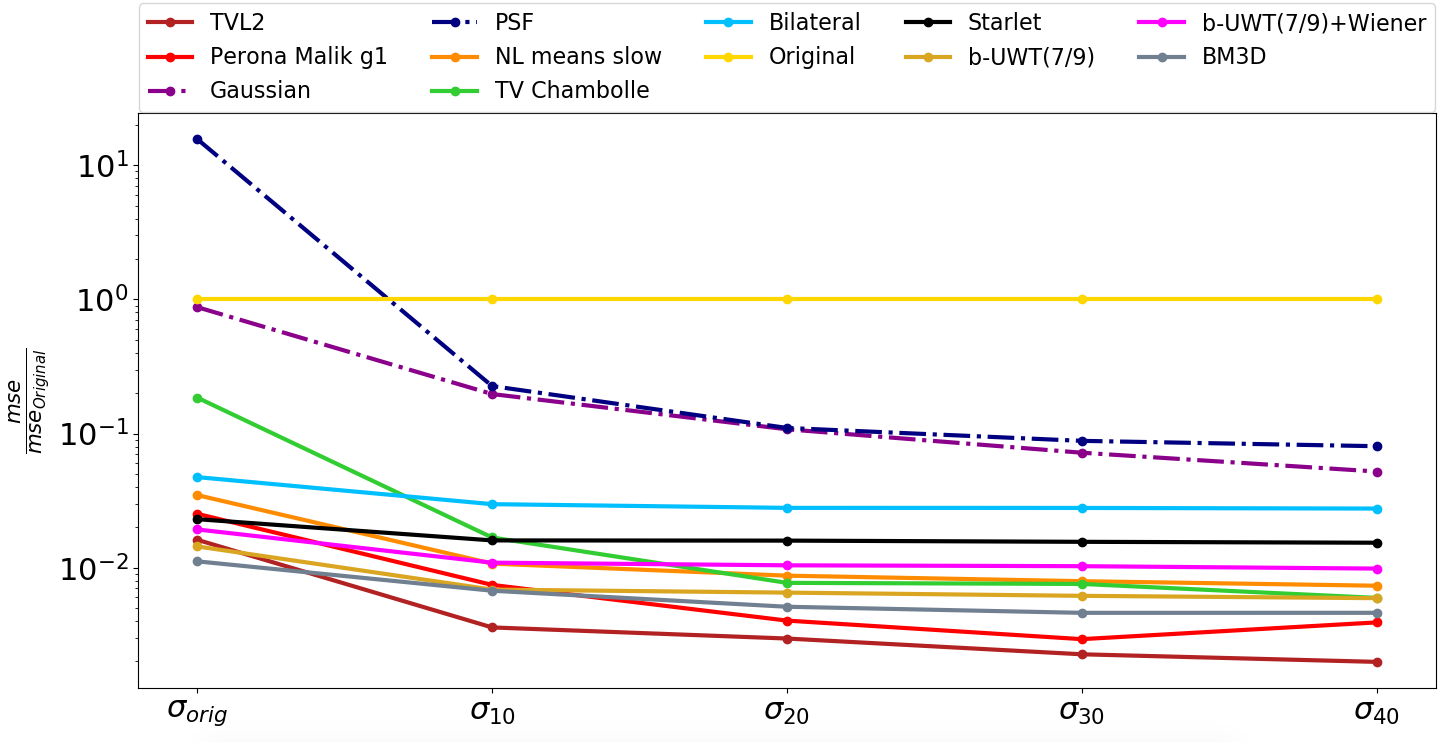}}
        \caption{\label{fig:noised_f160} H160 depth variation comparison plot. On the x-axis the H160 images with Gaussian noise standard deviation equal to 1, 10, 20, 30 and 40 times the original value, whereas on the y-axis $\frac{mse}{mse_{Original}}$.}
\end{figure}
\newpage
\onecolumn
\section{VIS crops visual comparison}
\begin{figure*}[h]
        \centering
        \includegraphics[width=17cm]{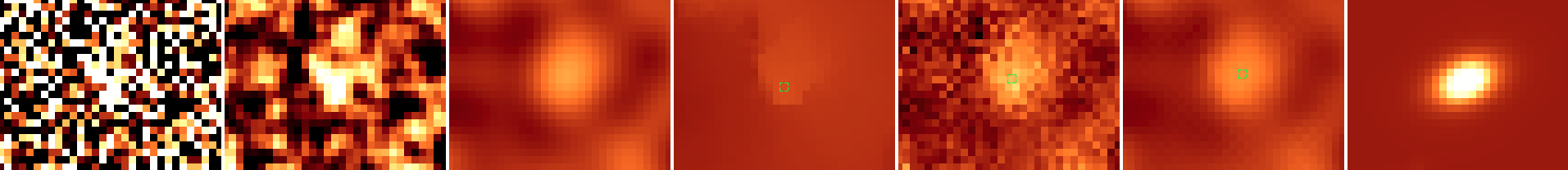}
        \caption{\label{fig:vis121} VIS crops visual comparison: Original, PSF, Perona-Malik, TVL2, Bilateral, TV Chambolle, Noiseless. The green boxes are the detected objects regions. The central object has been detected with a \textit{SNR} of 38.8 with magnitude of 25.79}
\end{figure*}
\begin{figure*}[h]
        \centering
        \includegraphics[width=17cm]{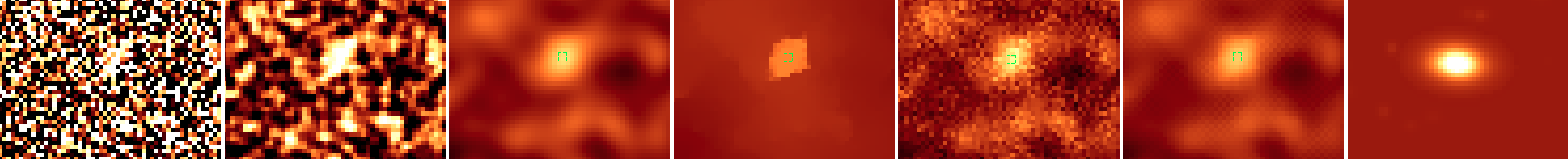}
        \caption{\label{fig:vis144} VIS crops visual comparison: Original, PSF, Perona-Malik, TVL2, Bilateral, TV Chambolle, Noiseless. The green boxes are the detected objects regions. The central object has been detected with a \textit{SNR} of 48.2 with magnitude of 24.76}
\end{figure*}
\begin{figure*}[h]
        \centering
        \includegraphics[width=17cm]{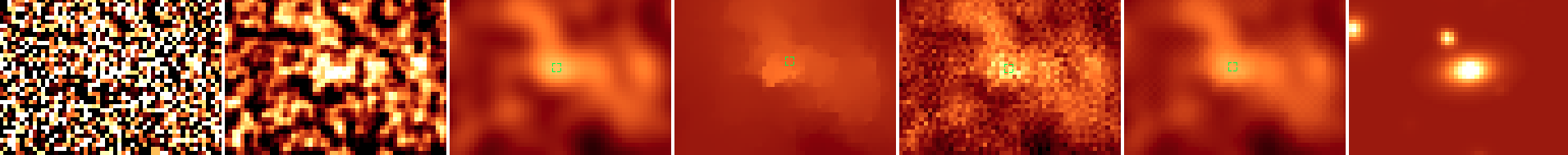}
        \caption{\label{fig:vis1845} VIS crops visual comparison: Original, PSF, Perona-Malik, TVL2, Bilateral, TV Chambolle, Noiseless. The green boxes are the detected objects regions. The central object has been detected with a \textit{SNR} of 72.9 with magnitude of 23.82}
\end{figure*}
\begin{figure*}[h]
        \centering
        \includegraphics[width=17cm]{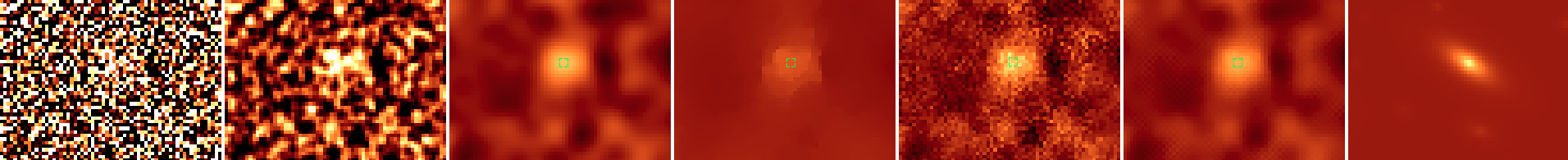}
        \caption{\label{fig:vis2311} VIS crops visual comparison: Original, PSF, Perona-Malik, TVL2, Bilateral, TV Chambolle, Noiseless. The green boxes are the detected objects regions. The central object has been detected with a \textit{SNR} of 47.5 with magnitude of 25.01}
\end{figure*}
\begin{figure*}[h]
        \centering
        \includegraphics[width=17cm]{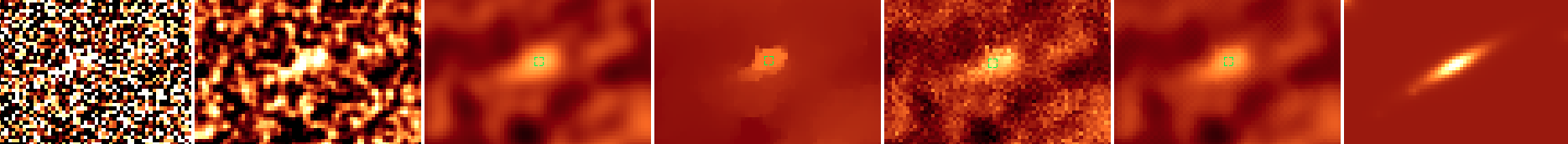}
        \caption{\label{fig:vis2446} VIS crops visual comparison: Original, PSF, Perona-Malik, TVL2, Bilateral, TV Chambolle, Noiseless. The green boxes are the detected objects regions. The central object has been detected with a \textit{SNR} of 35.44 with magnitude of 25.39}
\end{figure*}
\begin{figure*}[h]
        \centering
        \includegraphics[width=17cm]{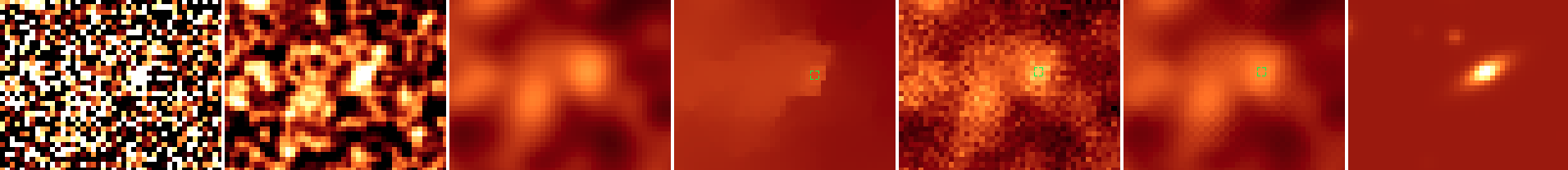}
        \caption{\label{fig:vis2569} VIS crops visual comparison: Original, PSF, Perona-Malik, TVL2, Bilateral, TV Chambolle, Noiseless. The green boxes are the detected objects regions. The central object has been detected with a \textit{SNR} of 21.26 with magnitude of 26.48}
\end{figure*}
\begin{figure*}[h]
        \centering
        \includegraphics[width=17cm]{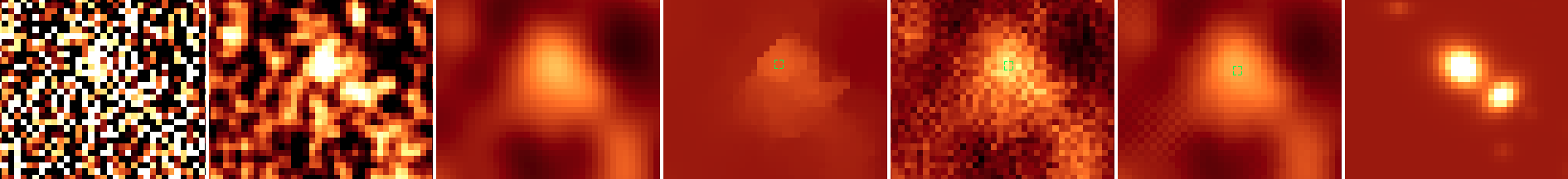}
        \caption{\label{fig:vis3133} VIS crops visual comparison: Original, PSF, Perona-Malik, TVL2, Bilateral, TV Chambolle, Noiseless. The green boxes are the detected objects regions. The central object has been detected with a \textit{SNR} of 27.70 with magnitude of 25.72}
\end{figure*}
\begin{figure*}[h]
        \centering
        \includegraphics[width=17cm]{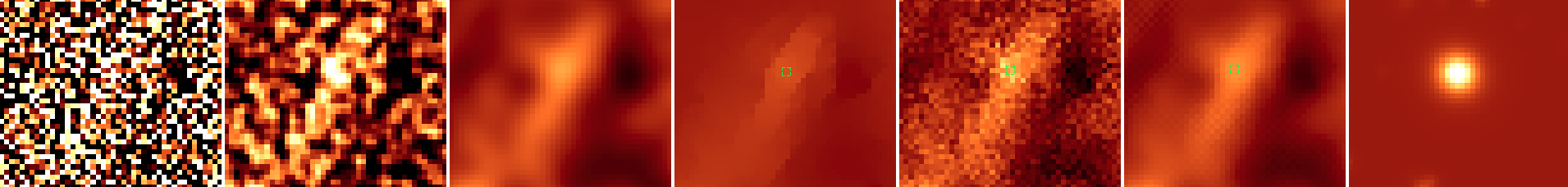}
        \caption{\label{fig:vis3358} VIS crops visual comparison: Original, PSF, Perona-Malik, TVL2, Bilateral, TV Chambolle, Noiseless. The green boxes are the detected objects regions. The central object has been detected with a \textit{SNR} of 45.57 with magnitude of 24.97}
\end{figure*}
\begin{figure*}[h]
        \centering
        \includegraphics[width=17cm]{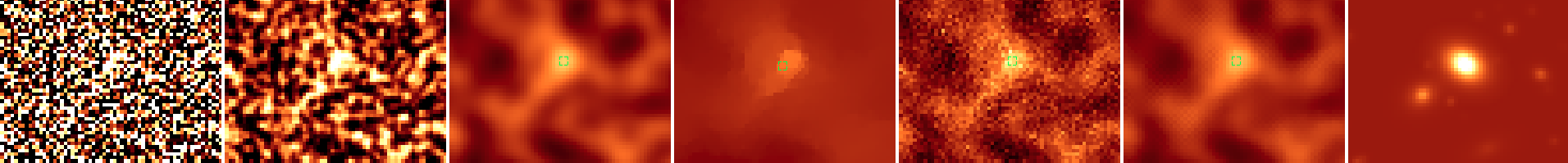}
        \caption{\label{fig:vis379} VIS crops visual comparison: Original, PSF, Perona-Malik, TVL2, Bilateral, TV Chambolle, Noiseless. The green boxes are the detected objects regions. The central object has been detected with a \textit{SNR} of 68.29 with magnitude of 24.14}
\end{figure*}
\begin{figure*}[h]
        \centering
        \includegraphics[width=17cm]{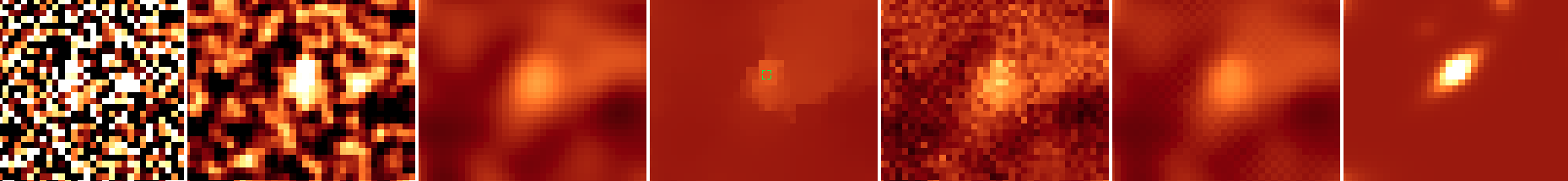}
        \caption{\label{fig:vis3924} VIS crops visual comparison: Original, PSF, Perona-Malik, TVL2, Bilateral, TV Chambolle, Noiseless. The green boxes are the detected objects regions. The central object has been detected with a \textit{SNR} of 26.74 with magnitude of 26.13}
\end{figure*}
\begin{comment}
\begin{figure*}[h]
        \centering
        \includegraphics[width=17cm]{images/vis_4493-106_51-971_24-5565.png}
        \caption{\label{fig:vis4493} VIS crops visual comparison: Original, PSF, Perona-Malik, TVL2, Bilateral, TV Chambolle, Noiseless. The green boxes are the detected objects regions. The central object has been detected with a \textit{SNR} of 51.97 with magnitude of 24.56}
\end{figure*}
\end{comment}
\begin{figure*}[h]
        \centering
        \includegraphics[width=17cm]{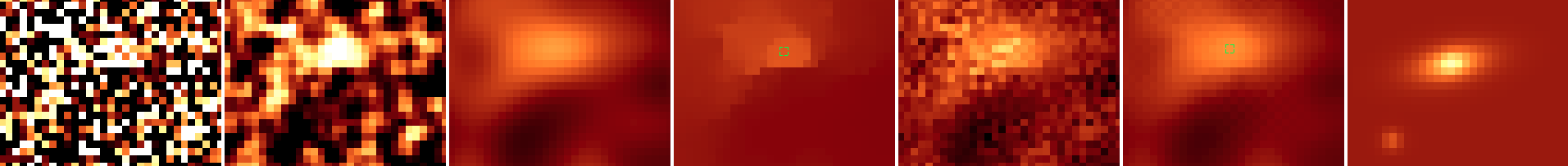}
        \caption{\label{fig:vis4495} VIS crops visual comparison: Original, PSF, Perona-Malik, TVL2, Bilateral, TV Chambolle, Noiseless. The green boxes are the detected objects regions. The central object has been detected with a \textit{SNR} of 25.44 with magnitude of 26.19}
\end{figure*}
\begin{figure*}[h]
        \centering
        \includegraphics[width=17cm]{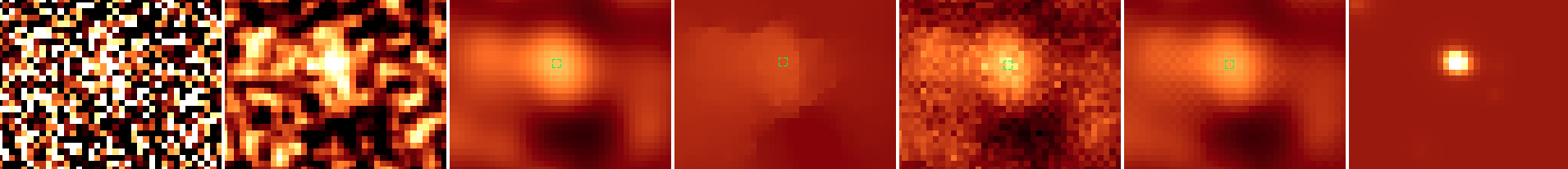}
        \caption{\label{fig:vis4673} VIS crops visual comparison: Original, PSF, Perona-Malik, TVL2, Bilateral, TV Chambolle, Noiseless. The green boxes are the detected objects regions. The central object has been detected with a \textit{SNR} of 36.99 with magnitude of 25.71}
\end{figure*}
\begin{figure*}[h]
        \centering
        \includegraphics[width=17cm]{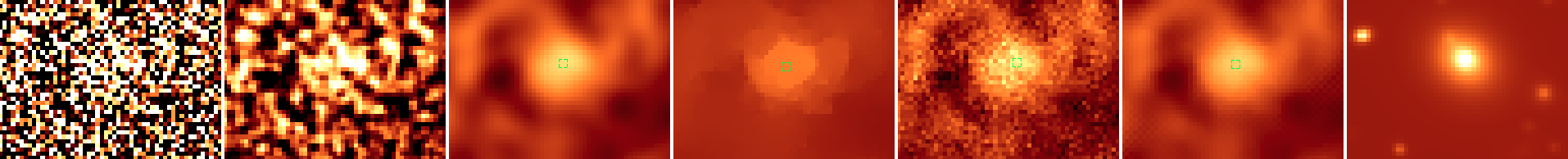}
        \caption{\label{fig:vis4712} VIS crops visual comparison: Original, PSF, Perona-Malik, TVL2, Bilateral, TV Chambolle, Noiseless. The green boxes are the detected objects regions. The central object has been detected with a \textit{SNR} of 90.35 with magnitude of 23.34}
\end{figure*}

\begin{figure*}[h]
        \centering
        \includegraphics[width=17cm]{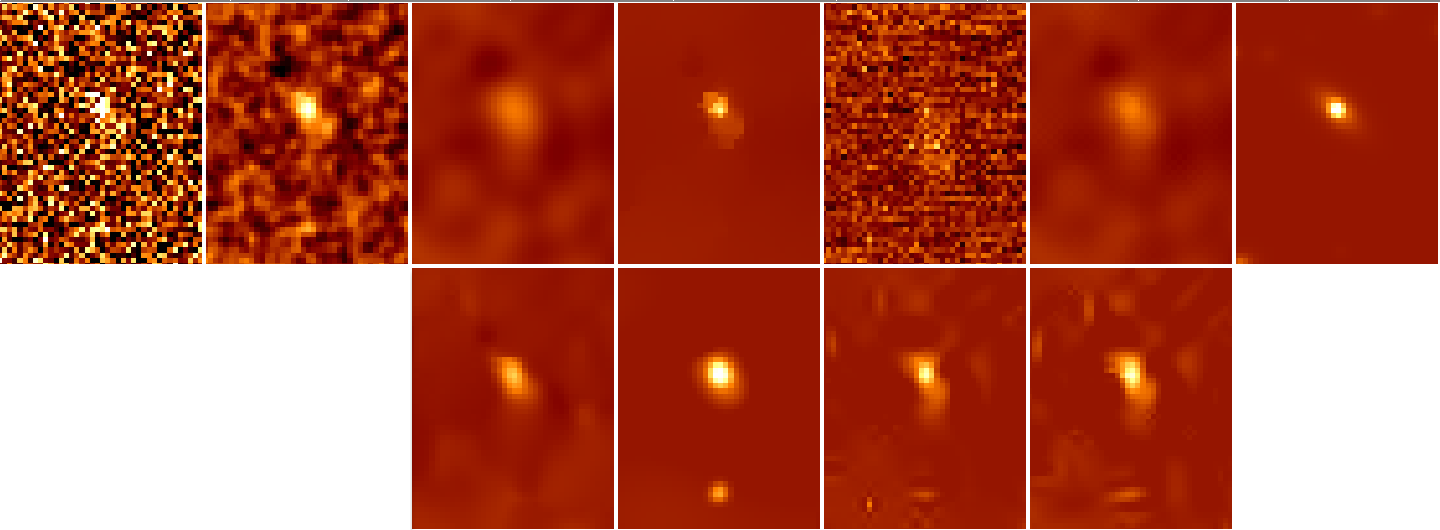}
        \caption{\label{fig:vis153} VIS crops visual comparison. On the first row: Original, PSF, Perona-Malik, TVL2, Bilateral, TV Chambolle, Noiseless; On the second row: BM3D, Starlet, b-UWT(7/9), and b-UWT(7/9)+Wiener. The central object has been detected with a \textit{SNR} of 7.99 with magnitude of 25.21}
\end{figure*}

\begin{figure*}[h]
        \centering
        \includegraphics[width=17cm]{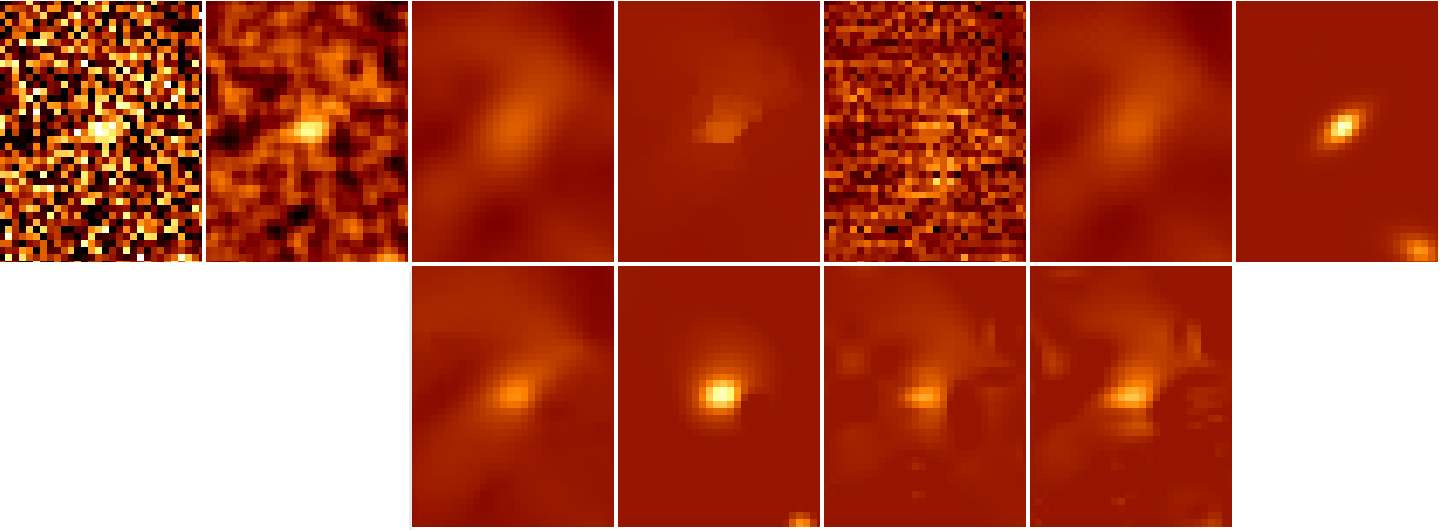}
        \caption{\label{fig:vis2783} VIS crops visual comparison. On the first row: Original, PSF, Perona-Malik, TVL2, Bilateral, TV Chambolle, Noiseless; On the second row: BM3D, Starlet, b-UWT(7/9), and b-UWT(7/9)+Wiener. The central object has been detected with a \textit{SNR} of 3.80 with magnitude of 26.01}
\end{figure*}

\begin{figure*}[h]
        \centering
        \includegraphics[width=17cm]{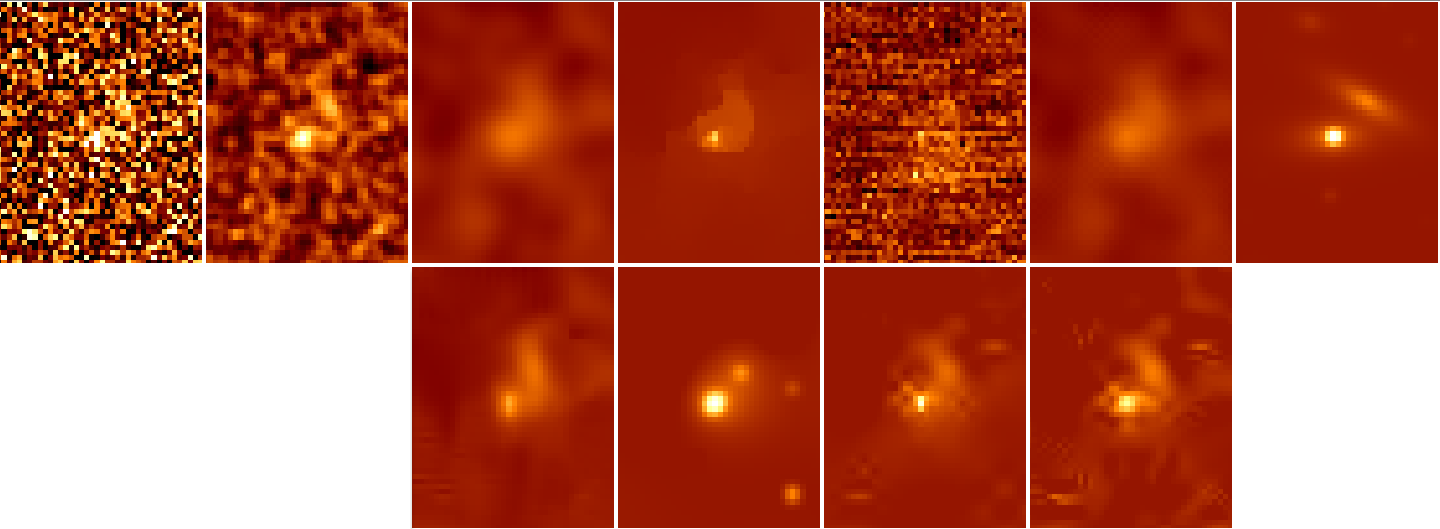}
        \caption{\label{fig:vis3616} VIS crops visual comparison. On the first row: Original, PSF, Perona-Malik, TVL2, Bilateral, TV Chambolle, Noiseless; On the second row: BM3D, Starlet, b-UWT(7/9), and b-UWT(7/9)+Wiener. The central object has been detected with a \textit{SNR} of 2.23 with magnitude of 26.59}
\end{figure*}

\begin{figure*}[h]
        \centering
        \includegraphics[width=17cm]{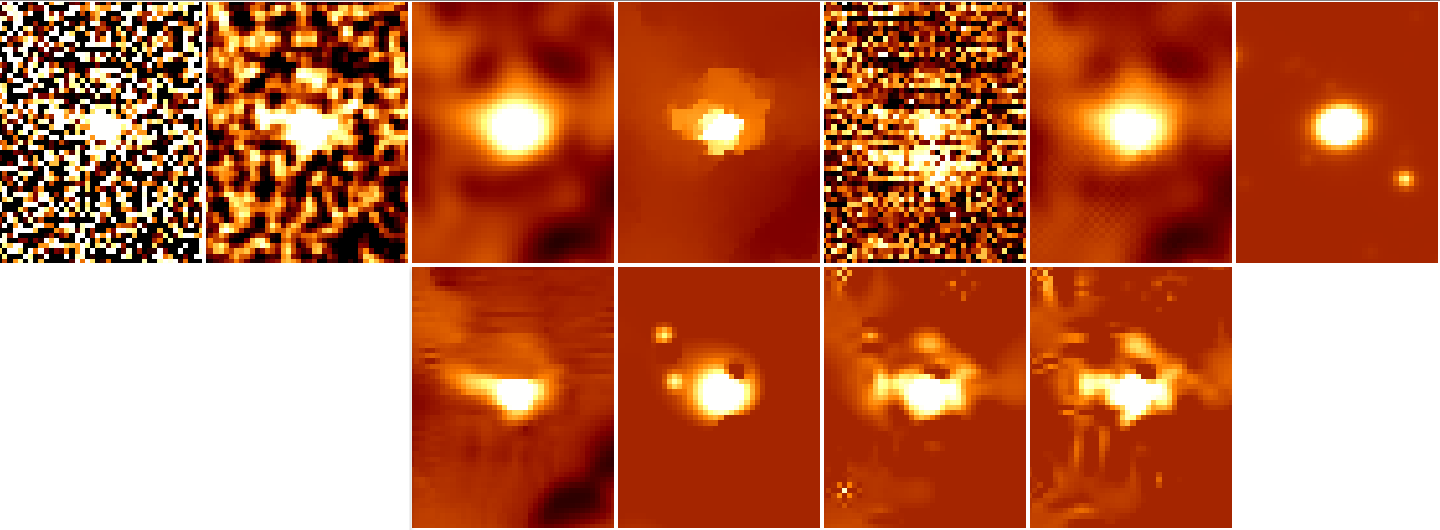}
        \caption{\label{fig:vis402} VIS crops visual comparison. On the first row: Original, PSF, Perona-Malik, TVL2, Bilateral, TV Chambolle, Noiseless; On the second row: BM3D, Starlet, b-UWT(7/9), and b-UWT(7/9)+Wiener. The central object has been detected with a \textit{SNR} of 15.73 with magnitude of 24.48}
\end{figure*}

\begin{figure*}[h]
        \centering
        \includegraphics[width=17cm]{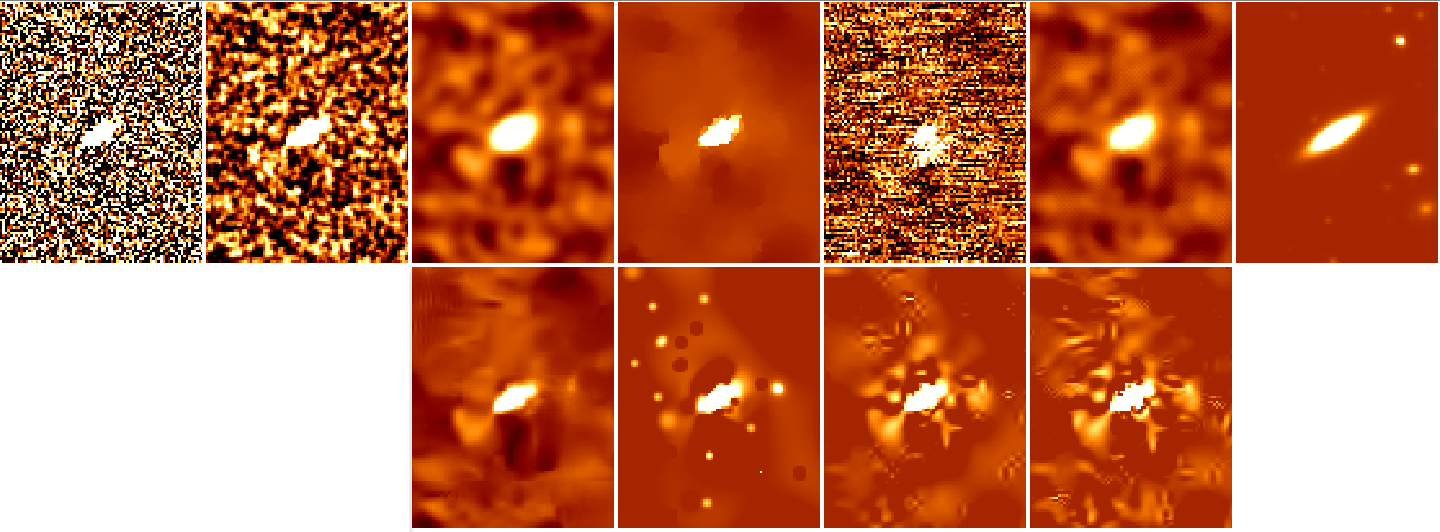}
        \caption{\label{fig:vis799} VIS crops visual comparison. On the first row: Original, PSF, Perona-Malik, TVL2, Bilateral, TV Chambolle, Noiseless; On the second row: BM3D, Starlet, b-UWT(7/9), and b-UWT(7/9)+Wiener. The central object has been detected with a \textit{SNR} of 56.23 with magnitude of 23.09}
\end{figure*}
\nopagebreak
\onecolumn
\nopagebreak
\section{GSDEEP crops visual comparison}
\begin{figure*}[h]
        \centering
        \includegraphics[width=17cm]{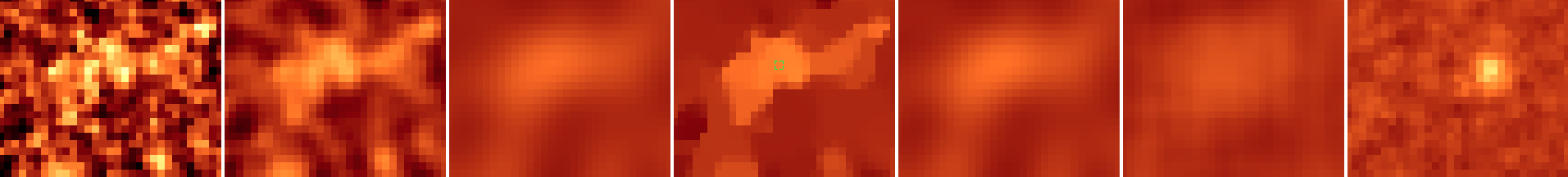}
        \caption{\label{fig:gsdeep1081} GSDEEP crops visual comparison: Original, PSF, Perona-Malik, TVL2, Bilateral, NL means, HUDF09. The green boxes are the detected objects regions. The central object has been detected with a \textit{SNR} of 6.71 with magnitude of 27.47}
\end{figure*}
\begin{figure*}[h]
        \centering
        \includegraphics[width=17cm]{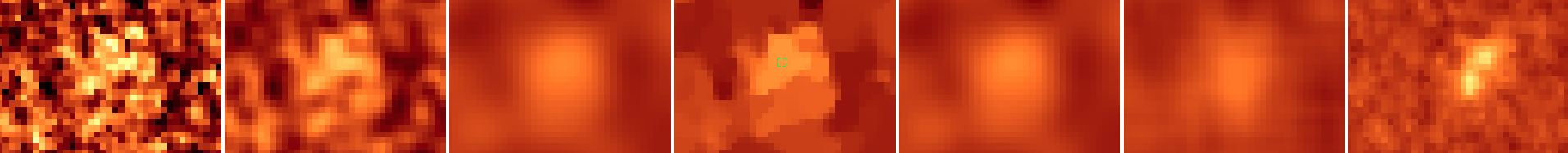}
        \caption{\label{fig:gsdeep1467} GSDEEP crops visual comparison: Original, PSF, Perona-Malik, TVL2, Bilateral, NL means, HUDF09. The green boxes are the detected objects regions. The central object has been detected with a \textit{SNR} of 7.37 with magnitude of 27.20}
\end{figure*}
\begin{figure*}[h]
        \centering
        \includegraphics[width=17cm]{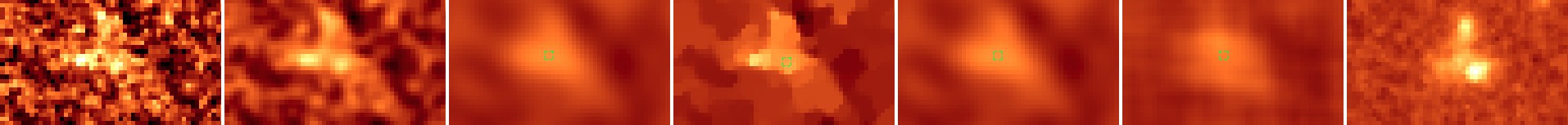}
        \caption{\label{fig:gsdeep1487} GSDEEP crops visual comparison: Original, PSF, Perona-Malik, TVL2, Bilateral, NL means, HUDF09. The green boxes are the detected objects regions. The central object has been detected with a \textit{SNR} of 8.44 with magnitude of 26.80}
\end{figure*}
\begin{figure*}[h]
        \centering
        \includegraphics[width=17cm]{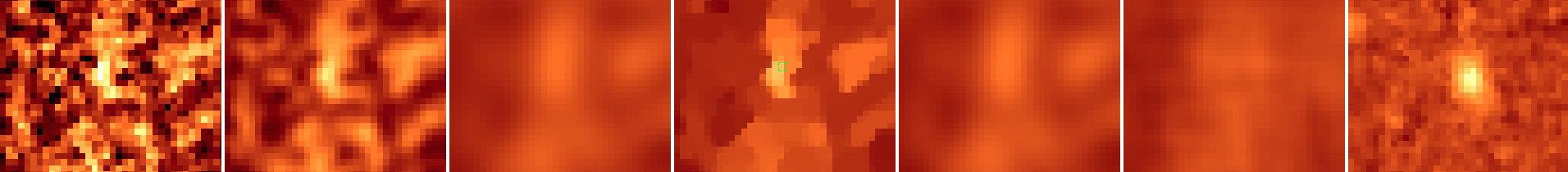}
        \caption{\label{fig:gsdeep1802} GSDEEP crops visual comparison: Original, PSF, Perona-Malik, TVL2, Bilateral, NL means, HUDF09. The green boxes are the detected objects regions. The central object has been detected with a \textit{SNR} of 6.88 with magnitude of 27.18}
\end{figure*}
\begin{figure*}[h]
        \centering
        \includegraphics[width=17cm]{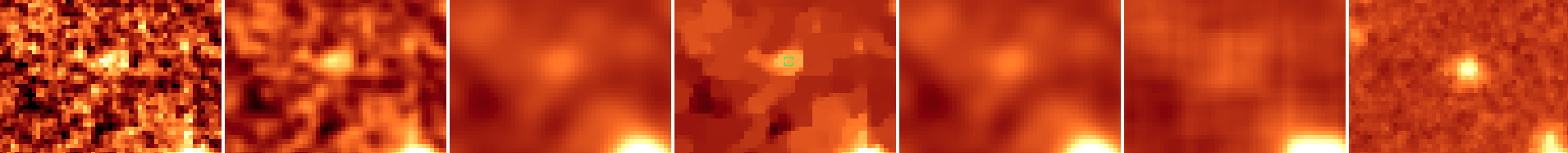}
        \caption{\label{fig:gsdeep2014} GSDEEP crops visual comparison: Original, PSF, Perona-Malik, TVL2, Bilateral, NL means, HUDF09. The green boxes are the detected objects regions. The central object has been detected with a \textit{SNR} of 5.82 with magnitude of 27.48}
\end{figure*}
\begin{figure*}[h]
        \centering
        \includegraphics[width=17cm]{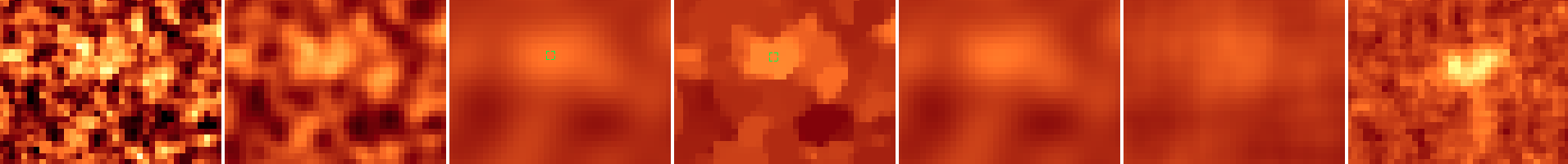}
        \caption{\label{fig:gsdeep2129} GSDEEP crops visual comparison: Original, PSF, Perona-Malik, TVL2, Bilateral, NL means, HUDF09. The green boxes are the detected objects regions. The central object has been detected with a \textit{SNR} of 5.08 with magnitude of 27.48}
\end{figure*}
\begin{figure*}[h]
        \centering
        \includegraphics[width=17cm]{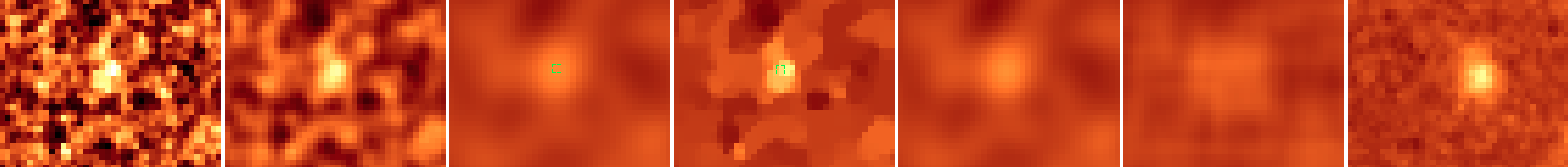}
        \caption{\label{fig:gsdeep509} GSDEEP crops visual comparison: Original, PSF, Perona-Malik, TVL2, Bilateral, NL means, HUDF09. The green boxes are the detected objects regions. The central object has been detected with a \textit{SNR} of 6.51 with magnitude of 27.58}
\end{figure*}
\begin{figure*}[h]
        \centering
        \includegraphics[width=17cm]{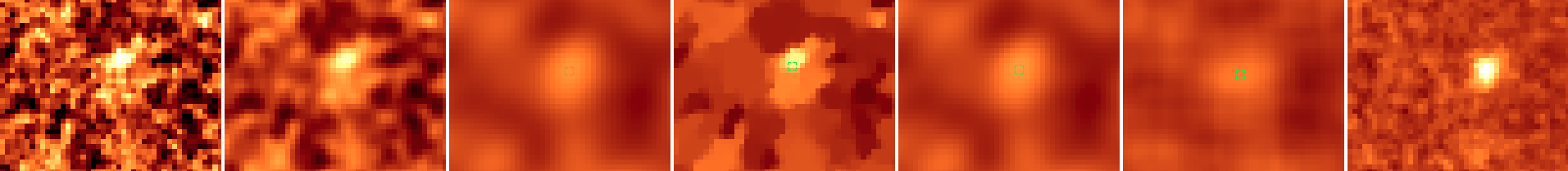}
        \caption{\label{fig:gsdeep73} GSDEEP crops visual comparison: Original, PSF, Perona-Malik, TVL2, Bilateral, NL means, HUDF09. The green boxes are the detected objects regions. The central object has been detected with a \textit{SNR} of 12.48 with magnitude of 27.17}
\end{figure*}
\begin{figure*}
        \centering
        \includegraphics[width=17cm]{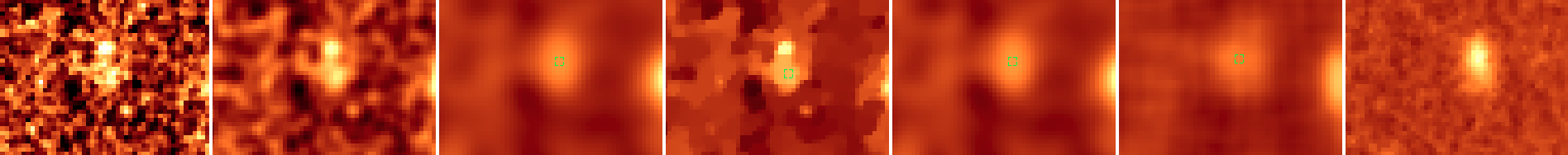}
        \caption{\label{fig:gsdeep909} GSDEEP crops visual comparison: Original, PSF, Perona-Malik, TVL2, Bilateral, NL means, HUDF09. The green boxes are the detected objects regions. The central object has been detected with a \textit{SNR} of 9.77 with magnitude of 27.36}
\end{figure*}

\begin{figure*}
        \centering
        \includegraphics[width=17cm]{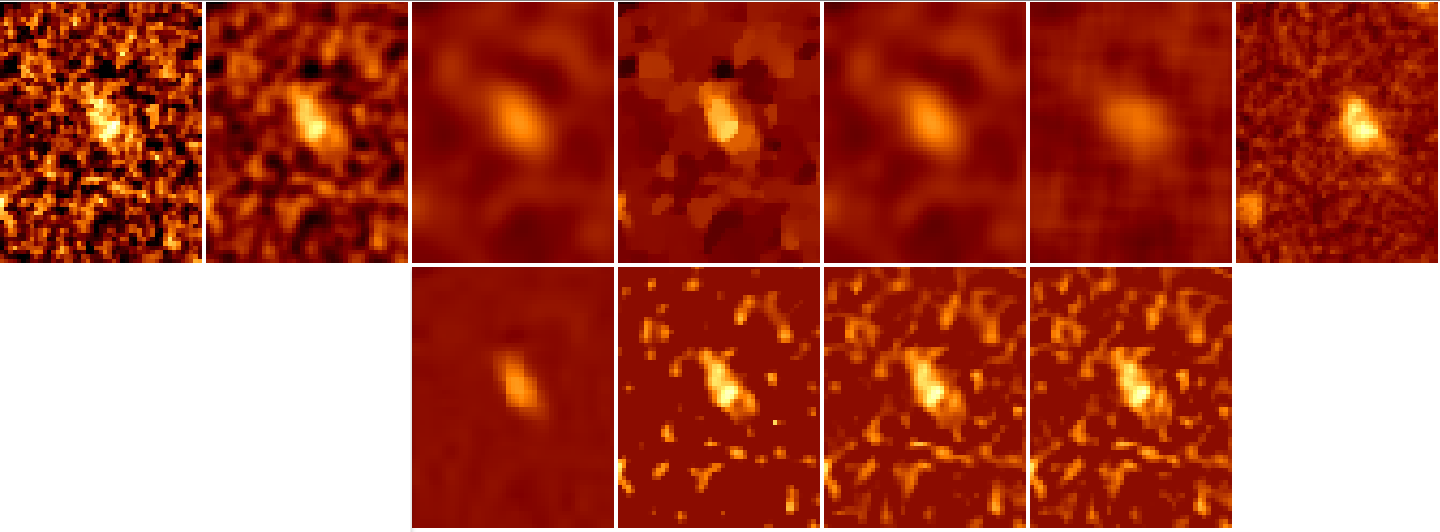}
        \caption{\label{fig:gsdeep100} GSDEEP crops visual comparison. On the first row: Original, PSF, Perona-Malik, TVL2, Bilateral, NL means, HUDF09. On the second row: BM3D, Starlet, b-UWT(7/9), and b-UWT(7/9)+Wiener. The central object has been detected with a \textit{SNR} of 11.98 with magnitude of 28.34}
\end{figure*}

\begin{figure*}
        \centering
        \includegraphics[width=17cm]{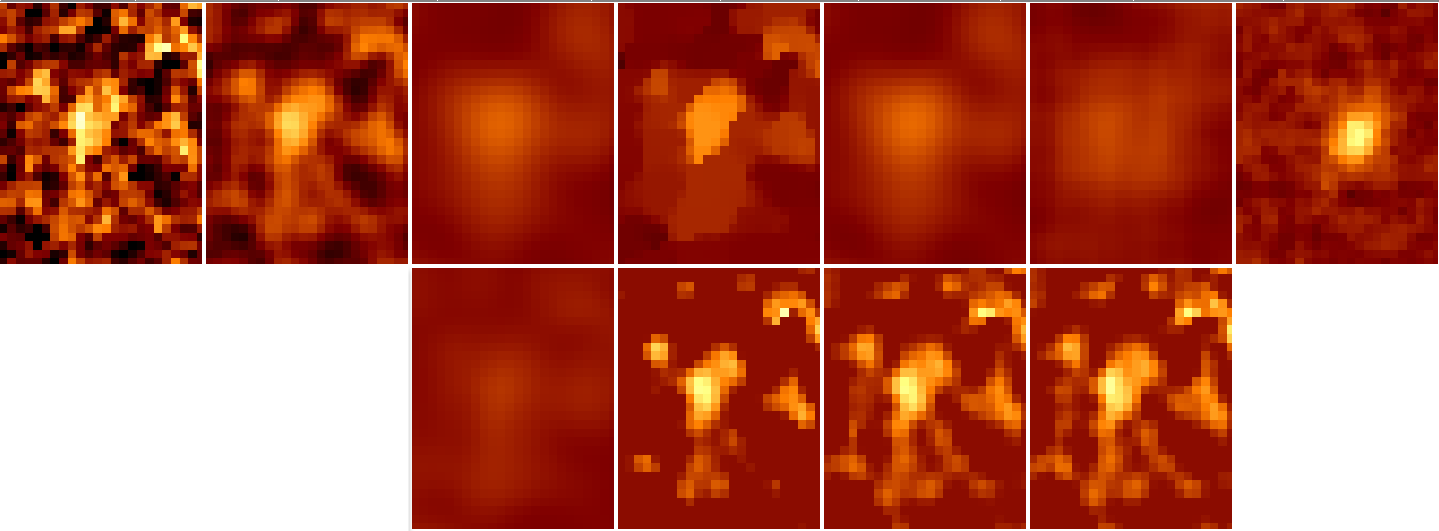}
        \caption{\label{fig:gsdeep1014} GSDEEP crops visual comparison. On the first row: Original, PSF, Perona-Malik, TVL2, Bilateral, NL means, HUDF09. On the second row: BM3D, Starlet, b-UWT(7/9), and b-UWT(7/9)+Wiener. The central object has been detected with a \textit{SNR} of 9.85 with magnitude of 28.45}
\end{figure*}

\begin{figure*}
        \centering
        \includegraphics[width=17cm]{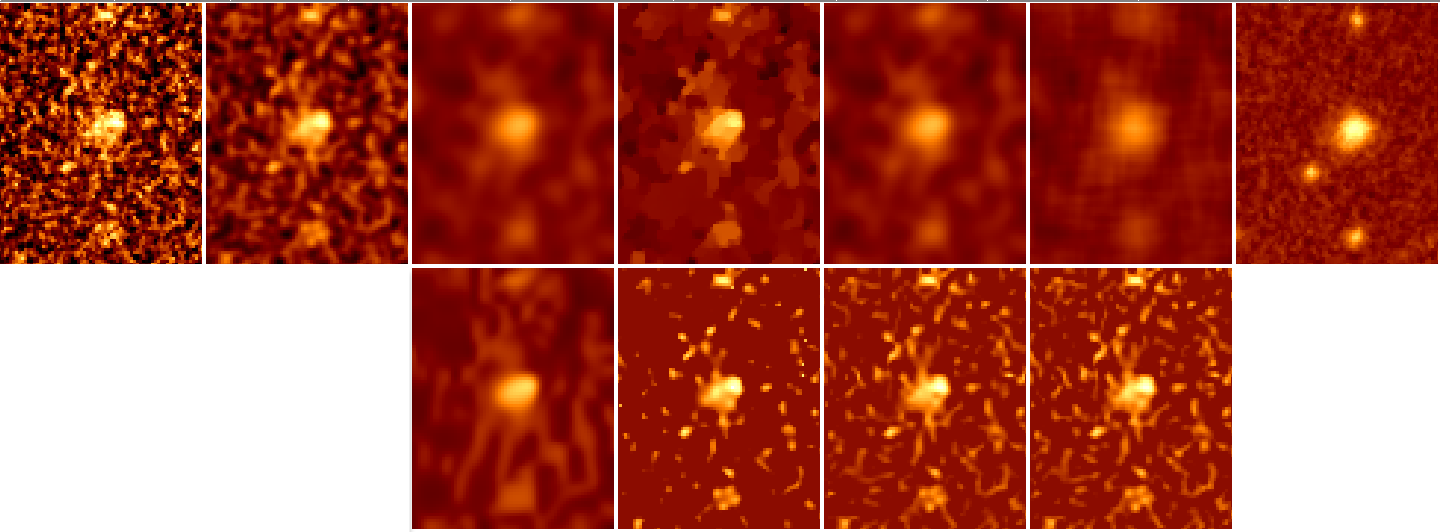}
        \caption{\label{fig:gsdeep1772} GSDEEP crops visual comparison. On the first row: Original, PSF, Perona-Malik, TVL2, Bilateral, NL means, HUDF09. On the second row: BM3D, Starlet, b-UWT(7/9), and b-UWT(7/9)+Wiener. The central object has been detected with a \textit{SNR} of 11.26 with magnitude of 28.37}
\end{figure*}

\begin{figure*}
        \centering
        \includegraphics[width=17cm]{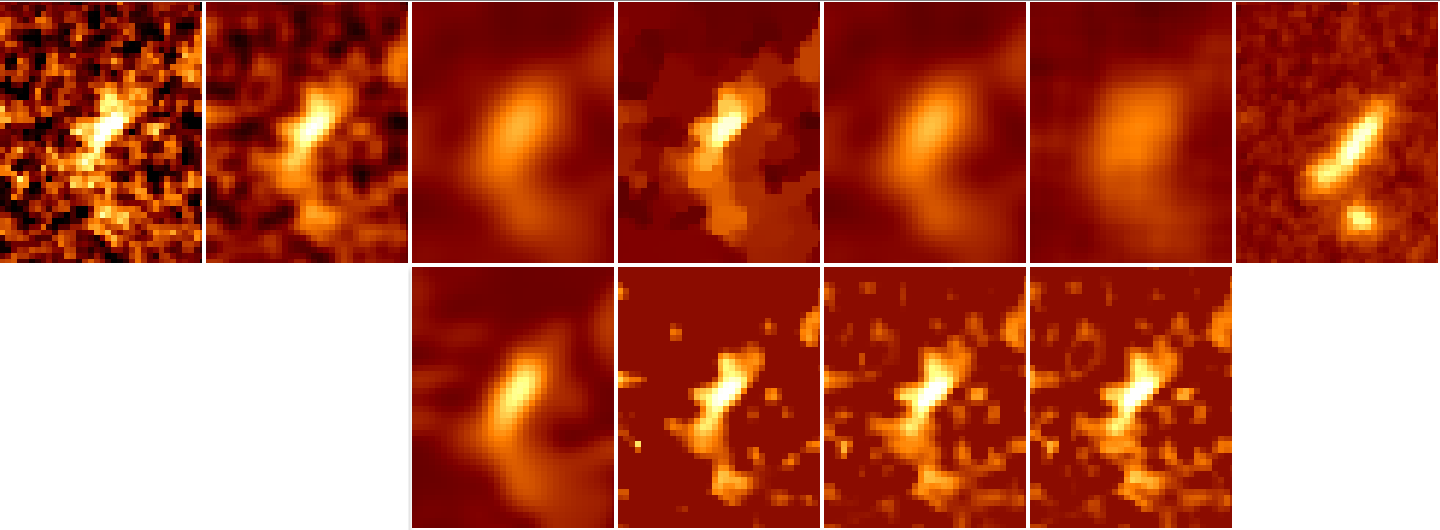}
        \caption{\label{fig:gsdeep2051} GSDEEP crops visual comparison. On the first row: Original, PSF, Perona-Malik, TVL2, Bilateral, NL means, HUDF09. On the second row: BM3D, Starlet, b-UWT(7/9), and b-UWT(7/9)+Wiener. The central object has been detected with a \textit{SNR} of 17.70 with magnitude of 27.86}
\end{figure*}

\begin{figure*}
        \centering
        \includegraphics[width=17cm]{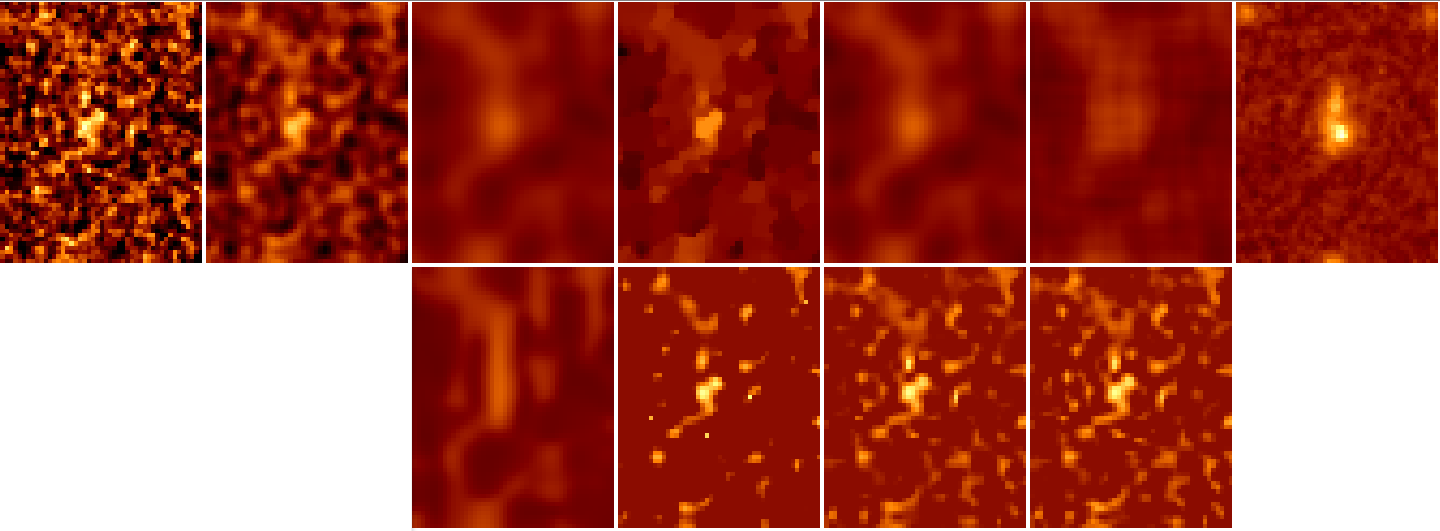}
        \caption{\label{fig:gsdeep607} GSDEEP crops visual comparison. On the first row: Original, PSF, Perona-Malik, TVL2, Bilateral, NL means, HUDF09. On the second row: BM3D, Starlet, b-UWT(7/9), and b-UWT(7/9)+Wiener. The central object has been detected with a \textit{SNR} of 10.40 with magnitude of 28.48}
\end{figure*}

\end{document}